\documentclass[aps,prd,eqsecnum,11pt,showpacs,preprintnumbers,nofootinbib]{revtex4-1}
\pdfoutput=1
\def\nbox#1#2{\vcenter{\hrule \hbox{\vrule height#2in
\kern#1in \vrule} \hrule}}
\def\sq{\,\raise.5pt\hbox{$\nbox{.09}{.09}$}\,}
\def\sqb{\,\raise.5pt\hbox{$\overline{\nbox{.09}{.09}}$}\,}

\newcommand{\bea}{\begin{eqnarray}}
\newcommand{\eea}{\end{eqnarray}}
\newcommand{\be}{\begin{equation}}
\newcommand{\ee}{\end{equation}}
\newcommand{\bes}{\begin{subequations}}
\newcommand{\ees}{\end{subequations}}
\def\lag{\langle}
\def\rag{\rangle}

\def\nn{\nonumber \\}

\usepackage{amssymb,amsmath}
\usepackage{graphicx}
\usepackage{graphics}

\begin{document}

\preprint{LA-UR-11-10115}
\preprint{CERN-PH-TH/2011-057}

\title{\hspace{1cm}  \\ \Large
Conformal Invariance, Dark Energy, and CMB Non-Gaussianity\vspace{2cm}}

\author{Ignatios Antoniadis\footnote{On leave from CPHT (UMR CNRS 7644) 
Ecole Polytechnique, 91128 Palaiseau Cedex, France}}
\affiliation{Department of Physics\\
CERN, Theory Division\\
CH-1211 Geneva 23, Switzerland\vspace{1cm}}
\email{ignatios.antoniadis@cern.ch}

\author{Pawel O. Mazur}
\affiliation{Department of Physics and Astronomy,\\
University of South Carolina\\ Columbia SC 29208 USA\vspace{1cm}}
\email{mazur@physics.sc.edu,\ mazurmeister@gmail.com}

\author{Emil Mottola}
\affiliation{Theoretical Division, MS B285\\
Los Alamos National Laboratory\\ Los Alamos, NM 87545 USA}
\email{emil@lanl.gov}

%\pacs{\ 04.62.+v,\ 95.36.+x,\ 98.80.Qc}

\maketitle

\vfill\eject

\topmargin -1.5cm
\textheight 9in
\textwidth 6.8in

\centerline{\vspace{2cm}{\bf Abstract}}
In addition to simple scale invariance, a universe dominated by dark energy naturally gives rise 
to correlation functions possessing full conformal invariance. This is due to the mathematical 
isomorphism between the conformal group of certain three dimensional slices of de Sitter space 
and the de Sitter isometry group $SO(4,1)$. In the standard homogeneous, isotropic cosmological 
model in which primordial density perturbations are generated during a long vacuum energy 
dominated de Sitter phase, the embedding of flat spatial ${\mathbb R}^3$ sections in de Sitter 
space induces a conformal invariant perturbation spectrum and definite prediction for the shape 
of the non-Gaussian CMB bispectrum. In the case in which the density fluctuations are generated 
instead on the de Sitter horizon, conformal invariance of the ${\mathbb S}^2$ horizon embedding 
implies a different but also quite definite prediction for the angular correlations of CMB non-Gaussianity 
on the sky. Each of these forms for the bispectrum is intrinsic to the symmetries of de Sitter space, 
and in that sense, independent of specific model assumptions. Each is different from the predictions 
of single field slow roll inflation models, which rely on the breaking of de Sitter invariance. We 
propose a quantum origin for the CMB fluctuations in the scalar gravitational sector from the 
conformal anomaly that could give rise to these non-Gaussianities without a slow roll inflaton field, 
and argue that conformal invariance also leads to the expectation for the relation $n_{_S} - 1 = n_{_T}$ 
between the spectral indices of the scalar and tensor power spectrum. Confirmation of this prediction 
or detection of non-Gaussian correlations in the CMB of one of the bispectral shape functions predicted 
by conformal invariance can be used both to establish the physical origins of primordial density fluctuations, 
and distinguish between different dynamical models of cosmological vacuum dark energy. 
\vfil\eject

\tableofcontents

\section{Introduction}
\label{Sec:Intro}

The Cosmic Microwave Background (CMB) provides an essentially unique window on 
the universe at very great distances from our local neighborhood, or equivalently at 
very early times before the present epoch. The fact that the CMB exists at all, with a high 
degree of isotropy and a thermal spectrum, is evidence that the primordial universe was 
to high accuracy at some point in a nearly uniform state of thermal equilibrium, and therefore 
in causal contact at a time and place prior to the last scattering of the CMB photons. The small
but measurable anisotropy in the CMB presents the most compelling clues to the possible
quantum origin of the universe, as well as the source of the complex large scale structure of
matter we observe today. Determining the statistical properties of CMB anisotropies, and in
particular, non-Gaussian primordial correlations, offers the possibility of discriminating
between different cosmological models of the very early and very distant universe
\cite{nongauss, Komatsu}. 

In parallel to the CMB measurements by COBE \cite{COBE}, WMAP \cite{WMAP} and the Planck 
satellite \cite{Planck},  the discovery of cosmological dark energy at the present epoch \cite {SNI} 
emphasizes the need for a more fundamental understanding of vacuum energy and the 
cosmological `constant.'  A comprehensive theory of dark energy may have consequences 
also for the origin of the CMB anisotropy. In this paper we explore the possible relation between
cosmological dark energy and CMB properties. In order to investigate  this connection in a way 
which is as free as possible from dynamical assumptions, we follow an essentially kinematical 
approach, deriving the form of the non-Gaussian correlations of CMB anisotropies directly 
from the inherent symmetry group $SO(4,1)$ of a dark energy dominated de Sitter universe. 
In particular, we show that embedding the three dimensional surface on which the fluctuations 
giving rise to the CMB  anisotropies originate in four dimensional de Sitter spacetime leads 
naturally to conformal invariance and determines the angular shape of the three-point CMB 
bispectrum, which is different for different embeddings, but otherwise independent of particular 
dynamical assumptions.

The standard cosmological model is based on the assumption of a globally spatially homogeneous 
and isotropic universe with uniform expansion. This is described by a 
Friedmann-Lema\^itre-Robertson-Walker (FLRW) line element with flat spatial sections,
\be
ds^2 = - d\tau^2 + a^2\, d\vec x \cdot d\vec x
\label{flatRW}
\ee
in which the FLRW scale factor $a(\tau)$ obeys the Einstein-Friedmann eq.,
\be
H^2 \equiv \frac{1}{a^2}\left(\frac{d a}{d\tau}\right)^2 = \frac{8 \pi G \rho}{3}
\label{FRW}
\ee
in cosmological comoving time $\tau$. The total energy density of the universe 
$\rho(\tau)$ and pressure $p(\tau)$ of the cosmological fluid are also assumed 
spatially homogeneous and isotropic on average at large scales
and satisfy the covariant conservation eq. for a perfect fluid, 
\be
\frac{d\rho}{d\tau} + 3 H (p + \rho) = 0\,,
\label{cons}
\ee
which assumes no entropy production or viscosity. In the FLRW cosmological model all that 
is left then to specify is the eq. of state relationship between $p_i$ and $\rho_i$ (with $p = \sum_i p_i, \ 
\rho = \sum_i \rho_i$) for the various components of the cosmological fluid at different epochs 
of the universe's expansion. CMB anisotropies are treated as small ($\sim 10^{-5}$) linear 
perturbations away from this exactly homogeneous and isotropic background, whose effect 
on the mean geometry is negligible and may be neglected at early times.

If at some epoch the energy density and pressure are dominated by a
cosmological constant term $\Lambda > 0$, for which 
\be
\rho_{\Lambda} =  \frac{\Lambda}{8\pi G} = - p_{\Lambda} > 0
\label{constLamb}
\ee
is a constant,  then it follows from (\ref{FRW}) and (\ref{cons}) that  
the expansion rate $H = \sqrt {\Lambda/3}$ is a also constant, and the scale factor,
\be
a(\tau) = e^{H\tau}
\label{expand}
\ee
describes an exponentially expanding phase. The line element (\ref{flatRW}) with 
(\ref{expand}) then is a particular coordinatization of de Sitter spacetime, the
maximally symmetric solution of Einstein's equations with a positive cosmological 
term. 

This exponential inflationary de Sitter expansion takes regions that were 
in close causal contact initially and expands them to very great distances apart, 
potentially accounting for the thermal spectrum and high degree of isotropy observed 
in the CMB emitted from different directions in the sky. Primordial quantum fluctuations 
in this de Sitter inflationary epoch are held to be responsible for the observed small 
anisotropy in the CMB, and naturally gives rise to the approximately scale invariant 
CMB power spectrum consistent with the COBE and WMAP data \cite{COBE,WMAP}. 
This agreement is one of the principal arguments in favor of inflationary models, broadly 
defined as the hypothesis that the universe passed through a de Sitter phase early in 
its history, during which an effective and nearly constant cosmological constant energy
density dominated all other forms of energy, and the cosmological expansion was 
exponential according to (\ref{expand}) for at least several dozen $e$-foldings \cite{LidL}.

Although this simple picture is quite appealing on general grounds, a fundamental 
theory for this exponential expansion, the physical origin of the effective cosmological 
constant term $\Lambda$ associated with the energy density of the `vacuum' itself, 
its value today, and vacuum energy generally is lacking. In its place a number of different 
phenomenological models of inflation have been suggested \cite{LidL}. 
The simplest of these models postulates a single scalar field $\phi$ (called the 
``inflaton") which has a non-negative potential energy function $V(\phi)$. If $\phi (\tau)$ 
and $V(\phi (\tau))$ are nearly constant for a long enough period of time, $V$ acts as 
a nearly constant $\rho = V/8\pi G = -p$ source in (\ref{FRW}) which drives the 
exponential expansion (\ref{expand}) for many $e$-foldings. Eventually the inflaton 
field is supposed to roll down its potential, preserving spatial homogeneity and
isotropy in the large, and settle to a value $\phi_0$ with $V(\phi_0) = 0$
(or nearly zero) at the minimum of the potential. Although the mean value of the 
inflaton field is assumed to be spatially homogeneous and isotropic, small
spatially inhomogeneous fluctuations in this field couple (linearly) to gravity. During the 
inflationary epoch these generate fluctuations in the metric geometry of space and 
the local gravitational potential that eventually induce corresponding spatiotemporal 
fluctuations in the local temperature of the cosmological fluid, in particular at the 
epoch of matter-radiation decoupling. In this picture these primordial fluctuations of 
quantum origin are imprinted on the radiation at the time of last scattering and lead 
to the small angular anisotropies observed in the CMB today \cite{LidL,MukFelBra}.

In this class of single field ``slow roll" inflaton models the primordial CMB 
fluctuations have a (nearly) scale invariant Harrison-Zel'dovich spectrum \cite{HZ}
and obey (nearly) Gaussian statistics, with small corrections coming from the
slow roll dynamics itself, which are probably too small to be observed even by
Planck \cite{BKMRio}. It follows that any detection of non-Gaussian three-point 
correlations on the microwave sky of primordial origin will immediately falsify the 
simplest single field slow roll inflation model, and provide an important observational 
constraint on the physical origins of the primordial fluctuations themselves.

Several years ago the present authors suggested that the approximate scale 
invariance of the CMB power spectrum might be a hint of a more general conformal 
invariance property of the primordial fluctuations, that doesn't rely on any slow
roll inflaton field(s) or similar specific dynamical models of inflation \cite{sky}. Since 
scale and conformal invariance arises in other branches of physics, such as turbulent 
flows and critical phenomena, with universal behavior independent of the particular 
dynamical details or short distance interactions of the system, the approximate scale 
invariance of the primordial CMB power spectrum may be connected to rather 
different and universal physics than the slow roll hypothesis assumes in the simplest 
models of inflation.  We suggested that the conformal invariance hypothesis could be 
tested by detection of non-Gaussian statistics in the CMB and in particular,
the measurement of the angular correlations in the three-point bispectrum.
The generic assumption of conformal invariance fixes the form of the bispectrum 
without any reference to specialized models of inflation, provided only that a conformally 
invariant phase of the universe's spacetime history existed during which the CMB 
fluctuations were generated. Although the overall magnitude of the CMB correlation 
functions cannot be determined by the conformal invariance hypothesis, and requires 
more specific dynamical information about the origin of the fluctuations, the {\it shape} 
of the non-Gaussian correlations is highly constrained by the essentially {\it kinematic} 
requirements of conformal symmetry alone, and hence can act as a powerful probe 
of a possible conformal phase of the universe.

In this paper we will take this idea one step further, demonstrating that full conformal 
invariance and conformal field theory (CFT) behavior is in fact a natural consequence 
of the isometries of de Sitter spacetime, which is generated by a period of vacuum 
cosmological dark energy dominance. Specifically the de Sitter group $SO(4,1)$ is 
isomorphic to the group of conformal transformations of certain three dimensional 
foliations (slicings) of de Sitter space. Moreover, the several mathematical possibilities 
of different slicings imply different realizations of conformal symmetry and determine 
different shapes of the CMB bispectrum, which can be distinguished observationally. 
Since these different realizations correspond to quite different physical origins of
CMB anisotropies, and are related to the dominance of a de Sitter phase
in different cosmological models of vacuum energy, there is the intriguing 
possibility of elucidating the physics of cosmological dark energy by future 
measurement of the angular form of CMB non-Gaussian correlations.

The first way that conformal invariance can arise is by a long period of
vacuum energy dominance in the usual FLRW coordinates (\ref{flatRW}) with
flat spatial sections. The limiting behavior of de Sitter invariant correlation 
functions in the asymptotic region $H\tau \gg 1$, after many de Sitter $e$-foldings,
stretches all distance scales by the exponential expansion factor (\ref{expand}) 
and only the terms which fall off most slowly with distance survive. In this
limit the correlation functions of any anisotropic density fluctuations exhibit
conformal behavior \cite{astph}. This is the mathematical basis for one form
of dS/CFT correspondence \cite{Starus,MazMot,Strom,Skend}, by which we mean
conformal behavior of fields and Green's functions in $d$ dimensional
(bulk) de Sitter space at its $d-1$ dimensional boundary $I_+$ at spacelike infinity,
{\it c.f.} Fig. \ref{Fig:dSCarPen}. The late time limit of three dimensional spacelike 
sections of (\ref{flatRW}) embedded in four dimensional de Sitter spacetime leads 
to simple conformal behavior with a single conformal weight, and a  specific form 
of the non-Gaussian  bispectrum given by eqs. (\ref{G3corr})-(\ref{Zdef}) below. 

No particular dynamical mechanism of the origin of the fluctuations needs to be 
specified, and in that sense this prediction (\ref{G3corr}) of the bispectral shape 
is model independent, following from purely kinematic embedding of the conformal 
group of ${\mathbb R}^3$ in the de Sitter group. It is necessary only that the 
fluctuations arose at a constant FLRW cosmic time $\tau$ late in the de Sitter expansion 
(so that any effects of non-de Sitter invariant initial conditions are suppressed), 
and that the fluctuations are {\it intrinsic} to de Sitter space, respecting all of its built-in 
symmetries. This is a quite different assumption than that made in current inflationary 
models based on slow roll {\it out} of de Sitter space, and as we discuss in detail
in Sec. 3, eqs. (\ref{G3corr})-(\ref{Zdef}) give a {\it different form} of the CMB angular 
bispectrum than that predicted in single field slow roll models. Thus any detection of 
the non-Gaussian CMB bispectrum and determination of its angular dependence 
(regardless of its overall amplitude) has the ability to distinguish between an intrinsic 
origin of the primordial fluctuations of the kind we discuss in this paper, and specific 
phenomenological dynamical models of their origin, such as that of the slow roll 
scenario or its variants.

Freeing the origin of primordial fluctuations from a particular class of models
based on a (so far unobserved) scalar inflaton field also allows for quite
different global cosmologies than strictly homogeneous, isotropic form 
(\ref{flatRW}). A second and quite distinct way that conformal invariance 
is mathematically realized in de Sitter spacetime is through the 
presence of a cosmological horizon. In inflationary models the primordial
fluctuations leave the de Sitter horizon of the inflationary epoch and 
re-enter the cosmological Hubble sphere much later in the history of the universe.
However no direct local observation of the CMB can exclude the possibility
that the fluctuations arose on or near the cosmological horizon itself,
and that the global geometry of the universe outside our cosmological
horizon is quite different than that described by the FLRW line element 
(\ref{flatRW}). The fact that cosmological dark energy is some $72$\% of 
the energy density of the {\it present} universe is a powerful reminder 
of the lack of a basic understanding of vacuum energy, much less a 
complete theory of its dynamical role in cosmology. It is quite possible 
that a more fundamental understanding of vacuum dark energy will require 
a radical revision of the basic assumptions of cosmology {\it in toto}, including
global spatial homogeneity and isotropy, and that the CMB anisotropies may 
have arisen in a quite different way from a nearly de Sitter invariant conformal 
phase possibly related with the presence of cosmological dark energy at more 
recent epochs.

In this paper we shall also describe an alternate embedding of the conformal group of 
directions in the sky (the celestial sphere ${\mathbb S}^2$), appropriate if the fluctuations 
are generated near the de Sitter cosmological horizon itself, rather than on fixed ${\mathbb R}^3$ 
sections. In this case the bulk space is the three dimensional space of constant {\it static}
time $t$ in (\ref{dSstat}), and ${\mathbb S}^2$ is its two dimensional horizon boundary.
This slicing leads to yet a third, non-Gaussian bispectral form (\ref{G3sph}),
different than either that determined by a slow roll model or conformal invariance 
considerations based on (\ref{flatRW}) and ${\mathbb R}^3$. If the third form of the 
non-Gaussian bispectrum is observed in the CMB, it will lead to quite different cosmological 
consequences, possibly related to a more fundamental understanding of the spacetime
dependent dynamics of vacuum energy associated with the cosmological horizon, and 
the residual cosmological dark energy observed in the present epoch. Thus the 
observation of a non-Gaussian bispectrum by WMAP or Planck of a definite angular 
form has the potential to distinguish between quite different global geometries of the 
universe and quite different loci and physical mechanisms for the origin of CMB 
anisotropies, both testing the homogeneity and isotropy hypothesis of the standard 
FLRW cosmology, and potentially leading to dynamical models for cosmological
dark energy on the largest scales.

The paper is organized as follows. In the next section we review the basic kinematics
and isometries of de Sitter space, and the isomorphism between generators of
the de Sitter isometries and the conformal transformations of ${\mathbb R}^3$,
which is the mathematical basis for dS/CFT correspondence. In Section 
\ref{Sec:Bispectrum} we derive the conformal Ward identities that follow from 
this correspondence and use them to determine the form of the three-point 
non-Gaussian bispectrum for scalar fluctuations at a fixed FLRW time $\tau$. 
In Section \ref{Sec:GravPert} we give the general solution for the gauge invariant 
gravitational potentials of the linearized Einstein eqs. around de Sitter space
in terms of the stress energy tensor perturbations, necessary to connect the
density perturbations to the observable CMB anisotropies.
In Section \ref{Sec:Static} we consider the static time slicing of de Sitter space 
and the induced conformal invariance of correlation functions on the cosmological
${\mathbb S}^2$ horizon. The conformal Ward identities of the spherical
horizon lead to a completely different form again of the non-Gaussian bispectrum
in angular coordinates on the sky. In Sec. \ref{Sec:Horizon} we discuss a possible origin 
for this ${\mathbb S}^2$ non-Gaussian conformal correlation functions from the 
fluctuations of the cosmological horizon modes derived from the trace anomaly
of any massless quantum matter fields in curved space, and show that they can give
rise to the scale invariant Harrison-Zel'dovich power spectrum. In Sec. \ref{Sec:GravWaves}
we extend these considerations of conformal invariance to the spectrum of
transverse, traceless spin-$2$ fields, {\it i.e.} gravitational waves, which will
influence CMB polarization measurements. We conclude in Sec. \ref{Sec:DarkEnergy} 
with a summary of our main results and a discussion of the possibility of detection of 
CMB non-Gaussianities becoming a discriminating probe of the physics of dark energy 
and the large scale geometry of the universe.

For the convenience of the reader we have collected in the seven Appendices a 
number of known and some less well known mathematical properties of de Sitter 
spacetime used in the text. These are (A) Geometry, Coordinates and dS/CFT Correspondence 
of de Sitter Space; (B) Killing Vectors of de Sitter Space in Flat FLRW and (C) Static Coordinates;
(D) Invariant Distance and Correlation Functions; (E) The $SO(3,1)$ Conformal Group on 
${\mathbb S}^2$; (F) Exact Formulae for the Bispectral Shape 
Function (\ref{bispectrum}); and (G) Differential Operators in de Sitter Space.
Some of these results have been reported previously in Refs. \cite{sky,MazMot,MazMor,Moriond}.

\section{Conformal Invariance of Flat Spatial Sections in de Sitter Space}
\label{Sec:Flat}

The essential kinematical feature of a vacuum dark energy dominated de Sitter universe
is that the {\it conformal} group of certain embeddings of {\it three} dimensional
hypersurfaces in de Sitter spacetime may be mapped (either one-to-one or 
multiple-to-one) to the geometric {\it isometry} group of the full {\it four} dimensional
spacetime into which the hypersurfaces are embedded \cite{Dirac}. The first 
example of such an embedding is that of flat Euclidean ${\mathbb R}^3$ in 
de Sitter spacetime in coordinates (\ref{flatRW}). The conformal group of the 
three dimensional spatial ${\mathbb R}^3$ sections is in fact identical 
(isomorphic) to the isometry group $SO(4,1)$ of the four 
dimensional de Sitter spacetime, as we now review.

Since (eternal) de Sitter space is maximally symmetric, it posseses the maximum 
number of isometries for a spacetime in $d=4$ dimensions, namely $\frac{d(d+1)}{2} =10$, 
corresponding to the $10$ solutions of the Killing equation,
\be
\nabla_{\mu} K^{(\alpha)}_{\nu} + \nabla_{\nu} K^{(\alpha)}_{\mu} = 0\,, 
\qquad \mu,\nu = 0, 1, 2, 3;
\quad \alpha = 1, \dots , 10.
\label{Kil}
\ee
Each of the $10$ linearly independent solutions to this eq. (labelled by $\alpha$) is
a vector field in de Sitter space corresponding to an infinitesimal coordinate transformation, 
$x^{\mu} \rightarrow x^{\mu} + K^{\mu}(x)$ that leaves the de Sitter geometry and line element 
(\ref{flatRW}) with (\ref{expand}) invariant. These are the $10$ generators of the de Sitter 
isometry group, the non-compact Lie group $SO(4,1)$. The geometry and frequently
used coordinates of de Sitter space are reviewed in Appendix \ref{App:Geometry}.

The isomorphism with conformal transformations of ${\mathbb R}^3$ is that each of these $10$ 
solutions of (\ref{Kil}) may be placed in one-to-one correspondence with the $10$ solutions 
of the {\it conformal} Killing eq. of three dimensional flat space ${\mathbb R}^3$, {\it i.e.}
\be
\partial_i\xi^{(\alpha)}_j + \partial_j\xi^{(\alpha)}_i 
= \frac{2}{3}\, \delta_{ij}\, \partial_k \xi^{(\alpha)}_k
\,, \qquad i,j, k = 1, 2 ,3;\quad \alpha = 1, \dots 10
\label{confKilflat}
\ee
In (\ref{Kil}) the spacetime indices $\mu, \nu$ range over $4$ values and $\nabla_{\nu}$ 
is the covariant derivative with respect to the full $4$ dimensional metric of de Sitter 
spacetime, whereas in (\ref{confKilflat}), $i,j$ are $3$ dimensional spatial indices of
the three Cartesian coordinates $x^i$ of Euclidean ${\mathbb R}^3$ of one
dimension lower with flat metric $\delta_{ij}$. Solutions to the conformal Killing 
eq. (\ref{confKilflat}) are transformations of $x^i \rightarrow x^i + \xi^i(\vec x)$ 
which preserve angles in ${\mathbb R}^3$. This isomorphism between geometric
isometries of $3 + 1$ dimensional de Sitter spacetime and conformal transformations 
of $3$ dimensional flat space embedded in it is the origin of conformal invariance of 
correlation functions generated in a de Sitter phase of the universe. 

The $10$ solutions of (\ref{confKilflat}) for vector fields in flat ${\mathbb R}^3$ are easily
found. They are of two kinds. First there are $6$ solutions of (\ref{confKilflat}) with 
$\partial_k \xi_k = 0$, corresponding to the strict isometries of ${\mathbb R}^3$, namely 
$3$ translations and $3$ rotations. Second, there are also $4$ solutions of (\ref{confKilflat}) 
with $\partial_k \xi_k \neq  0$. These are the $4$ conformal transformations of flat
space that are not strict isometries but preserve all angles. They consist of $1$ global 
dilation and $3$ special conformal transformations. The explicit form of these solutions 
and demonstration of their one-to-one correspondence with the solutions of (\ref{Kil}) in 
de Sitter spacetime is given in Appendix \ref{App:KillingFlat}. Here we focus on 
deriving the consequences for the CMB.

The Killing eq. (\ref{Kil}) for de Sitter space in the flat FLRW coordinates 
(\ref{flatRW}) becomes the set of eqs. 
\bes
\bea
\partial_{\tau}K_{\tau} &=& 0\,,\\
\partial_{\tau} K_i + \partial_iK_{\tau} &=& 2H K_i \,,\\
\partial_iK_j + \partial_jK_i &=& 2Ha^2 \delta_{ij} K_{\tau}\,.
\label{cKv}
\eea
\label{KilldS}\ees\noindent 
The global dilational Killing solution $K_{\mu}^{(D)}$ of eqs. (\ref{KilldS}) is particularly easy to grasp. 
By fixing an arbitrary normalization constant, we may write this solution in the form
\bes
\bea
K_{\tau}^{(D)} = H^{-1}\qquad &{\rm or}& \qquad K^{\tau\, (D)} = - H^{-1}\,,\\
K_i^{(D)} = a^2\, \xi_i^{(D)} = a^2\,x_i \qquad &{\rm or}& \qquad K^{i\, (D)} = x^i\,,
\label{dil}
\eea
\label{Kildil}
\ees
which satisfies (\ref{KilldS}). This is the infinitesimal form of the finite dilational symmetry,
\bes
\bea
&&\vec x \rightarrow \lambda \vec x\,,\label{spatdil}\\
&& a(\tau) \rightarrow \lambda^{-1} a(\tau)\,,\label{ascale}\\
&&\tau \rightarrow \tau - H^{-1}\ln \lambda \label{taushift}
\eea
\label{dilxa}\ees
which clearly leaves the metric of de Sitter spacetime (\ref{flatRW}) with (\ref{expand}) invariant. 
Considering only a fixed time spatial slice, (\ref{dil}) or (\ref{spatdil}) shows that this is a 
transformation of ${\mathbb R}^3$ which does not leave the geometry invariant but which 
rather rescales all distances uniformly, so that $\partial_i \xi^{i\,(D)} = \partial_i x_i = 3$ is 
independent of $\vec x$. However in de Sitter spacetime this scale transformation of spatial 
coordinates in (\ref{flatRW}) can be compensated by a shift in the cosmological time 
(\ref{taushift}) such that the full transformation (\ref{Kildil}) is an exact symmetry with 
$K_{\mu}^{(D)}$ obeying (\ref{KilldS}). The consequence of this exact symmetry of the 
full de Sitter spacetime is that Fourier modes of different comoving wavenumber $\vec k$ 
leaving the cosmological horizon at a different FLRW time $\tau$ in a de Sitter invariant state 
give rise to a primordial fluctuation spectrum which has simple power law scaling behavior 
$|\vec k|^w$ with $|\vec k|$. If the conformal weight $w$ of the field generating the 
primordial fluctuations is chosen appropriately, namely $w \approx 0$, then this
essentially {\it kinematic} feature of de Sitter space embodied in the dilation symmetry
(\ref{dilxa}) leads to the prediction of a scale invariant Harrison-Zel'dovich power spectrum 
for the CMB  at largest scales, otherwise {\it independently} of how those fluctuations are 
generated.

In addition to the global scaling symmetry (\ref{dilxa}), de Sitter spacetime possesses a 
larger $SO(4,1)$ symmetry group and $3$ additional solutions of (\ref{Kil}), given by 
(\ref{speconformal}), whose spatial components are the special conformal transformations 
of ${\mathbb R}^3$. The existence of this additional conformal symmetry implies that any 
$SO(4,1)$ de Sitter invariant correlation function must decompose into representations 
of the conformal group of $3$ dimensional flat space. As we show in the next section 
this imposes non-trivial constraints on the shape of non-Gaussianities generated during 
a de Sitter phase of the universe's expansion. In general, the representations of the conformal 
group of ${\mathbb R}^3$ induced by de Sitter invariance need not be simple or irreducible 
representations. However because of the exponential expansion (\ref{expand}) the conformal 
representations become simple at times and distances large compared to the horizon scale $1/H$.
This can be seen most simply at the level of the scalar two-point correlation function
which we discuss first.

The de Sitter invariance of the two-point correlation function of scalar fields of arbitrary
mass $G(x,x'; M^2) = i \langle \Phi(x) \Phi(x')\rangle$ implies that $G$ must in fact depend 
only upon the $SO(4,1)$ invariant distance between the two spacetime points $x^{\mu}$
and $x^{\prime\,\mu}$, which can be expressed in terms of the de Sitter invariant scalar function
\be
1- z(x,x') = \frac{e^{H(\tau+\tau')}}{4}\,\left[ H^2 \vert\vec x - \vec x'\vert^2
-\left(e^{-H\tau} - e^{-H\tau'}\right)^2 \right] 
\label{zflatRW}
\ee
in coordinates (\ref{flatRW}). That this is the appropriate distance invariant
can be seen most readily by transforming to conformal time
\be
\eta = \int^{\tau} \frac{d\tau}{a(\tau)} = -H^{-1} e^{-H\tau} = - \frac{1}{H a(\tau)}\,,
\ee
in de Sitter space, so that (\ref{flatRW}) becomes
\be
ds^2 = \Omega^2\left( -d\eta^2 + d\vec x^2\right),\qquad {\rm with} \qquad 
\Omega(\eta) = -\frac{1}{H\eta} = a(\tau)\,,
\label{conftime}
\ee
with $\Omega$ the conformal factor. Then
\be
1-z(x,x') = \frac{1}{4\eta\eta'} \left[ -(\eta - \eta')^2 + |\vec x - \vec x'|^2\right]
= \frac{H^2\, \Omega(\eta)\Omega(\eta')}{4}\ (x - x')^2
\label{distconf}
\ee
is the Lorentz invariant flat spacetime distance $(x - x')^2 = -(\eta - \eta')^2 + (\vec x - \vec x')^2$,
rescaled by the local conformal factor $\Omega$ at both $x^{\mu}$ and $x^{\prime \mu}$. 

The two-point correlation function $G\left(z(x,x'); M^2\right)$ of a massive scalar field in an 
$SO(4,1)$ de Sitter invariant state satisfies the scalar wave eq.
\be
\left(-\sq + M^2\right) G[z(x,x'); M^2] = -H^2 \left[ z(1-z) \frac{d^2}{dz^2} + 2 (1 -2z) \frac{d}{dz} -
\frac{M^2}{H^2}\right]G[z; M^2] = 0\,,\quad x \neq x'\,,z\neq 1\,,
\label{massiveprop}
\ee 
When $M^2 = 2H^2 = \frac{R}{6}$ takes on the value for a conformally, coupled massless field,
then the solution of (\ref{massiveprop}) in the fully $O(4,1)$ de Sitter invariant state is
the two-point function \cite{BD},
\be
G_{conf}(z) \equiv G[z; 2H^2] =\frac{H^2}{16\pi^2}\, \frac{1}{1-z} 
= \frac{1}{\Omega(\eta)}\,\left[\frac{1}{4\pi^2}\,
\frac{1}{(x-x')^2}\right]\,\frac{1}{\Omega(\eta')}\,,
\label{conscalar}
\ee
which is the flat spacetime two-point function of a massless scalar conformally scaled 
by single powers of $\Omega^{-1}$ at $\eta$ and $\eta'$. On any fixed time slice $\tau=\tau'$ 
or $\eta=\eta'$ the propagator (\ref{conscalar}) is a simple power of $|\vec x - \vec x'|^2$ (namely $-1$), 
describing a scale and conformal invariant two-point correlation function of a conformal field 
of conformal weight one, corresponding exactly to the engineering mass dimension of the 
scalar field $\Phi$.

For general $M^2$ (\ref{massiveprop}) is the standard form of the hypergeometric eq. 
with the solution \cite{BD}
\be
G[z; M^2] = \frac{H^2}{16\pi^2}\, \Gamma(\alpha) \Gamma(\beta)\, F(\alpha,\beta; 2; z)
\label{hypergeom}
\ee
in terms of the Gauss hypergeometric function $_2F_1 = F$, with parameters
\bes
\bea
&&\alpha = \frac{3}{2} + \nu \,,\\
&&\beta = \frac{3}{2} - \nu\,,
\eea
\label{abdef}\ees\noindent
and $\nu$ defined by 
\be
\nu \equiv \sqrt{\frac{9}{4} - \frac{M^2}{H^2}}\,,
\label{nudef}
\ee
for $0 < M^2 \le \frac{9}{4}\,H^2$. If $M^2$ exceeds $\frac{9}{4}\,H^2$, $\nu$ becomes pure imaginary
but (\ref{hypergeom})-(\ref{nudef}) remain valid. At $x=x'$ (\ref{hypergeom}) is correctly 
normalized for (\ref{massiveprop}) to give $\delta^4(x,x')$ with unit weight. 

As might have been anticipated, (\ref{hypergeom}) does not have simple scaling behavior
in $|\vec x - \vec x'|$ at fixed time for arbitrary $M$. However in the limit of large times 
$\tau \sim \tau' \gg H^{-1}$, the behavior of $G[z;M^2]$ again becomes simple.
Inspection of (\ref{distconf}) shows that in this late time limit the de Sitter invariant
\be
1 - z(x,x') \rightarrow \frac {|\vec x - \vec x'|^2}{4\eta \eta'} \rightarrow \infty
\label{zlim}
\ee
as $\eta, \eta' \rightarrow 0$ for any fixed finite spatial  separation $|\vec x - \vec x'|$. Then the 
known asymptotic  form for $_2F_1(z)$ in (\ref{hypergeom}) for $z \rightarrow -\infty$ gives  \cite{astph}
\be
G[z; M^2]  \rightarrow A_{+\nu}\,\vert\vec x - \vec x'\vert^{- 3 + 2\nu} + A_{-\nu}\,
\vert\vec x - \vec x'\vert^{-3 - 2\nu}
\label{confmass}
\ee
where
\be
A_{\pm\nu}(\eta,\eta') = \frac{H^2}{2\pi^2}\ 
\frac{2^{\mp 2\nu}\Gamma(\frac{3}{2} \mp \nu)\Gamma(\pm 2\nu)}{\Gamma(\frac{1}{2} \pm \nu)} 
\ (\eta\eta')^{\frac{3}{2} \mp\nu}\,,
\label{Acoeff}
\ee
valid for $\alpha - \beta =2 \nu$ not equal to an integer. Hence after many $e$-foldings, the de Sitter 
invariant scalar field propagator of {\it arbitrary} mass exhibits conformal behavior and simple 
power law scaling. Moreover if  $M^2 < \frac{9}{4}\,H^2$ so that $\nu$ is real, then only the leading 
power law term $|\vec x - \vec x'|^{-3 + 2 \nu}$ of (\ref{confmass}) survives in the asymptotic limit. 

The power law behavior (\ref{confmass}) can be compared with the definition of the correlation 
function of a conformal operator ${\cal O}_{w}$ with conformal weight $w$, namely
\be
\langle{\cal O}_{w} (\vec x) {\cal O}_{w} (\vec x')\rangle
\sim \vert \vec x-\vec x' \vert^{-2w}\,.
\label{scalingtwo}
\ee
The conformal weight of the scalar field $\Phi$ at late times in de Sitter space inferred 
from (\ref{confmass}) and (\ref{scalingtwo}) is 
\be
w_{_{\Phi}} = \frac{3}{2} - \nu\,,
\label{DelPhi}
\ee
which differs from its canonical dimension of unity if $M^2$ differs from its conformal value 
of $2H^2$. At $M^2 = 2H^2$, from (\ref{nudef}) $\nu = \frac{1}{2}$, the second term in
(\ref{confmass}) drops out entirely and one obtains only the single power corresponding 
to (\ref{conscalar}), with the conformal dimension or weight (\ref{DelPhi}) equal to unity.

The conformal weights at infinity may be identifed in an even simpler way by examining 
the behavior of the individual mode solutions to the massive scalar wave eq. 
\be
(-\sq + M^2) \Phi = 0\,,
\label{massPhi}
\ee
with no spatial coordinate $\vec x$ dependence, {\it i.e.}
\bes\bea
&&\left( \frac{d^2}{d\tau^2} + 3 H \frac{d}{d\tau} + M^2 \right) \Phi(\tau) = 0\,\\
&& \Phi (\tau) = \frac{1}{a^w} = e^{-wH\tau} \propto \eta^w \,,\qquad   w_{\pm} = \frac{3}{2} \mp \nu \,,
\label{Phiscal}\eea\label{timedep}\ees
This is the $\eta$ dependence of the coefficients $A_{\pm \nu}(\eta,\eta')$ of (\ref{Acoeff}), associated 
with the conformal weights $w_{\pm}$ of the two terms in (\ref{confmass}). That the conformal weight(s) 
can be inferred either from the correlator (\ref{hypergeom}), (\ref{confmass}) or more simply from (\ref{timedep}) 
is a consequence of the fact that as $\eta, \eta' \rightarrow 0$ the de Sitter invariant $1 - z$ behaves as 
in (\ref{zlim}), so the powers of $\eta\eta'$ are necessarily tied to those of $|\vec x - \vec x'|$ in any de Sitter
invariant correlation function. Since the two-point function satisfies the sourcefree scalar wave equation 
(\ref{massiveprop}) for non-coincident points, and the spatial dependence of $\vec \nabla^2/a^2$ 
vanishes at late times due to the cosmological redshift, it follows that the power of the invariant $1-z$ 
can be fixed by simply examining the power dependence of the spatially independent solutions 
to (\ref{timedep}), as we have just verified explicitly. Thus {\it any} $SO(4,1)$ invariant Green's function 
of  $z(x,x')$ (not necessarily just that of massless fields) exhibits conformal behavior of correlators 
in the scaling region $|\vec x - \vec x'| \gg \eta, \eta'$, since it automatically incorporates
the connection between spatial scaling (\ref{spatdil}) and readjustment of the FLRW scale factor 
in (\ref{ascale}) which is an exact symmetry of de Sitter space, and all finite distance scales in
$|\vec x - \vec x'|$ are scaled to arbitrarily large values by (\ref{zlim}) in this region.

The lesson of this simple example of a free scalar field is that even non-conformal 
fields in de Sitter space exhibit simple conformal power law scaling behavior on the flat 
${\mathbb R}^3$ spatial sections after sufficiently long exponential expansion in a de Sitter 
phase, with an effective conformal weight given by $w_{_{\Phi}}$ in eq. (\ref{DelPhi}). The conformal 
behavior of de Sitter invariant correlation functions (\ref{confmass}) is an example of  the dS/CFT 
correspondence described in Refs. \cite{Starus,astph,MazMot,Skend,CasMazSta}, and the result of the 
mathematical isomophism between the conformal group of the flat ${\mathbb R}^3$ sections and 
the four dimensional de Sitter group $SO(4,1)$. Because of the larger symmetry of de Sitter space, 
expressed by the $3$ additional solutions to the Killing eq. (\ref{Kil}), given by (\ref{speconformal}), 
which are in one-to-one correspondence with the $3$ additional special conformal transformations 
of flat space satisfying (\ref{confKilflat}), de Sitter invariant correlation functions possess full 
conformal invariance (not just invariance under global dilations). At future spacelike infinity $I_+$ 
only the leading lowest weight real irreducible representation of the conformal group of ${\mathbb R}^3$ 
on the boundary is selected for $M^2 < \frac{9}{4}\,H^2$, in the late time limit $H\tau \sim H\tau' \gg 1$.

In the massless, minimally coupled case, corresponding to a scalar inflaton field, 
$M^2$ approaches zero and $\nu$ approaches $\frac{3}{2}$. In this case (\ref{hypergeom})
exhibits a divergence, and strictly speaking there is no de Sitter invariant
state or correlation function for $M^2 = 0$ \cite{Allen}. For small $M^2 \rightarrow 0^+$ 
(\ref{nudef}) and (\ref{DelPhi}) give
\be
w_{_{\Phi}} \simeq \frac{1}{3}\frac{M^2}{H^2} \rightarrow 0^+\,,
\label{minmass}
\ee
so the effective conformal weight of the inflaton goes to zero from positive values.
To see how this scaling behavior (\ref{scalingtwo}) with (\ref{minmass}) can lead directly to a 
Harrison-Zeld'ovich scale invariant primordial CMB spectrum in an essentially model independent 
way, consider that the stress tensor of a free field is a bilinear in the $\Phi$ field (multiplying
$w_{_{\Phi}}$ by $2$) and involves two derivatives (shifting the conformal weight by two units).
If the vacuum expectation value $\langle \Phi \rangle = 0$, the effective conformal weight of scalar 
energy density fluctuations $\delta \rho$ of the scalar field $\Phi$ is then
\be
w_{\rho} = 2 + 2\, w_{_{\Phi}}\,.
\label{Delrho}
\ee
From (\ref{scalingtwo}) a general operator with conformal weight $w$ has a power spectrum in Fourier space
\be
\tilde G_2 (|\vec k|; w) = \langle\tilde{\cal O}_{w} (\vec k) \tilde {\cal O}_{w}(-\vec k)\rangle \sim 
\int d^3 \vec x \  e^{i \vec k\cdot \vec x}\,\vert \vec x\vert^{-2w}\,
\sim \vert \vec k \vert^{2w - 3} \,.
\label{Fourier}
\ee
Thus from (\ref{minmass})-(\ref{Fourier}), the spectral index of the energy density fluctuations 
in Fourier space is
\be
n_{_S} = 2 w_{\rho} - 3 = 1 + 4\, w_{_{\Phi}} \simeq 1 + \frac{4}{3}\frac{M^2}{H^2} \,.
\label{specindex}
\ee
The value $n_{_S}=1$ for $M^2 \rightarrow 0$ is the classical Harrison-Zel'dovich (HZ) result \cite{HZ}. 
Notice also that $n_{_S}=1$ is the minimal value for a positive $M^2$ field. Negative $M^2$ corresponds 
to unstable tachyon fields in de Sitter space. The classical HZ spectral index for the CMB is 
the minimal one allowed by unitarity of quantum field theory in de Sitter space.

In Sec. \ref{Sec:GravPert} we show by an exact treatment of gauge invariant linearized perturbations 
of Einstein's eqs. in de Sitter space that the perturbations of the gravitational potential(s) $\Upsilon$ 
are related to the energy density perturbations $\delta \rho$ by an effective Poisson eq.  
$\nabla^2 \Upsilon \sim  4\pi G a^2 \delta \rho$. This fluctuation in the gravitational potential is `
`frozen" outside of the cosmological Hubble sphere and provides the primordial initial condition for perturbations 
at Hubble sphere re-entry \cite{LidL,MukFelBra}. Since the simple conformal behavior (\ref{confmass}) occurs at 
late times in the de Sitter phase,  the simple primordial power spectrum of energy density fluctuations 
(\ref{Fourier}) applies only at the very largest angular scales or lowest multipole moments. For large 
angular separations, or equivalently for the lower $\ell$ multipoles of the CMB power spectrum the 
Sachs-Wolfe effect dominates. Thus the observable temperature fluctuation in the CMB in a given 
direction on the sky is related to the gravitational potential perturbation $\Upsilon$ and energy density 
perturbation in the de Sitter phase by
\be
\left(\frac{\delta T}{T}\right)_{now} =  \left(\frac{\delta T}{T}\right)_{decoupl} + \ \Upsilon \quad \sim \quad
\frac{1}{k^2}\frac{\delta\rho}{\rho}\,.
\label{Poisson}
\ee
In (\ref{Poisson}) the first term is the intrinsic temperature fluctuation of a CMB photon at decoupling while 
the second is the gravitational potential perturbation $\Upsilon$ it encounters in travelling to us to be 
observed. The sum is gauge invariant, and is most readily evaluated in the rest frame of the cosmological fluid, 
where the first term $\left(\frac{\delta T}{T}\right)_{decoupl}$ is negligible \cite{SaWolf,WhiteHu}. The last 
proportionality to $\delta \rho/k^2$ is not a Newtonian approximation, but in fact follows from the gauge 
invariant analysis of Sec. {\ref{Sec:GravPert}, relating the gravitational potential to the energy density 
perturbation $\delta \rho$ in the primordial de Sitter phase, where it originates. 

It follows from the relative factor of $k^2$ in (\ref{Poisson}) that if the conformal weight of the density 
perturbations is $w_{\rho}$, then the scaling weight of the perturbations in the gravitational potential is
reduced by $2$ and given by
\be
s = w_{\rho} - 2 = 2 w_{_{\Phi}}\,,
\label{scals}
\ee
where (\ref{Delrho}) has been used in the last equality. Hence the two-point function of CMB temperature 
fluctuations is determined by the scaling dimension $s$ to be
\bea
G_2^{CMB}(\hat n\cdot\hat n'; s) &\equiv &\left\langle\frac{\delta T}{T}(\hat n)
\frac{\delta T}{T}(\hat  n')\right\rangle
\sim \int d^3 \vec k\left(\frac{1}{|\vec k|^2}\right)^2 \tilde G_2 (|\vec k|; s + 2) e^{i \vec k\cdot 
(\vec x- \vec x')}\nn
&=& C_{s} \Gamma (-s) (1 - \hat n \cdot\hat n')^{- s}\,,
\label{C2}
\eea
for some (generally $s$ dependent) constant $C_s$. The emission points 
of the CMB photons are at equal distance $|\vec x| = |\vec x'|$ from the observer by the 
assumption that the photons were emitted at the last scattering surface at equal cosmic time $\tau$. 
Notice that the power behavior of the density perturbations in momentum space (\ref{Fourier}) and definition
of the spectral index by (\ref{specindex}) corresponds to the power behavior $|\vec k|^{2s-3} = |\vec k|^{n_{_S} - 4}$
which is $|\vec k|^{-3}$ if $s=0, n_{_S} = 1$ for the integrand of (\ref{C2}). This is the standard power behavior 
for the spectrum of the curvature perturbation (denoted by $\cal R$ or $\zeta$ by other authors, {\it c.f.} 
\cite{LidL,MukFelBra}) in momentum space. Thus, in order for the CMB temperature anisotropies to be scale
invariant at large angular separations $s=0$, the primordial energy density fluctuations must have
conformal weight $w_{\rho} = 2$, leading to the classical Harrison-Zel\'dovich spectral index $n_{_S} = 1$
for those energy/mass density fluctuations, {\it c.f.} (\ref{specindex}). 

Expanding the function $G_2^{CMB}$ in multipole moments,
\be
G_2^{CMB}(\hat n \cdot \hat n'; s) = \frac{1}{4\pi} \sum_{\ell =1}^{\infty} (2\ell + 1)
c_{\ell}(s) P_{\ell} (\hat n \cdot \hat n')\ ,
\label{c2m}
\ee
gives 
\be
c_{\ell}(s) \sim \Gamma(-s) \sin\left(\pi s\right)
\frac{\Gamma (\ell + s)}{\Gamma (\ell + 2 - s)}\ ,
\ee
for general $s$, with a pole singularity at $s =0$ appearing in the $\ell = 0$ 
monopole moment. This is just the reflection of the fact that the ${\mathbb R}^3$ Laplacian
$\nabla^2$ cannot be inverted on constant functions, which should be excluded from the
power spectrum. Since the CMB anisotropy is defined by removing the isotropic 
monopole moment (as well as the dipole moment), the $\ell =0$ term does not appear 
in the moment sum (\ref{c2m}) in any case. The higher moments of the anisotropic two-point 
correlation are well-defined for $s$ near $0$. Normalizing to the quadrupole moment
$c_2(s)$, we find
\be
c_{\ell}(s) = c_2(s)\,
\frac{\Gamma (4 - s)} {\Gamma (2+ s)} \,
\frac{\Gamma (\ell + s)}{\Gamma(\ell + 2 - s)}\,,
\label{cell}
\ee
which is a standard result \cite{Peeb}. Indeed, if $w_{\rho} = 2, s =0$ and $n_{_S}=1$ 
by (\ref{Delrho}) (with $w_{_{\Phi}} = 0$), we obtain $\ell (\ell + 1) c_{\ell}^{(2)} = 6 c_2^{(2)}$, 
for the classical Harrison-Zeld'ovich power spectrum, for the lower moments of the CMB 
anisotropy assuming the Sachs-Wolfe effect dominates. This approximation holds 
reasonably well for $\ell \lesssim 40$. For larger values of $\ell$ a detailed transfer function 
analysis is necessary, and the moments corrected to include other effects on the 
propagation of the CMB photons from emission to observation point by the standard line 
of sight integration of the relevant kinetic equations over the past light cone \cite{SelZal,Durr}. 
The propagation of the temperature perturbations within the cosmological horizon lead to the 
acoustic peaks in the CMB power spectrum at larger $\ell$. 

In the real space ${\mathbb S}^2$ angular variables of directions on the sky, (\ref{C2}) 
corresponds to a conformal weight of $w_{\rho}/2 -1 = w_{_{\Phi}}$. In the limit 
$w_{\rho} \rightarrow 2, n_{_S} \rightarrow 1$ of the classical Harrison-Zel'dovich power 
spectrum, we find from (\ref{C2})
\be
G_2^{CMB}(\hat n\cdot\hat n')\big\vert_{HZ} = \lim_{s \rightarrow 0} 
\left[ G_2^{CMB} (\hat n\cdot\hat n' ; s)
+ \frac{C_2}{s}\right] = C_2\, \ln (1 - \hat n \cdot \hat n') + const.
\label{G2log}
\ee
for the angular correlations on the sky, after subtracting the pole contribution which
contributes only to the $\ell = 0$ moment. Thus, the classical scale invariant HZ
spectral function also corresponds to a logarithmic zero conformal weight
distribution in the real space ${\mathbb S}^2$ angular directions on the sky.

We emphasize that no slow roll approximation of any kind has been assumed in our dS/CFT 
considerations to arrive at (\ref{specindex}), and $\langle \Phi\rangle = 0$ so that we have not 
assumed any spatially homogeneous classical inflaton field, slow rolling or otherwise to expand 
around. Instead (\ref{specindex}) applies to scalar density fluctuations 
$\langle \delta \rho_{\vec k} \, \delta \rho_{\!-\vec k}\rangle$ from {\it any} source, 
provided they are generated in a de Sitter invariant state by {\it some} scalar quantity with 
(approximately) zero conformal weight (\ref{DelPhi}), giving rise to fluctuations in the
energy density with conformal weight (approximately) equal to $2$. and their correlator is 
observed on spatial sections at late times where the asymptotic CFT behavior (\ref{confmass}) 
applies. These are fully quantum fluctuations in which intrinsic energy density fluctuations in 
de Sitter space are responsible for the CMB, and not by expansion about a classical inflaton field
expectation value which breaks de Sitter invariance. Any fluctuations with these conformal properties
will give rise to (\ref{C2}), (\ref{c2m}) or (\ref{G2log}), consistent with the observational evidence 
that the primordial CMB power spectrum at large angular scales has a spectral index $n_{_S} \simeq 1$ 
close to its classical Harrison-Zel'dovich value. The simple conformal behavior of the correlation 
functions in an $SO(4,1)$ de Sitter invariant state follows automatically from (\ref{confmass}), which 
is itself a consequence of the mathematical isomorphism between the isometry group of de Sitter 
spacetime and the conformal group of ${\mathbb R}^3$, together with the kinematical exponential 
redshift of all distance scales in de Sitter space at late times. 

The result (\ref{specindex}) is superficially similar to that obtained in slow roll models with a 
nearly flat potential, $V = V_0 + \frac{1}{2}M^2\Phi^2$, and with the slow roll parameter
$\eta(\Phi) = V''/8\pi G V_0 = M^2/3H^3 \ll 1$. Thus it also starts with a scalar field
of conformal weight approximately equal to zero (and typically slightly {\it negative}
since $V'' < 0$). However, the route to (\ref{specindex}) in slow roll inflation is completely 
different than the derivation given above, based on general considerations of conformal 
invariance, which we shall refer to as the dS/CFT approach. In order to emphasize the 
the differences, let us review the main steps in arriving at the CMB power spectrum in 
slow roll inflation models, comparing and contrasting them with the dS/CFT approach 
presented here.

First, although the energy-momentum tensor is bilinear {\it i.e.} quadratic in the field $\Phi$, 
in slow roll inflation models one factor of $\Phi$ is taken by the spatially homogeneous 
classical field $\langle \Phi\rangle = \phi_{cl}(\tau)$ slowly rolling in a nearly flat potential, 
and as a result the energy-momentum tensor of the fluctuations is {\it linear} in the spatially 
inhomogeneous perturbations $\delta \phi$ about $\phi_{cl}(\tau)$. Since the perturbation 
$\delta \phi$ couples linearly to gravity in standard slow roll inflation, and the perturbation in the 
gravitational potential $\Upsilon$ is given directly by
\be
\Upsilon \simeq  - H\,  \frac{\delta \phi}{\,\dot\phi_{cl}}
\label{Upsinfl}
\ee
at horizon exit in the inflationary phase \cite{LidL,MukFelBra}, which is also linear in the
perturbation $\delta \phi$, the conformal weight of the scalar field fluctuations $\delta \phi$,
is transferred directly to the gravitational potential perturbation. The gravitational perturbation (\ref{Upsinfl})
becomes effectively frozen and remains constant outside the horizon with an approximately scale 
invariant spectrum, without any need to consider the energy density perturbation explicitly. 
Strictly speaking in the standard picture the inflaton is not a conformal field with a well-defined 
conformal weight, and the full $SO(4,1)$ conformal invariance of de Sitter space plays no role 
in the slow roll model. The simple scale invariance of $\delta\phi$ follows directly from the 
existence of only the scale symmetry (\ref{Kildil})-(\ref{dilxa}) in the de Sitter inflationary phase, 
provided the fluctuations $\delta \phi$ are independent of the origin of cosmic time $\tau$. 
Likewise the Harrison-Zel'dovich spectrum follows from simple scale invariance under  
(\ref{Kildil})-(\ref{dilxa}) alone, while the larger $SO(4,1)$ de Sitter invariance is explicitly 
broken by the non-zero expectation value of the inflaton background field $\phi_{cl}(\tau)$.

In contrast, in the dS/CFT approach there is {\it no} classical inflaton field to expand around, 
and hence no relation of the kind (\ref{Upsinfl}). The full $SO(4,1)$ invariance of de Sitter
space is not broken by any scalar inflaton field expectation value. Instead the fully quantum 
fluctuations of the energy-momentum tensor in de Sitter space are the fundamental quantity.
With $\langle\Phi\rangle = \phi_{cl} = 0$, the energy-momentum tensor remains quadratic 
in the quantum field $\Phi$. As long as this quantum field has zero conformal weight 
$w_{_{\Phi}} \rightarrow 0^+$, and its energy-momentum tensor has two derivatives, 
according to (\ref{Delrho}) the energy density fluctuation $\delta \rho$ has
conformal weight $2$. Indeed (\ref{specindex}) shows that the energy density fluctuations 
{\it must} have a conformal weight equal to $2$ in order to give rise to a classical 
Harrison-Zel'dovich power spectrum for the gravitational potential perturbations $\Upsilon$ 
and CMB temperature anisotropy $\delta T$, consistent with observations \cite{sky}. This is
equally true in the standard picture, although it is usually not emphasized since the density 
fluctuations are not needed explicitly if (\ref{Upsinfl}) is used to bypass computation of 
$\delta \rho$. In the deS/CFT approach the energy density perturbation $\delta \rho$ plays 
the central role. Its intrinsic quantum fluctuations are conformal and are related 
to the potential perturbations $\Upsilon$ by gravitational perturbation theory of the linearized 
Einstein eqs. around de Sitter space, discussed in detail in Sec. \ref{Sec:GravPert}. 
The gravitational potential fluctuations in the linearized Einstein theory are derived quantities
which need not be conformal fields with a well-defined conformal weight, which is why we use the
notation $s$ (rather than $w$ reserved for fields of definite conformal weight) for their scaling 
dimensions under dilations. In both approaches the perturbations are adiabatic, and generate 
no entropy. Despite the quite different physical origins of the fluctuations, whether by a scalar 
inflaton or full $SO(4,1)$ invariance in the dS/CFT approach, the final result for the two-point 
power spectrum of the temperature anisotropies (\ref{c2m})-(\ref{cell}) is unchanged if $s \approx 0$, 
and the CMB power spectrum data is consistent with either hypothesis of their physical origin.

A second point about which we should like to be very clear concerns the {\it statistics} 
(Gaussian or not) of the temperature anisotropies. In slow roll inflation the relations (\ref{Poisson})
and (\ref{Upsinfl}) also fixes the statistics of the CMB temperature fluctuations to be exactly that
of the scalar inflaton field fluctuations $\delta \phi$. Since in the dS/CFT approach
there is no scalar inflaton field expectation value, the statistics of the energy density, 
temperature or observable gravitational potential perturbations cannot be related to
$\delta \phi$ but must be determined independently. Indeed in conformal field theory
cubic and higher non-Gaussian correlation functions of the stress tensor such as 
$\langle T_{ab}(x_1)T_{cd}(x_2)T_{ef}(x_3) \rangle$ are generally non-zero even in a {\it free}, 
Gaussian field theory. In $4$ dimensions this non-Gaussian three-point correlation function
is determined by conformal Ward identities in terms of $3$ dimensionless parameters which are 
model dependent \cite{OsbPetErd}. These non-Gaussian parameters may have {\it any value}, 
and could in principle be much larger than the predictions of inflationary models, yet if not too large,
have escaped detection to this point. In any case the statistics and magnitude of the non-Gaussian
energy density fluctuations and hence that of the CMB anisotropies have nothing to do with the 
statistics of $\delta \phi$ in the dS/CFT approach. 

In either the standard inflationary slow roll picture or the present dS/CFT approach
the CMB temperature fluctuations at large angular scales inherit the nearly scale invariant 
spectral index $w_{_{\Phi}} \approx 0$ in angular directions $\hat n$. In dS/CFT the equality 
(\ref{C2}) for the CMB power on the sky applies with $s = 2 w_{_{\Phi}}$ replacing 
by  combination of slow roll parameters $\eta - 3 \epsilon$ in slow roll models \cite{LidL}. 
Since the exact value of $w_{_{\Phi}}$, $s$ or the slow roll parameters are presumed close 
to zero in any case, the two predictions for the CMB two-point power spectrum can hardly 
be distinguished observationally. Note however that whereas general dS/CFT arguments 
of the kind we have been considering lead to a spectral index $n_{_S} \ge 1$ in 
(\ref{specindex}), tilted (if at all) slightly to the {\it blue} if $M^2 \ge 0$ for a unitary 
representation of the $SO(4,1)$ de Sitter group, single field slow roll models are typically 
tilted slightly to the red, corresponding physically to the slightly unstable mode of the inflaton 
rolling away from pure de Sitter space. The WMAP data currently favors a red spectral index 
$n_{_S} \simeq 0.96 \pm 0.036$ \cite{WMAP}, about one standard deviation less than unity, 
although that fit is based on a number of model dependent assumptions (such a $\Lambda$CDM) 
which still require direct confirmation. 

To summarize, the slow roll scenario assumes a particular de Sitter breaking dynamics governed 
by the form of the inflaton potential $V$, which determines also the magnitude of the CMB temperature 
fluctuations through the slow roll parameters, as well as their statistics which are Gaussian to a high degree
of accuracy. In the dS/CFT approach general considerations of conformal invariance intrinsic to 
the symmetries of de Sitter space determine the spectral index shape of the power spectrum, from 
simply keeping track of conformal dimensions, independently of any particular dynamical model, although 
the overall amplitude or magnitude of non-Gaussian correlations cannot be determined from these 
symmetry considerations alone. Since as $w_{_{\Phi}} \rightarrow 0$, $n_{_S} \rightarrow 1$ 
in either case, the important conclusion is that a (nearly) Harrison-Zel'dovich CMB power spectrum 
cannot distinguish between slow roll inflation and a physically quite different origins of the CMB
temperature anisotropies rooted in conformal invariance of nearly weight zero conformal fields, 
descended from full $O(4,1)$ invariance of the instrinsically quantum energy density fluctuations 
of conformal weight $w_{\rho} \approx 2$ of those fields in a de Sitter phase. We shall see in the next 
section how this degeneracy is lifted when one considers the non-Gaussian bispectrum.

\section{Conformal Invariance and the Non-Gaussian Bispectrum}
\label{Sec:Bispectrum}

Adopting the point of view that conformal invariance of correlation functions in de Sitter 
space is a kinematical consequence of de Sitter invariance itself, and of the intrinsically
quantum fluctuations of the stress tensor in a de Sitter phase, not necessarily
the effect of a scalar slow roll scenario, the implications for higher point correlation functions 
may be derived independently of dynamical assumptions or specific models as well. 
One has simply to require that the correlation functions on the ${\mathbb R}^3$ spatial 
sections of de Sitter spacetime obey the Ward identities of a primary conformal field of 
a general conformal weight $w$, namely \cite{Poly,CFT}
\be
G_N (\vec x_1, \dots, \vec x_N; w) \equiv \langle {\cal O}_{w} (\vec x_1) \dots 
{\cal O}_{w} (\vec x_N)\rangle = [\Omega(\vec x_1)\dots \Omega(\vec x_1)]^{w}\,
G_N (\vec x_1', \dots, \vec x_N'; w)
\label{Ward}
\ee
where $\vec x'$ is the conformally transformed value of the coordinate $\vec x$. For the
conformal Killing vector solutions of (\ref{confKilflat}) the infinitesimal conformal variations may 
be expressed as
\bes\bea
&&\delta \vec x = \vec\xi (\vec x) \,\\
&&\delta \Omega (\vec x) = \frac{1}{3}\vec \nabla \cdot \vec \xi(\vec x) \label{varOmega}\,.
\eea\label{varxiOm}\ees
Substituting these first order variations into (\ref{Ward}) gives a differential eq. for $G_N$,
\be
\left[ (w\delta \Omega_1 + \vec \xi_1\cdot \vec \nabla_1) + 
\dots + (w\delta \Omega_N + \vec \xi_N\cdot \vec \nabla_N)\right]
G_N (\vec x_1, \dots, \vec x_N; w) = 0\,,
\label{identity}
\ee
where $\vec\nabla_n$ is the flat space gradient operator with respect to the coordinate $\vec x_n$,
$\vec \xi_n \equiv \vec \xi(\vec x_n)$, and $\delta\Omega_n \equiv \delta\Omega(\vec x_n)$
are eqs. (\ref{varxiOm}) evaluated at $\vec x_n$ for $n = 1,\dots, N$. 

A differential identity of the form (\ref{identity}) is obtained for each of the $10$ solutions of 
(\ref{confKilflat}). The $3$ translations and $3$ rotations for which $\delta \Omega = 0$ require 
$G_N$ to be a function only of the translational and rotational invariant distances,
\be
r_{mn} \equiv \vert \vec x_m - \vec x_n\vert = r_{nm}\,.
\ee
Since the dilational mode (\ref{dil}) has $\xi^{(D)}(\vec x) = \vec x$, $\delta \Omega^{(D)} = 1$
and $\vec \nabla_1 r_{12} = (\vec x_1 - \vec x_2)/r_{12} = - \vec \nabla_2 r_{12}$, the dilation 
identity for the two-point function, $N=2$ is
\be
\left[ 2 w + r_{12} \frac{\partial}{\partial r_{12}} \right] G_2 = 0\,,
\label{F2identity}
\ee
which leads immediately to (\ref{scalingtwo}). The identity corresponding to the three special
conformal transformations of ${\mathbb R}^3$,
\bes\bea
&&\vec\xi^{(C)} (\vec x) = 2 \,(\vec C \cdot \vec x) \, \vec x - \vec C\, |\vec x|^2\,\\
&&\delta \Omega^{(C)} (\vec x) = 2\, \vec C \cdot \vec x
\eea\label{speconf}\ees
gives (\ref{F2identity}) multiplied by $\vec C \cdot (\vec x_1 + \vec x_2)$ and hence no additional
constraints on the two-point power spectrum, which is thus automatically both scale and 
conformal invariant.

For the three-point correlation function $N=3$, the dilational identity is
\bea
&&\left[3 w + \vec x_1 \cdot \vec \nabla_1 + \vec x_2 \cdot \vec \nabla_2 
+ \vec x_3 \cdot \vec \nabla_3\right]G_3 
\nonumber\\
&& = \left[3w + r_{12}\frac{\partial}{\partial r_{12}} + r_{13}\frac{\partial}{\partial r_{13}} 
+ r_{23}\frac{\partial}{\partial r_{23}}\right]G_3 = 0\,,
\label{identitydil}
\eea
since $G_3$ is a function of $(r_{12}, r_{23}, r_{13})$. This condition does not
completely fix the form of $G_3$. However, the Ward identity
corresponding to the special conformal transformations (\ref{speconf}),
\be
\vec C\cdot \left[ 2 (\vec x_1 + \vec x_2 + \vec x_3)\, w + (\vec x_1 + \vec x_2)\,
r_{12}\frac{\partial}{\partial r_{12}} + (\vec x_2 + \vec x_3)\,
r_{23}\frac{\partial}{\partial r_{23}} +  (\vec x_3 + \vec x_1)\,
r_{31}\frac{\partial}{\partial r_{31}}\right] G_3 = 0
\label{identitycon}
\ee
together with (\ref{identitydil}) implies that all of the $3$ partial derivative
terms in (\ref{identitydil}) must be equal to each other. Hence each
must be equal to $-w G_{3}$. Therefore, we find that the general
solution of (\ref{identitydil}) with (\ref{identitycon}) is
\be
G_3(r_{12}, r_{23}, r_{13}; w)  = \frac{C_3 (w)}{(r_{12}r_{23}r_{13})^{w}}
= \frac{C_3 (w)}{|\vec x_1-\vec x_2|^{w} 
|\vec x_2- \vec x_3|^{w} |\vec x_3 - \vec x_1|^{w}}\,,
\label{G3form}
\ee
with $C_3(w)$ an arbitrary constant. Thus the three-point correlator is determined up
to an arbitrary normalization constant by conformal invariance, while the two-point
function $G_2$ was fixed already by dilational invariance alone.

In Fourier space the three-point correlator (\ref{G3form}) becomes
\bea
&&\tilde G_3 (\vec k_1 ,\vec k_2, \vec k_3; w) = C_3(w) \left[\frac{2^{3-w} 
\pi^{\frac{3}{2}} \Gamma\left(\frac{3 -w}{2}\right)}{\Gamma\left(\frac{w}{2}\right)}\right]^3
\delta^3(\vec k_1 + \vec k_2 + \vec k_3) \int\!\frac{d^3\vec p}{|\vec p - \vec k_1|^{3-w} 
|\vec p + \vec k_2|^{3-w} |\vec p|^{3-w}}\nonumber\\
&& \qquad\qquad = C_3(w)\,\frac{2^{3-3w}}{\left[\Gamma\left(\frac{w}{2}\right)\right]^2}
\, (2\pi)^6 \,\delta^3(\vec k_1 + \vec k_2 + \vec k_3)\,(k_1)^{3w-6} \,S \left(X,Y; w\right)
\label{G3corr}
\eea
where $k_i \equiv |\vec k_i|$ and the shape function $S(X,Y;w)$ of the ratios
\be
X \equiv \frac{k_2^2}{k_1^2}\,,\qquad\qquad Y \equiv \frac{k_3^2}{k_1^2}
\label{XYdef}
\ee
may be expressed in the form ({\it c.f.} Appendix \ref{App:Bispectral})
\bea
&&S(X,Y;w) = \frac{\Gamma\left(3 - \frac{3w}{2}\right)}{\Gamma\left(\frac{w}{2}\right)}
\int_0^1 du \int_0^1 dv\ \frac{\left[u(1-u)v\right]^{\frac{1}{2}-\frac{w}{2}}(1-v)^{\frac{w}{2}- 1}}
{\left[u(1-u)(1-v) + (1-u)vX + uvY\right]^{3-\frac{3w}{2}}} \label{bispectrum}\\
&& = \frac{2}{\sqrt{\pi}}\,
\Gamma\left(3 - \frac{3w}{2}\right)\,\Gamma\left(\frac{3}{2} - \frac{w}{2}\right)\,\int_0^1du\ 
\frac{\left[u(1-u)\right]^{\frac{1}{2} - \frac{w}{2}}}{\left[(1-u)X + uY\right]^{3 - \frac{3w}{2}}}
\,F\left(3 - \frac{3w}{2}, \frac{w}{2} ; \frac{3}{2}; {\cal Z}(X,Y;u)\right)\,. \nonumber
\eea
valid for Re $w > 0$. In the last expression the argument ${\cal Z}(X,Y;u)$ 
of the Gauss hypergeometric function $F=\,_2F_1$ is 
\be
{\cal Z}(X,Y;u) \equiv 1- \frac{u(1-u)}{(1-u)X + u Y}\,.
\label{Zdef}
\ee
The original expression (\ref{G3corr}) is symmetric under permutations of the $\vec k_i$, so it 
is understood that $S$ must be symmetrized among the six permutations of $(k_1, k_2, k_3)$.
A closed form for the momentum integral in (\ref{G3corr}) in terms of the generalized
hypergeometric function $F_4$ (Appell's function) is also available \cite{Davydychev},
leading to the result for the bispectral shape function given by eq. (\ref{SF4}) of Appendix \ref{App:Bispectral}.

According to (\ref{Poisson}) and following the same reasoning leading to (\ref{C2}) for the two-point CMB power
spectrum, the three-point CMB bispectrum in real space is obtained from the Fourier transform (\ref{G3corr})
of the three-point conformal energy density correlation function by forming
\be
G_3 ^{CMB} (\vec x_1, \vec x_2, \vec x_3) \sim  \int \, d^3 \vec k_1 \, d^3 \vec k_2 \, d^3 \vec k_3\, 
\left( \frac{e^{i \vec k_1\cdot \vec x_1 + i\vec k_2\cdot \vec x_2 + i\vec k_3\cdot \vec x_3}}
{ \vert\vec k_1\vert^2\ \vert\vec k_2\vert^2\ \vert\vec k_3\vert^2} \right)
\tilde G_3 (\vec k_1 ,\vec k_2, \vec k_3; w=w_{\rho})\,,
\label{bispectCMB}
\ee
and then setting the magnitudes equal, $\vert \vec x_1\vert = \vert \vec x_2\vert = \vert \vec x_3\vert$
on the assumption that the CMB photons were emitted at equal cosmic time. Comparing to (\ref{G3corr})-(\ref{XYdef})
we see that this introduces an overall factor of $(k_1k_2k_3)^{-2} = (k_1)^{-6} (XY)^{-1}$ for the Fourier
transform of the observable CMB Fourier transform and 
\be
S^{CMB}(X,Y) = \frac{S(X,Y;w_{\rho})}{XY}
\label{SCMB}
\ee
is the effective bispectral shape function for the CMB.

The shape function $S(X,Y;w)$ is plotted in Figs. \ref{Fig:Bispectrum96}-\ref{Fig:Bispectrum104} 
for values of the conformal weight $w=1.98, 2.02$, corresponding by (\ref{specindex}) to
CMB spectral indices of $n=0.96, 1.04$ respectively. Because of the pole in 
$\Gamma\left(3 - \frac{3w}{2}\right)$ as $w \rightarrow 2$, we have plotted the shape
function (\ref{bispectrum}) with this constant pole contribution subtracted, {\it i.e.}
\be
S_{sub}(X,Y;w) \equiv S(X,Y;w) - \frac{4\pi}{3(2-w)}\,,
\label{Ssub}
\ee
which has the finite limit as $w \rightarrow 2$. Any constant subtraction affects only the $\ell =0$
moment or overall magnitude which is removed in any case from the CMB anisotropy correlation functions.

\begin{figure}
\begin{center}
\includegraphics[height=8cm,width=8.5cm, trim=0cm 1cm 0cm 0cm, clip=true]{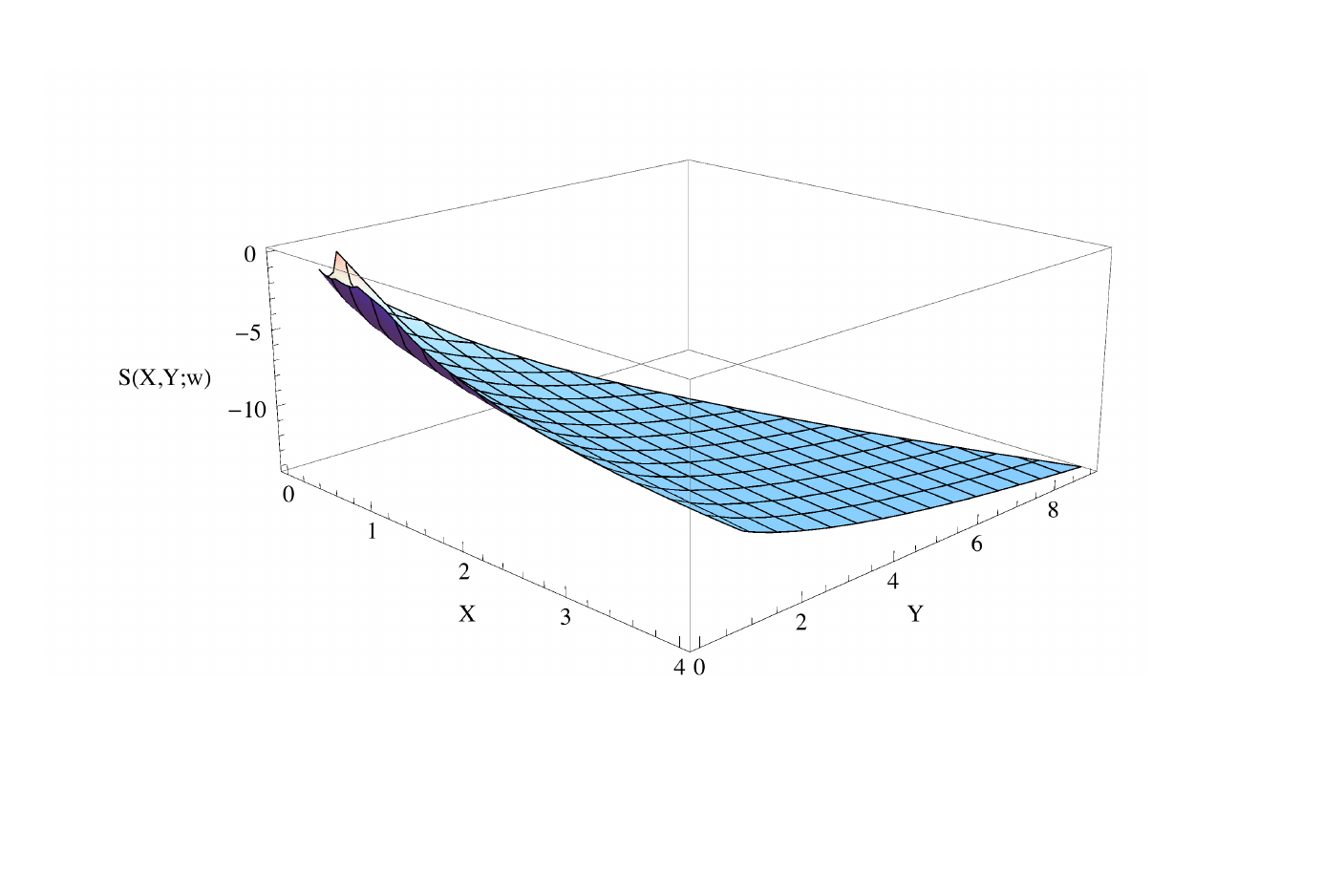}
\includegraphics[height=6.5cm,width=8.5cm, trim=1cm .5cm 1cm 2cm, clip=true]{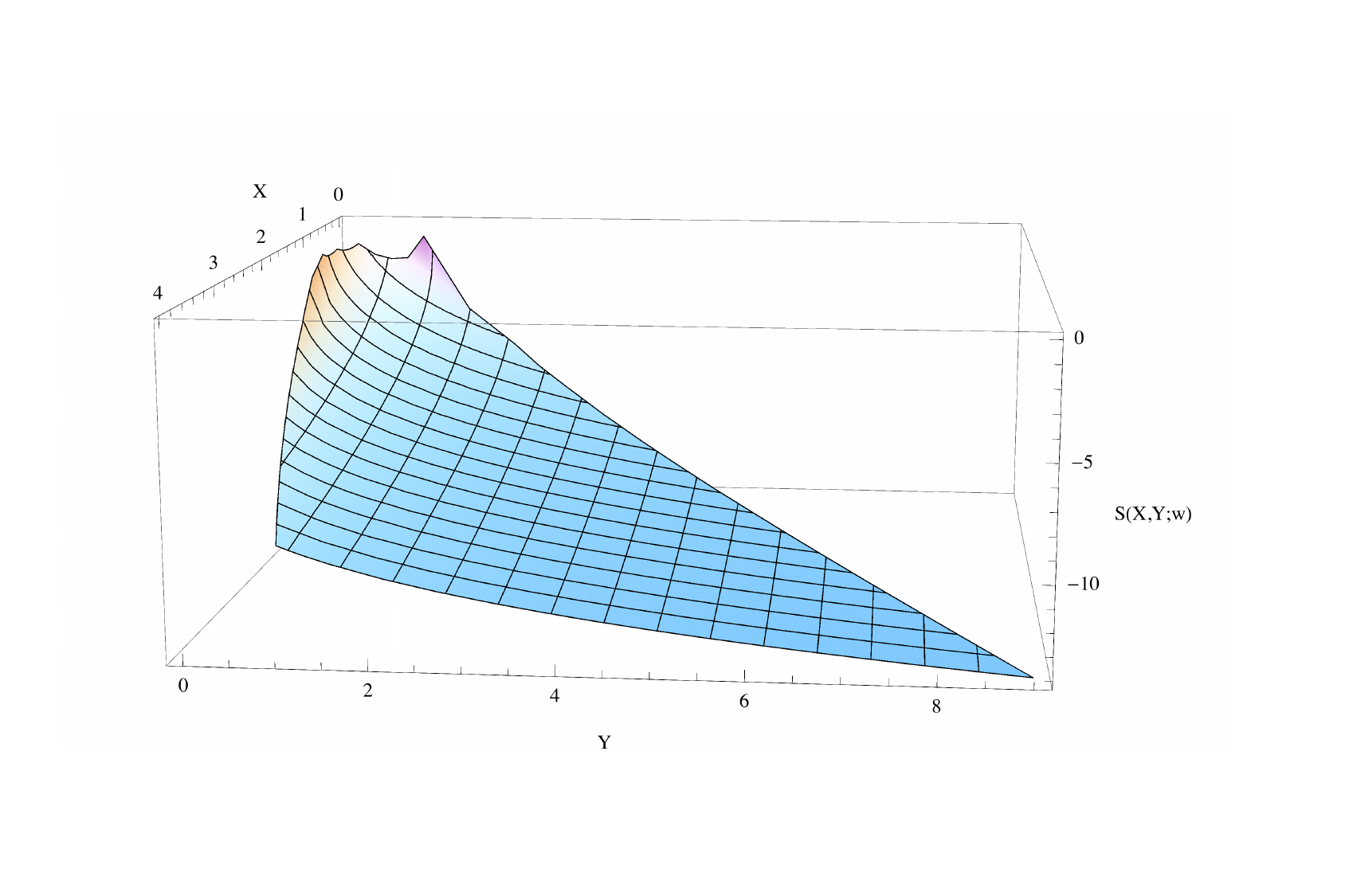}
\vspace{-1cm}
\caption{The bispectral shape function (\ref{bispectrum}) with pole contribution subtracted as in (\ref{Ssub}),
for conformal weight $w=1.98$, corresponding to a CMB spectral index $n=0.96$, as a function of 
$X = \frac{k_2^2}{k_1^2}$ and $Y=\frac{k_2^2}{k_1^2}$.} 
\label{Fig:Bispectrum96}
\end{center}
\vspace{-5mm}
\end{figure}

\begin{figure}
\begin{center}
\includegraphics[height=7.5cm,width=8.5cm, trim=0cm 0cm 0cm 0cm, clip=true]{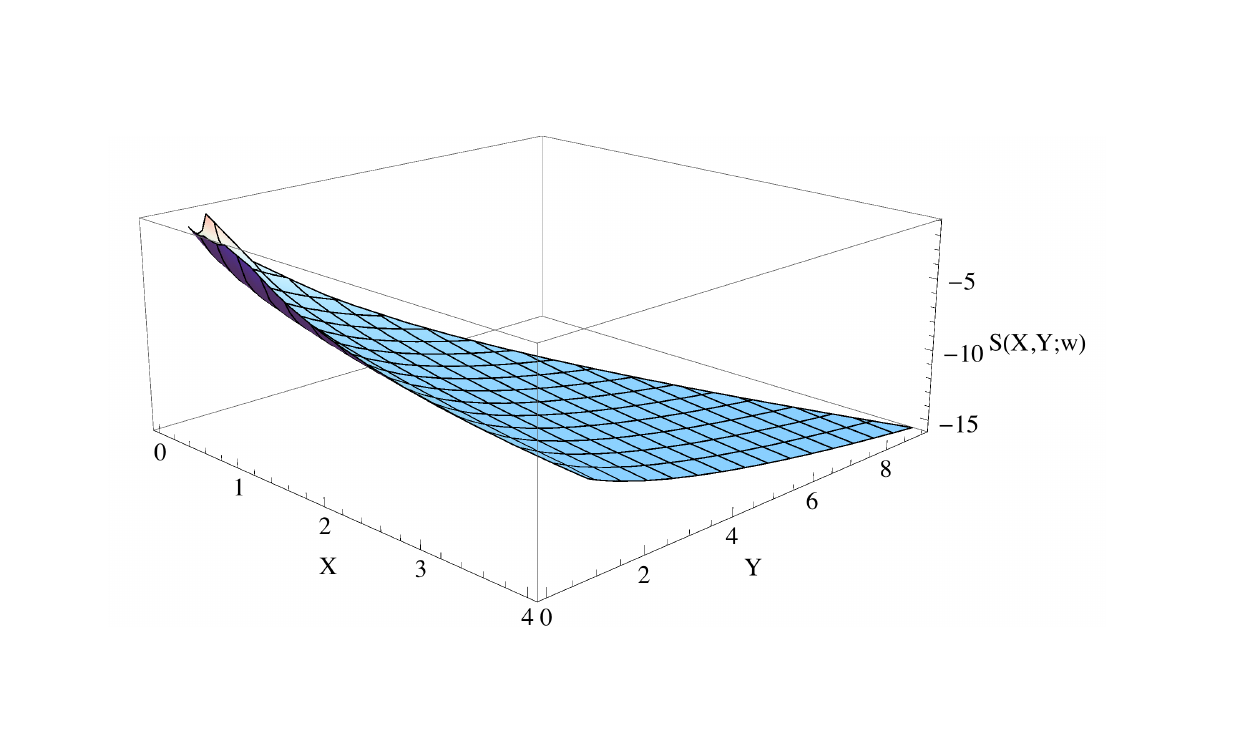}
\includegraphics[height=7cm,width=8.5cm, trim=0cm 0cm 0cm 0cm, clip=true]{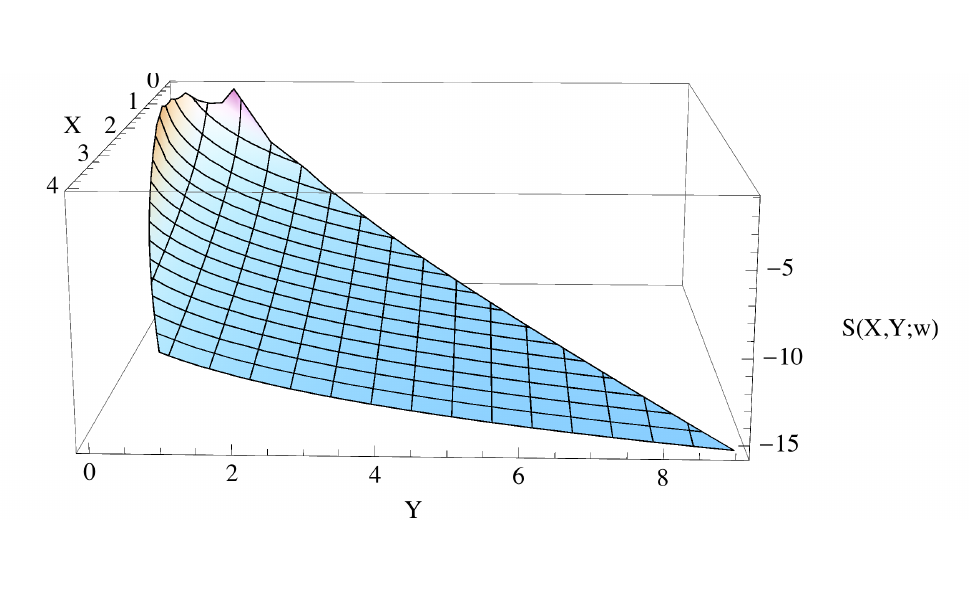}
\vspace{-8mm}
\caption{The bispectral shape function (\ref{bispectrum}), for conformal weight $w=2.02$, corresponding 
to a CMB spectral index $n=1.04$, as a function of $X = \frac{k_2^2}{k_1^2}$ and $Y=\frac{k_2^2}{k_1^2}$.
As in the previous figure, the pole contribution at $w=2$ has been subtracted: (\ref{Ssub})} 
\label{Fig:Bispectrum104}
\end{center}
\vspace{-5mm}
\end{figure}

The last form of (\ref{bispectrum}) is the most useful for examining special cases of $(k_1,k_2,k_3)$
for general $w$. Two important special configurations are:
\begin{itemize}
\item The Squeezed Case: For $k_1=0, k_2=k_3$, inspection of (\ref{bispectrum}) and (\ref{Zdef}) shows
that ${\cal Z}(X,Y;u)=1$ and 
\be
\lim_{k_1 \rightarrow 0} \left[(k_1)^{3w-6} S(X,Y; w)\right] = 
\frac{\Gamma\left(3 - \frac{3w}{2}\right)\,\left[\Gamma\left(\frac{3}{2} - \frac{w}{2}\right)\right]^2
\Gamma\left(w - \frac{3}{2}\right)}
{ \Gamma(3-w)  \Gamma\left(\frac{3w}{2} - \frac{3}{2}\right)}\ (k_2)^{3w -6}\,, \qquad \frac{3}{2} < {\rm Re}\, w < 3\,;
\label{squeezed}
\ee
\item The Equilateral Case: For $k_1=k_2=k_3\equiv k$, since ${\cal Z}(1,1;u) = 1-u(1-u)$,
\bea
&& \hspace{-1cm}k^{3w-6} S(1,1;w) =\frac{2}{\sqrt{\pi}}\,\Gamma\left(3 - \frac{3w}{2}\right)\,
\Gamma\left(\frac{3}{2} - \frac{w}{2}\right)\, k^{3w-6}\times\nonumber\\
&& \hspace{1cm}\int_0^1du\ 
\left[u(1-u)\right]^{\frac{1}{2} - \frac{w}{2}}
\,F\left(3 - \frac{3w}{2}, \frac{w}{2} ; \frac{3}{2}; 1- u(1-u)\right)\,,
\label{equil}
\eea
\end{itemize}
with a shape function independent of $k$. The remaining integral over $u$ in (\ref{equil}) may be 
performed and the result expressed in terms of a generalized hypergeometric function $_3F_2$, but as 
the result is not particularly illuminating, we omit this explicit form. In both the limiting cases of 
squeezed and equilateral triangles for the three momentum vectors $(\vec k_1, \vec k_2, \vec k_3)$
one obtains a simple power law for the bispectrum (\ref{squeezed}) or (\ref{equil}) respectively, in the 
remaining scalar variable.

It is also instructive to examine the limit $w \rightarrow 2$ for general $\vec k_i$. This limit is
finite for the pole subtracted bispectral shape function (\ref{Ssub}) and we find explicitly
\bea
\hspace{-1cm} &&S_{sub}(X,Y;2)
= -\int_0^1 \!du \int_0^1\! dv\ \left[u(1-u)v\right]^{-\frac{1}{2}}
\ln\left[u(1-u)(1-v) + (1-u)v X + uvY\right] + const. \nonumber\\
&& \quad = -2 \int_0^1 \frac{du}{\sqrt{u(1-u)}}\left\{\ln\left[(1-u)X + u Y\right] +
2\, \sqrt{\frac{1-{\cal Z}}{{\cal Z}}}\,\tan^{-1}\!\left(\sqrt{\frac{{\cal Z}}{1-{\cal Z}}}\right)\right\} + const.,
\label{S2}
\eea
up to an irrelevant finite constant, and ${\cal Z} = {\cal Z}(X,Y;u)$ is defined by (\ref{Zdef}).
The shape function $S_{sub}$ is plotted in Figs. \ref{Fig:BispectrumHZ}, and a contour plot is
also given in Fig. \ref{Fig:HZContourplot} for this case of classical HZ spectral index $n=1$.
As can be seen by comparing \ref{Fig:Bispectrum96}-\ref{Fig:BispectrumHZ},
the form of the shape function is relatively insensitive to the exact value of the
spectral index near $n=1$. 

\begin{figure}
\begin{center}
\includegraphics[height=6.5cm,width=9cm, trim=0cm 1cm 0cm .5cm, clip=true]{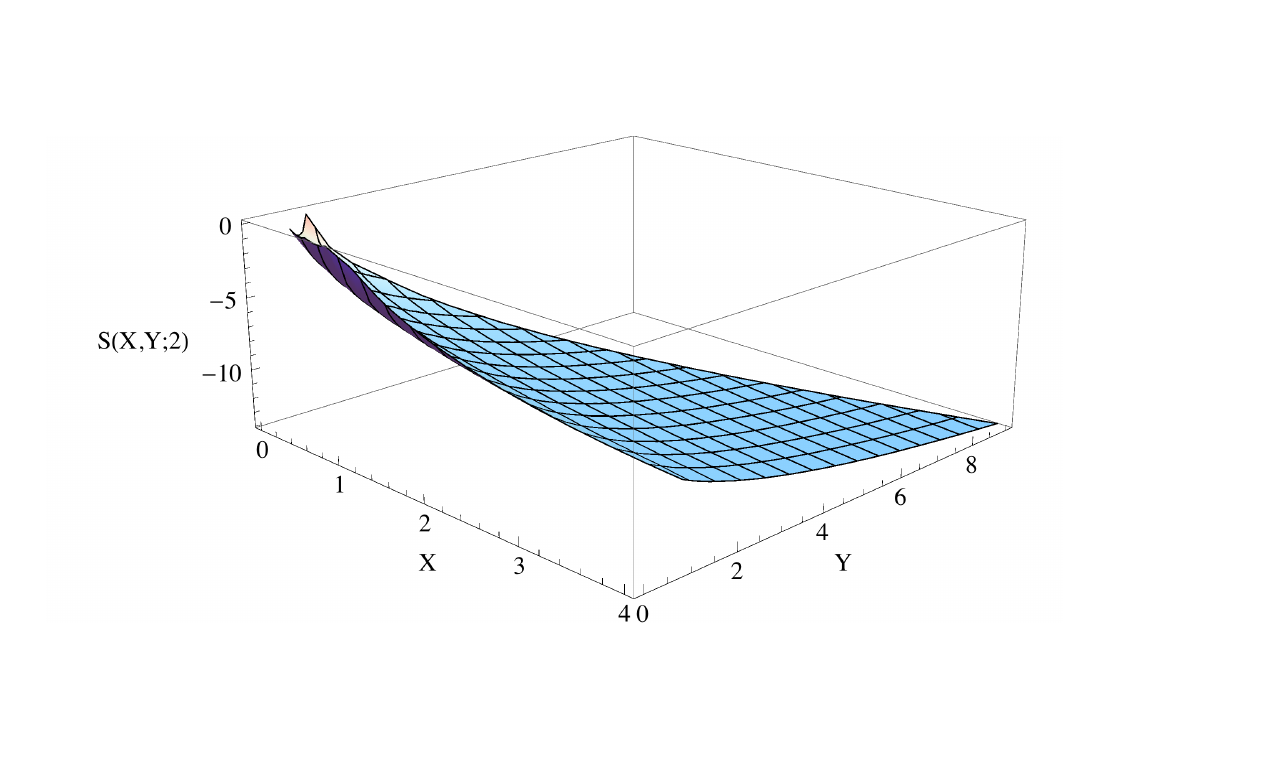}
\includegraphics[height=6cm,width=8cm, trim=0cm 0cm 0cm .5cm, clip=true]{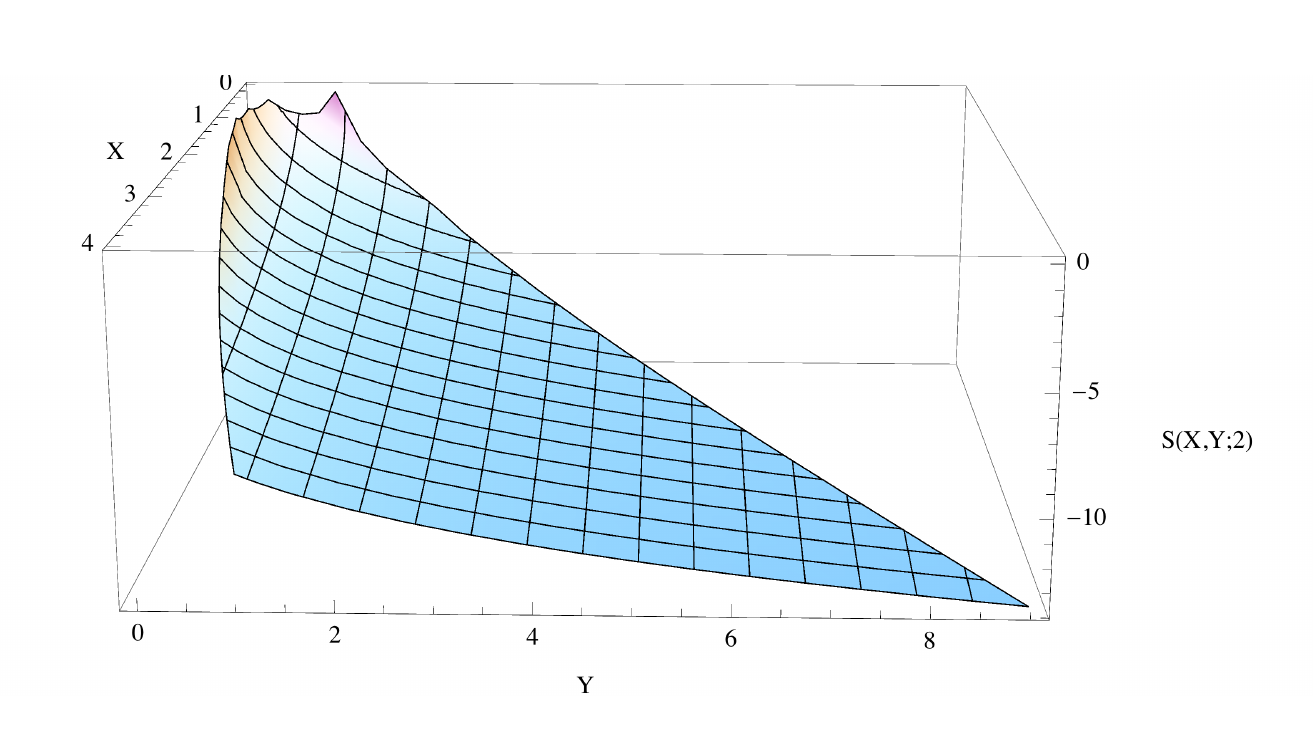}
\vspace{-7mm}
\caption{The subtracted bispectral shape function (\ref{S2}) for conformal weight $w=2.00$ corresponding
to a CMB spectral index $n=1.00$, as a function of $X = \frac{k_2^2}{k_1^2}$ and $Y=\frac{k_2^2}{k_1^2}$.} 
\label{Fig:BispectrumHZ}
\end{center}
\vspace{-7mm}
\end{figure}

\begin{figure}
\begin{center}
\includegraphics[height=8cm,width=8cm]{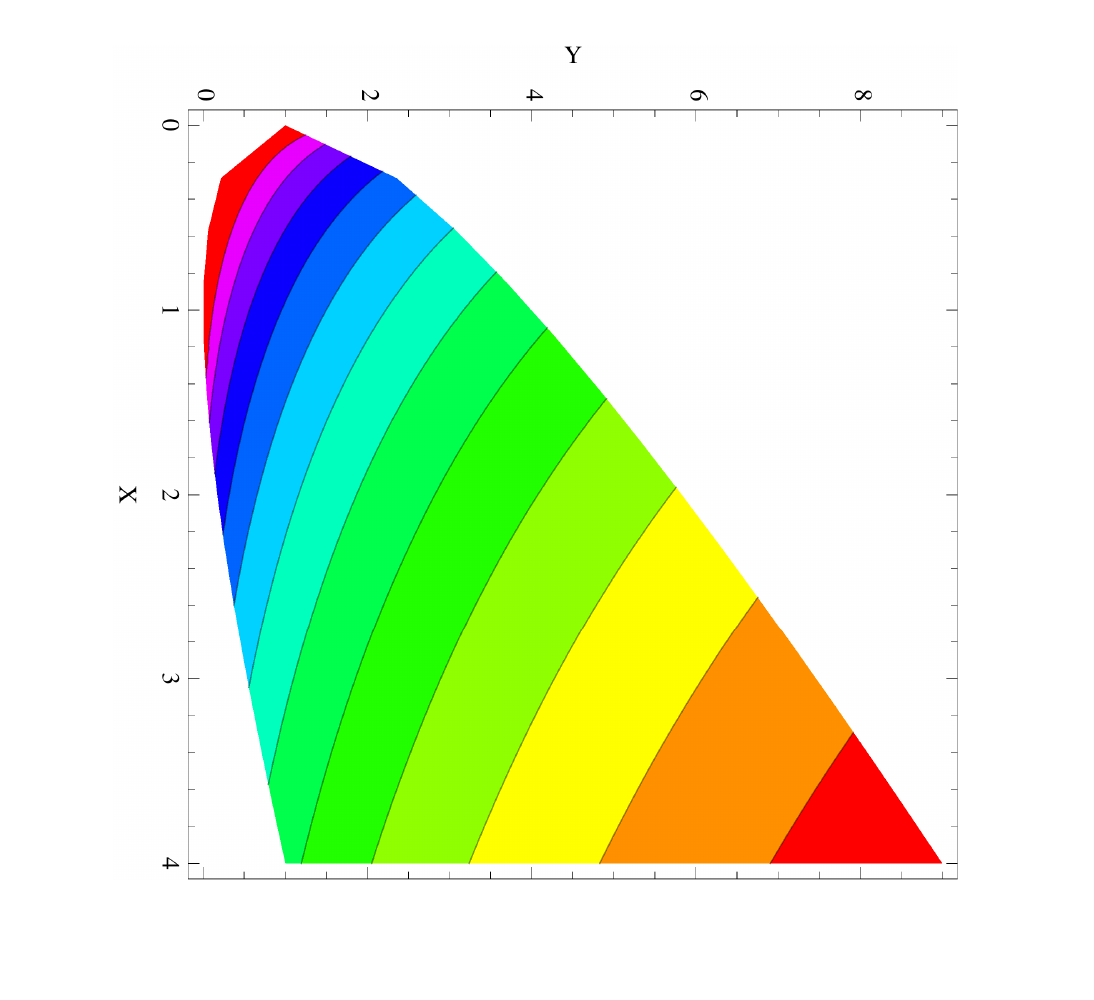}
\vspace{-1.3cm}
\caption{Contour plot for the bispectral shape function (\ref{S2}) corresponding to Figs. \ref{Fig:BispectrumHZ}
for a CMB spectral index of its classical HZ value $n=1.00$, as a function of $X = \frac{k_2^2}{k_1^2}$ and 
$Y=\frac{k_2^2}{k_1^2}$. The larger values of $S(X,Y;2)$ are for smaller $X,Y$ in the red region at the upper left, 
gradually decreasing toward larger $X$ and $Y$ at the lower right. }
\label{Fig:HZContourplot}
\end{center}
\vspace{-5mm}
\end{figure}

The form of the shape function in (\ref{bispectrum})-(\ref{S2}) determined from conformal invariance 
of flat ${\mathbb R}^3$ sections in de Sitter space is quite different from that derived from the simplest 
inflation models \cite{BKMRio,ABMR,LigSefFerShl}. For example, the parameterization of the non-Gaussian 
term in single field slow roll inflation by a small non-linear term, $\Phi(\vec x) = \Phi_L(\vec x) + f_{NL}
[\Phi_L^2(\vec x) - \langle \Phi_L^2(\vec x)\rangle ]$, where $f_{NL}$ is the non-linearity parameter, 
leads to the ``local" model of the dimensionless non-Gaussian shape function,
\bea
&&S_{local}(k_1, k_2, k_3) = 2 f_{NL} (k_1k_2k_3)^2 \left[P_0(k_1) P_0(k_2) 
+ P_0(k_2)P_0(k_3) + P_0(k_3)P_0(k_1)\right]
\nonumber\\
&& = 2f_{NL} N_0^2 \left[ \frac{k_1^2}{k_2k_3} + \frac{k_2^2}{k_1k_3} + \frac{k_3^2}{k_1k_2} \right]\,,
\label{Slocal}
\eea
where $N_0$ is the normalization of the two-point power spectrum $P_0(k) = N_0 /k^3$ for the inflaton
or gravitational potential fluctuations during slow roll. The ``equilateral" shape,
\be
S_{equil}(k_1,k_2,k_3) \propto \frac{(k_1 + k_2 - k_3)(k_2 + k_3 - k_1) (k_3 + k_1 - k_2)}{k_1k_2k_3}
\label{Sequil}
\ee
has also been considered by some authors, while the full prediction for the bispectral shape for
single field slow roll inflation is the linear combination \cite{LigSefFerShl}
\be
S_{slowroll} \propto (6 \epsilon - 2 \eta) S_{local}(k_1,k_2,k_3) +\frac{5 \epsilon}{3}S_{equil}(k_1,k_2,k_3)\,,
\label{slowroll}
\ee
with 
\be
\epsilon(\Phi) = \frac{1}{16 \pi G} \left(\frac{V'}{V}\right)^2\,\qquad \eta(\Phi) = \frac{V''}{8 \pi G V}
\ee
the usual slow roll parameters defined in terms of the scalar inflaton potential and its derivatives.

Clearly these shape functions are quite different from (\ref{bispectrum}) for general $w$ or $k_i$, 
or (\ref{S2}) for $w = w_{\rho} \rightarrow 2$, degenerating to each other only in special kinematic
configurations. The reason for this is that the standard assumptions of slow roll inflation models 
require a departure from de Sitter space and violation of conformal invariance. Only (approximate)
scale invariance is assumed, which as we have observed gives (\ref{identitydil}) which does
not completely fix the form of the bispectrum. In addition, in the usual picture the 
magnitude of non-Gaussianity of the CMB is tied to the violation of $SO(4,1)$ de Sitter and 
conformal invariance parameterized by the slow roll parameters, whereas (\ref{bispectrum}) 
or (\ref{S2}) is derived from the intrinsic conformal symmetries of de Sitter space itself, with 
an amplitude that is otherwise undetermined by conformal invariance considerations alone. 
It is therefore possible for both the magnitude and the shape of primordial non-Gaussianities 
in the CMB to be quite different from models tied to the slow roll of scalar inflaton field, and 
for observational detection of an alternate form of the shape function to point to a quite different 
dynamical origin for the CMB fluctuations. We note also that (\ref{Slocal}), (\ref{Sequil}) or 
(\ref{slowroll}) are quite singular as one of the $k_i \rightarrow 0$, whereas (\ref{bispectrum}) 
has at worst a logarithmic behavior in this limit. This implies that observational limits on the 
amplitude of the slow roll shape (\ref{slowroll}) will generally lead to less restrictive limits 
on the amplitude of the non-Gaussian bispectrum (\ref{bispectrum}).

For higher point correlations, the conformal invariance identities (\ref{identity}) do not 
completely determine the correlation functions $G_{N>3} (\vec x_1, \dots, \vec x_N; w)$ \cite{sky}. 
For $N=4$, the solution of (\ref{identity}) is 
\be
G_4 (\vec x_1, \vec x_2, \vec x_3, \vec x_4; w) = 
\langle{\cal O}_{w} (\vec x_1) {\cal O}_{w} (\vec x_2) 
{\cal O}_{w} (\vec x_3) {\cal O}_{w} (\vec x_4)\rangle = \frac{f_4(P, Q)} 
{\left[r_{12}r_{13} r_{14} r_{23}r_{24} r_{34}\right]^{\frac{2w}{3}} }\ ,
\ee
where $f_4$ is an arbitrary function of the two cross-ratios, 
$P= r_{13} r_{24}/r_{12} r_{34}$ and $Q = r_{14} r_{23}/r_{12}r_{34}$.
Analogous expressions hold for higher $N$-point functions.  In the equilateral case 
in Fourier space, $\vec k_1 = \dots =\vec k_N \equiv \vec k$, the coefficient amplitudes become 
constants and one again obtains a universal $\ln k$ dependence in the limit $w \rightarrow 2$.

\section{Gravitational Potentials and Their Scaling in de Sitter Space}
\label{Sec:GravPert}

In this section we derive the relationship between energy density perturbations and
scalar gravitational potential perturbations in de Sitter space, in a fully relativistic
gauge invariant formalism. In so doing we justify (\ref{scals}), showing how the density 
perturbations of conformal weight $w_{\rho} =2$ in the deS/CFT approach generate the 
scale invariant HZ  spectrum for the gravitational potentials, and hence the CMB 
temperature anisotropies at large angular scales, where the Sachs-Wolfe effect dominates

The fluctuations around any self-consistent solution of the semiclassical Einstein eqs. 
satisfy the linear response eqs.,
\be
\delta\left(R^a_{\ b} - \frac{R}{2} \,\delta^a_{\ b} + \Lambda \,\delta^a_{\ b} \right)
= 8\pi G\, \delta \lag T^a_{\ b}\rag\,.
\label{scEin}
\ee
These couple the fluctuations of the metric to the fluctuations of the quantum matter
stress tensor, at linear order in the deviation from the de Sitter metric $\bar g_{ab}$, {\it i.e.}
\be
g_{ab} = \bar g_{ab} + h_{ab}\,.
\ee
It is well known that the metric perturbations $h_{ab}$ which are scalar with respect to the background 
three-metric $\eta_{ij}$ in the flat FLRW coordinates (\ref{flatRW}) can be parameterized in 
terms of $4$ functions, $({\cal A, B, C, E})$, in the form \cite{Bar,Stew,MukFelBra},
\bes\bea
h_{\tau\tau}&=&-2\,{\cal A} \\
h_{\tau i}&=&a \,\partial_i {\cal B}\\
h_{ij}&=& 2\,a^2 \left[\eta_{ij}\, {\cal C} + \left(\partial_i\partial_j- \frac{\eta_{ij}}{3}\,\nabla^2\right) {\cal E}\right]\, \, .
\eea
\label{linmet}\ees
\noindent
Since the linearized metric perturbation under a linearized coordinate (gauge) transformation 
$x^a \rightarrow x^a + \xi^a(x)$ undergoes the change
\be
h_{ab} \rightarrow h_{ab} + \nabla_a \xi_b + \nabla_b\xi_a\,,
\label{hgauge}
\ee
and in the space plus time split the gauge transformation $(\xi^{\tau}, \xi^i) = (aT, \partial^i L)$ can 
be characterized by $2$ scalar functions $T$ and $L$, it follows that only $2$ linear combinations 
of the $4$ functions in (\ref{linmet}) are gauge invariant. These may be taken to be the $2$ 
gauge invariant gravitational potentials
\bes
\bea
&& \Upsilon_{\cal A} \equiv {\cal A} + \partial_{\tau}(a {\cal B})- \partial_{\tau}(a^2 \partial_{\tau}{\cal E})\,,\\
&&\Upsilon_{\cal C} \equiv {\cal C} - \frac{\stackrel{\rightarrow}{\nabla}\!\!^2{\cal E}}{3}
+ \dot a\, {\cal B} - a \dot a\, (\partial_{\tau} {\cal E})\,.
\eea
\label{defUpsAC}\ees 
\noindent
These are the gauge invariant gravitational potentials denoted by $\Phi_A$ and
$\Phi_H$ in ref. \cite{Bar}, and by $\Phi_A$ and $\Phi_C$ in Ref. \cite{Stew},
while the authors of ref. \cite{MukFelBra} employed the notation $(\Phi, -\Psi)$
for $(\Upsilon_{\cal A}, \Upsilon_{\cal C})$.

The general stress-energy tensor perturbation in flat FRLW coordinates in the scalar
sector may be decomposed according to the analog of (\ref{linmet}) for the metric 
tensor. Thus we specify the stress tensor components in the scalar sector in terms
of $4$ functions by
\bes
\bea
&&\hspace{1.5cm}\delta T^{\tau}_{\ \tau} = - \delta \rho\,,\\
&&\delta T^{\tau}_{\ i} = - \partial_iV\,,\qquad \delta T^i_{\ \tau} = \frac{\eta^{ij}}{a^2} \, \partial_j V\,,\\
&&\delta T^i_{\ j} = \delta p \, \delta^i_{\ j}  + \left(\frac{\delta^i_{\ j}}{3}- \, \partial^i \frac{1}{\nabla^2} \partial_j\right)W
\eea
\label{tendecomp}\ees
We note that the stress tensor perturbations around exact de Sitter space are 
gauge invariant \cite{MukFelBra}, and hence so are the functions $\delta\rho, \delta p, V, W$.
However, invariance under the coordinate gauge transformation (\ref{hgauge}) implies that the $4$ 
scalar functions $(\delta\rho, \delta p, V, W)$ are not all independent, but rather
are related by $2$ non-trivial constraints embodied in the conservation eqs.
\bes
\bea
&&\frac{\partial}{\partial \tau} \delta \rho + 3\, \frac{\dot a}{a} \,(\delta \rho + \delta p) = \frac{\nabla^2}{a^2}\, V\,,\\
\label{drhodp}
&&\partial_i \left( \frac{\partial}{\partial \tau} V + 3HV + \frac{2}{3}\,W - \delta p\right) =0\,,
\eea
\label{deltacons}\ees
corresponding to the $2$ coordinate gauge symmetries specified by $T, L$. There remain $2$ independent 
stress tensor components in the scalar sector and the information needed to determine the $2$ gauge 
invariant potentials in terms of these is contained completely in the $\tau\tau$ and total trace components 
of the linearized Einstein eqs., 
\bes
\bea
&&\delta G^{\tau}_{\ \tau} = 8 \pi G\, \delta T^{\tau}_{\ \tau} = -8 \pi G\, \delta \rho\,,\\
&&\delta R = - 8 \pi G\, \delta T^a_{\ a} = 8 \pi G\, (\delta \rho - 3\, \delta p)\,,
\eea
\label{enertr}\ees
the other components of the linearized Einstein eqs. being automatically satisfied if the Einstein eqs.  for the
background and stress tensor conservation eqs. (\ref{deltacons}) are satisfied \cite{DSAnom}. 

Specializing to de Sitter space where $\dot a/ a = H$ is a constant and the energy
density and pressure perturbations are themselves gauge invariant for arbitrary
spatial dependence \cite{MukFelBra}, the two relevant linearized Einstein eqs. (\ref{enertr})
may be written in the gauge invariant form  \cite{DSAnom}
\bes\bea
&&\hspace{3cm}H \frac{\partial}{\partial \tau} \Upsilon_{\cal C} - H^2 \Upsilon_{\cal A} 
- \frac{1}{3}\frac{\vec \nabla^2}{a^2} \Upsilon_{\cal C}
= \frac{4\pi G}{3} \, \delta \rho \,, \label{linrestt}\\
&& \frac{\partial^2}{\partial\tau^2} \Upsilon_{\cal C} 
- H \frac{\partial}{\partial \tau}\Upsilon_{\cal A} + 4H \frac{\partial}{\partial \tau} \Upsilon_{\cal C} 
- 4H^2 \Upsilon_{\cal A} - \frac{1}{3}\frac{\vec \nabla^2}{a^2} \Upsilon_{\cal A}
 - \frac{2}{3}\frac{\vec \nabla^2}{a^2} \Upsilon_{\cal C} = \frac{4\pi G}{3} \, (\delta \rho - 3\, \delta p)\,.
\label{linrestrace}\eea
\label{linres}\ees
By differentiating (\ref{linrestt}) with respect to $\tau$ and operating on it with $\nabla^2/a^2$,
and then taking appropriate linear combinations with (\ref{linrestrace}), these eqs. may be solved 
for the gravitational potentials in terms of $\delta \rho$ and $\delta p$ in the form
\bes\bea
&&\hspace{2cm} \left(\frac{\vec \nabla^2}{a^2}\right)^2 \Upsilon_{\cal C} =
- 12\pi G\,\left[H\frac{\partial}{\partial \tau} \delta\rho + 3H^2 (\delta\rho + \delta p) 
+ \frac{1}{3}\frac{\vec\nabla^2}{a^2}\,\delta \rho \right] \label{potC},\\
&&\hspace{-7mm}\left(\frac{\vec \nabla^2}{a^2}\right)^2 \Upsilon_{\cal A} = - 12\pi G \, 
\left[\frac{\partial^2}{\partial\tau^2} \delta\rho + 7H \frac{\partial}{\partial \tau} \delta \rho
+ 3 H \frac{\partial}{\partial \tau}  \delta p + 12H^2(\delta \rho + \delta p) 
- \frac{\vec \nabla^2}{a^2}\left(\frac{\delta \rho}{3} + \delta p\right)\right].
\label{potA}
\eea\label{Upsoln}\ees
When the conservation eqs.\! (\ref{deltacons}) are used, we see that the right sides of (\ref{Upsoln})
are proportional to total spatial Laplacians, one of which therefore may be cancelled from each
side of eqs. (\ref{Upsoln}), with the simple result
\bes
\bea
\frac{\vec \nabla^2}{a^2}\, \Upsilon_{\cal C} &=& - 4\pi G \, (\delta \rho + 3HV)\,,\qquad\qquad\ \vec k \neq 0\,,\\
\frac{\vec \nabla^2}{a^2}\, \Upsilon_{\cal A} &=&  \ 4\pi G \, (\delta \rho + 3HV + 2W)\,,\qquad \vec k \neq 0\,,
\eea
\label{Upsolndelrho}\ees
in the sector with non-vanishing spatial derivatives, or in Fourier space $\vec k \neq 0$. The sum
and difference of eqs. (\ref{Upsolndelrho}) give
\bes
\bea
&&\hspace{1.5cm} \frac{\vec \nabla^2}{a^2}\, (\Upsilon_{\cal A} + \Upsilon_{\cal C}) 
= 8\pi G \, W\,,\qquad \vec k \neq 0\,,\\
&&\frac{\vec \nabla^2}{a^2}\, (\Upsilon_{\cal A} - \Upsilon_{\cal C}) 
=  8\pi G \, (\delta \rho + 3HV + W)\,,\qquad \vec k \neq 0\,,
\eea
\label{Upsolnsumdiff}\ees
showing that if there are no anisotropic spatial stresses, $W=0$, the gravitational 
potentials are equal and opposite $\Upsilon_{\cal C} = - \Upsilon_{\cal A}$ in the spatially 
inhomogeneous $\vec k \neq 0$ sector, or $\Psi=\Phi$ in the notation of \cite{MukFelBra}. 

In the spatially homogeneous sector $\vec k =0$, we may set $V=W=0$, there is only one gauge 
invariant potential, namely $\partial_{\tau} \Upsilon_{\cal C} - H \Upsilon_{\cal A}$ which is 
given directly by (\ref{linrestt}), and (\ref{linrestrace}) is automatically satisfied because of the
remaining conservation eq. (\ref{drhodp}) with $V=0$. This degeneracy of the $\vec k = 0$ case 
is related to the existence of ($5$) conformal Killing vectors in de Sitter space, which being composed
of linear combinations of constant, linear ($\vec x$) and quadratic ($\vec x^2$) functions of the 
spatial coordinate, are non-Fourier normalizable at $\vec k =0$, where the Laplacian 
$\vec \nabla^2$ is non-invertible. The existence of these conformal Killing modes implies 
that there is an enhanced linear diffeomorphism symmetry and hence fewer gauge invariant 
potentials at $\vec k = 0$.  This connection can be shown most clearly by working in the spatially 
closed $\mathbb{S}^3$ coordinates of (\ref{hypermet}), for which the spatial Laplacian has a 
discrete spectrum, all the modes are normalizable on $\mathbb{S}^3$, and the degeneracy 
in the gravitational potentials and conformal Killing vectors appear together in the $k=0$ (singlet) 
and $k=1$ (quartet) harmonics on $\mathbb{S}^3$ \cite{SSBdS}.

The result (\ref{Upsolndelrho}) shows that the relativistic gravitational potentials
in linear perturbation theory around de Sitter space are related to the stress
tensor sources in the scalar sector in the same manner as in Newtonian theory, 
namely by the Laplacian $\nabla^2/a^2$ of the non-relativistic Poisson eq. The only
differences from the non-relativistic case are the appearance of two potentials in the 
relativistic theory rather than one when anisotropic stresses are present ({\it i.e.} $W \neq 0$), 
and the appearance of the velocity perturbation $V$ as an additional source of the 
relativistic potentials. Note that no assumption about the perturbations being sub-
or super-horizon has been made. Eqs. (\ref{Upsolndelrho}) are {\it exact} gauge invariant
relations in the flat FRLW coordinates (\ref{flatRW}) of de Sitter space.

From the discussion in Sec. \ref{Sec:Flat}, a density perturbation of weight $w$
redshifts at large cosmological time in the asymptotic conformal scaling region of
de Sitter space according to
\be
\delta \rho \propto a^{-w_{\rho}}\,.
\label{delrhoscal}
\ee
If the other components of the general stress tensor perturbation (\ref{tendecomp})
scale the same way at late cosmological time $H\tau \gg 1$,
then from the solution (\ref{Upsolndelrho}) under dilations the gravitational  potentials 
$\Upsilon_{\cal A}, \Upsilon_{\cal C}$ scale according to
\be
\Upsilon_{\cal A} \,, \Upsilon_{\cal C} \propto a^{-s}\,,\qquad {\rm with} \qquad  s = w_{\rho} - 2
\label{swrho}
\ee
in the asymptotic region of de Sitter space. This establishes the scaling weight
of the potentials in terms of the energy density perturbations in the scaling region.
In particular it shows that the scaling weight $s=0$ of the gravitational potentials
necessary to give rise to a scale invariant Harrison-Zel'dovich spectrum of CMB 
temperature anisotropies must arise from a conformal weight $w_{\rho}=2$
energy density perturbation in de Sitter space. Such $\delta \rho\propto  a^{-2}$
behavior has been found in the energy density perturbations 
of a massless, minimally coupled scalar field in de Sitter space \cite{Attractor},
exactly as expected on general grounds from (\ref{minmass}) and (\ref{Delrho})
for $M =0$ and $w_{\Phi} = 0$. Let us emphasize again that these are quantum 
fluctuations intrinsic to de Sitter space, with vanishing expectation value
of the scalar field $\langle \Phi\rangle = 0$, and no slow rolling inflaton.

We have used the notation $s$ for the gravitational potential scaling dimension (\ref{swrho}) 
because it is not a conformal weight in the same sense as $w_{\rho}$. The reason is
that the Laplacian $\nabla^2$ is {\it not} a conformal differential operator in $3$ dimensions 
(as it is in $2$ dimensions). Thus the gravitational potentials from the linearized Einstein eqs.
satisfying (\ref{Upsolndelrho}) are {\it not} conformal fields possessing well-defined transformation
properties under the special conformal transformations (\ref{speconf}). This has no effect on the
two-point power spectrum (\ref{C2}), whose form is determined by global dilations alone,
without the full $SO(4,1)$ conformal group. The power of $k^2$ needed to convert energy 
density fluctuations to gravitational potential fluctuations (hence CMB temperature fluctuations), 
which follows from (\ref{Upsolndelrho}) and global scaling under dilations is all that is needed
to convert the power spectrum of density perturbations to that of the gravitational potential
in de Sitter space. 

The non-convariant transformation properties of $\nabla^2$ in $3$ dimensions does mean 
that treating the energy density as a conformal field with conformal weight $w_{\rho} =2$,
whose three-point correlator is given by (\ref{G3form}) or (\ref{G3corr})-(\ref{Zdef}) in Fourier space,
satisfying the Ward identity (\ref{identitycon}), and then dividing by $k_1^2 k_2^2 k_3^2$ according 
to (\ref{weight0}), is {\it not} equivalent to treating the gravitational potentials and CMB 
temperature fluctuations themselves as conformal fields of weight $s = 0$. This is obvious from 
the fact that treating the energy density $\delta \rho$ as a conformal field with a definite weight gives 
\be
\frac{1}{|k_1|^2 \, |k_2|^2\, |k_3|^2}\, \int\!\frac{d^3\vec p}{|\vec p - \vec k_1|^{3-w_{\rho}} 
|\vec p + \vec k_2|^{3-w_{\rho}} |\vec p|^{3-w_{\rho}}} \qquad {\rm for } \ w_{\rho} \ {\rm near}\ 2\,,
\label{weight2}
\ee
leading to eq. (\ref{bispectCMB}) of the previous section for the CMB bispectrum, while the second 
prescription of treating the gravitational potential(s) and CMB temperature fluctuations themselves as 
conformal fields of weight $s$ would give
\be
\int\!\frac{d^3\vec p}{|\vec p - \vec k_1|^{3-s} 
|\vec p + \vec k_2|^{3-s} |\vec p|^{3-s}} \qquad {\rm for } \ s \ {\rm near}\  0\,.
\label{weight0}
\ee
Clearly (\ref{weight2}) and (\ref{weight0}) and their corresponding shape functions are quite 
different including in the squeezed limit, although they have the same {\it overall} scaling under
global dilations $\vec k_i \rightarrow \lambda \vec k_i$ of all $3$ vectors simultaneously if $s$
is given by (\ref{swrho}). This is the reason that the authors of \cite{KehRio} obtained a
different result for the squeezed limit of the non-Gaussian CMB bispectrum under the
second hypothesis than we have obtained under the first hypothesis of treating $\delta \rho$
as a conformal field of weight near $2$, rather than any error in evaluating the integral in 
terms of hypergeometric functions, either in Sec. \ref{Sec:Bispectrum} or \cite{KehRio}. 

\section{The Static Frame and Conformal Invariance on the de Sitter Horizon}
\label{Sec:Static}

The embedding of the flat spatial coordinates (\ref{flatRW}) is one way in which conformal 
invariance can be realized in de Sitter spacetime. The use of flat spatial sections is not 
essential in fact. The conformal group of the closed ${\mathbb S}^3$ and open hyperbolic 
${\mathbb H}^3$ spatial sections in cosmological coordinates is also the de Sitter isometry 
group $SO(4,1)$. Hence the main results apply to all cosmological homogeneous and 
isotropic metrics whether the spatial sections are open, closed or flat. 

In this section we consider a rather different realization of conformal invariance from 
embedding in de Sitter space based on the existence and kinematics of the cosmological 
horizon. The leads to a different set of conformal Ward identities and a different form 
for the CMB non-Gaussian bispectrum, as well as a rather different suggestion
for the physical origin of CMB anisotropies, and models of dark energy. As in the previous 
case, the basic kinematics is best illustrated by exhibiting explicit coordinates. In this case 
we make use of the static coordinates of de Sitter space (reviewed in Appendix \ref{App:Geometry}) .

In the static coordinates the line element of de Sitter spacetime can be expressed in the form
\be
ds^2 = - (1-H^2 r^2)\, dt^2 + \frac{dr^2}{(1-H^2r^2)} + r^2 d \Omega^2\,.
\label{dSstat}
\ee
These static coordinates are quite different from the spatially flat FLRW coordinates 
(\ref{flatRW}) globally. Rather than exponentially expanding, the metric in (\ref{dSstat}) 
is independent of time $t$. Rather than being spatially homogeneous, the
static coordinates have a preferred origin about which they are rotationally
invariant. The horizon for the observer at $r=0$ is apparent in (\ref{dSstat})
as the spherical surface at $r= H^{-1} \equiv r_{\!_H}$ where the line element
becomes singular. Notwithstanding their rather different forms, the line element 
(\ref{dSstat}) is a representation of de Sitter spacetime equally valid to the FLRW 
line element (\ref{flatRW}) in their region of common applicability within the
observer's cosmological horizon. 

The explicit coordinate transformations from flat FLRW coordinates (\ref{flatRW}) 
to static coordinates (\ref{dSstat}) are given by Eqs. (\ref{FLRWtostatic})-(\ref{statictoFLRW}) 
of Appendix \ref{App:Differential}. Note in particular from (\ref{FLRWtostatic}) 
\be
a(\tau) = e^{H\tau} = \sqrt{1-H^2r^2} \,e^{Ht}\,,
\label{tauHrt}
\ee
so that the distant past $\tau \rightarrow -\infty , a(\tau) \rightarrow 0$ corresponds to {\it either} 
$t \rightarrow -\infty$ or $r \rightarrow r_{\!_H}$. Hence in a de Sitter universe (or a universe
dominated by dark energy which approximates a de Sitter universe) primordial fluctuations
usually thought of as arising in a very early epoch in the history of the universe, may
in fact arise on or near the cosmological horizon at $r \rightarrow r_{\!_H}$ instead. 

Note next that in the static frame the metric factor
\be
f(r) \equiv 1 - H^2r^2 \rightarrow 0\qquad {\rm as} \qquad r \rightarrow r_{\!_H}\,,
\label{fdef}
\ee
so that the static time variable drops out completely at the horizon. The invariant distance 
function in de Sitter space defined previously by (\ref{zflatRW}) becomes in the static coordinates
(\ref{dSstat})
\be
1 - z(x,x') = \frac{1}{2} \left\{ 1 - H^2 rr' \hat n \cdot \hat n' - \sqrt{1 -H^2r^2}\sqrt{1 -H^2r^{\prime 2}}\,
\cosh \left[H(t-t')\right]\right\}\,,
\label{zstat}
\ee
where $\hat n, \hat n'$ are the two unit direction vectors on the sphere. Eq. (\ref{zstat}) shows 
that if the two points are on the horizon $r=r' = r_{\!_H}$ with respect to the origin of static coordinates, 
then this function becomes simply
\be
1 - z(x,x') \rightarrow \frac{1}{2}\, (1 - \hat n \cdot \hat n')\,,
\label{zhoriz}
\ee
and the de Sitter invariant function $z(x,x')$ when both points are on the horizon is a function only of the $O(3)$ 
invariant quantity on the sphere, namely $\hat n \cdot \hat n'$.

That the cosmological horizon is also the locus of conformal behavior may be seen in several 
different ways. On the static $t=$ const. hypersurfaces the remaining three dimensional line element 
of (\ref{dSstat}) takes the form
\be
ds^2\big\vert_{dt =0} = f(r)\, d \ell^2_{\mbox{\tiny \L}} = f(r)  \left[dr^{*\,2} + H^{-2} \sinh^2(Hr^*) \,d\Omega^2\right]\,,
\label{optical}
\ee
where $dr^* \equiv \frac{dr}{f}$ and
\be 
Hr^* = {\rm tanh}^{-1}(Hr) = \frac{1}{2} \ln \left(\frac{1 +Hr}{1-Hr}\right)\,.
\ee
Hence the horizon at $r = H^{-1} = r_{\!_H}$ is mapped to $r^* = \infty$ in these coordinates, and 
the line element within the square brackets of (\ref{optical}) may be recognized as a standard form 
of three dimensional \L obachewsky (\L ) or Euclidean Anti-deSitter (EAdS$_3$) space, consisting 
of one sheet of the doubled sheeted hyberboloid with isometry group $SO(3,1)$. Thus, by well-known 
arguments of AdS/CFT correspondence \cite{Witten}, one expects conformal behavior at the two dimensional 
horizon boundary, namely on the sphere ${\mathbb S}^2$ of directions $\hat n= (n_x, n_y, n_z)$ with 
$\hat n\cdot\hat n =1$ \cite{MazMor}. This conformal behavior can be seen most clearly by introducing 
Poincar\'e coordinates for the three dimensional optical metric, {\it viz.}
\bes\bea
&& \bar x \equiv \frac{r\, n_x}{1-Hr n_z}\,,\\
&& \bar y \equiv \frac{r\, n_y}{1-Hr n_z}\,,\\
&& \bar z \equiv \frac{r_{\!_H}\, f^{\frac{1}{2}}}{1-Hr n_z} 
= \frac{1}{H}\, \frac{\sqrt{1-H^2r^2}}{1-Hr n_z}\,,\label{zbardef}
\eea\label{Poincoor}\ees
so that the three dimensional \L obachewsky metric in (\ref{optical}) takes the standard 
Poincar\'e form
\be
d \ell^2_{\mbox{\tiny \L}} = \frac{1}{H^2 \bar z^2} \left( d\bar z^2 + d\bar x^2 + d\bar y^2\right)\,.
\label{Poinmetric}
\ee
In these Poincar\'e coordinates the two dimensional horizon boundary manifold ${\mathbb S}^2$
at $\bar z = 0$ has been mapped into the ${\mathbb R}^2$ plane, parameterized by $(\bar x, \bar y)$,
by stereographic projection \cite{MazMor}. The north pole at $n_z = 1$ and $r = r_{_H}$ is mapped 
to the point at $\bar z = \infty$. It is clear from (\ref{Poinmetric}) that scale transformations 
of the ${\mathbb R}^2$ plane $(\bar x, \bar y) \rightarrow \lambda\, (\bar x, \bar y)$ at fixed $\bar z$ 
can be compensated by a corresponding simple scale transformation on $\bar z \rightarrow \lambda \bar z$.
This is an exact symmetry of the three dimensional \L obachewsky metric. When this scale transformation 
is supplemented by the two special conformal transformations of ${\mathbb R}^2$, one obtains 
the full global conformal group of either ${\mathbb R}^2$ or ${\mathbb S}^2$, namely $SO(3,1)$,
which is exactly the isometry group of the three dimensional \L obachewsky space $d \ell^2_{\mbox{\tiny \L}}$.

On ${\mathbb S}^2$ the action of $SO(3,1)$ can also be represented as the group of proper Lorentz 
transformations acting on the directions of radial light rays in $3 + 1$ dimensions, {\it c.f.} 
Appendix \ref{App:Conformal}. In de Sitter space this is the subgroup of the de Sitter group
that leaves the static time $t$ and three dimensional metric (\ref{optical}) invariant. 
Again the conformal group of a dimension $d-1$ slicing is isomorphic to the symmetry 
group of the embedding $d$ dimensional space, with now $d=3$. Since $SO(3,1) \subset SO(4,1)$, 
the mapping of Killing vectors of de Sitter spacetime to conformal Killing vectors on ${\mathbb S}^2$ 
cannot be one to one, but there is a projective (several to one) mapping, which is described in detail 
in Appendix \ref{App:Killingstatic}. Hence conformal transformations of the spherical horizon
boundary metric at $r=r_{_H}$ can be viewed as coordinate transformations which leave the three 
dimensional de Sitter metric at constant $t$ invariant, and conformal invariance on the cosmological 
horizon follows from de Sitter invariance of the bulk, just as in the ${\mathbb R}^3$ embedding 
considered previously.

The conformal behavior at the cosmological horizon of de Sitter space in the static frame
may viewed in another way. In coordinates (\ref{dSstat}) because of the $r$
dependent compression of physical time $\sqrt{f}dt$, there is the usual
kinematic gravitational redshift/blueshift of the local frequency
\be
\omega_{loc}(r) = \frac{\omega}{\sqrt{f(r)}}  = \frac{\omega}{\sqrt{1-H^2r^2}}
\label{blueshift}
\ee
measured at $r$ relative to that at the origin. The corresponding energy $\hbar \omega_{loc}$
diverges as $r \rightarrow r_{\!_H}$ and therefore becomes much greater than any finite mass 
scale, which may be neglected at the horizon. As shown in Appendix \ref{App:Conformal} this 
is reflected in the fact that the wave equation for a scalar field of arbitrary finite mass (\ref{massPhi}) 
becomes indistinguishable from that of a {\it massless} conformal field in the horizon limit 
$r \rightarrow r_{\!_H}$. Thus conformal behavior in this limit is to be expected on physical
grounds of all mass scales becoming negligible.

The three dimensional line element (\ref{Poinmetric}) has a form quite analogous to the four 
dimensional de Sitter line element (\ref{conftime}) with the spacelike coordinate $\bar z$ taking 
the place of the conformal time $\eta$, and with the ${\mathbb R}^2$ boundary in (\ref{Poinmetric}) 
at $\bar z =0$ taking the place of the ${\mathbb R}^3$ boundary in (\ref{conftime}) and Figs. 
\ref{Fig:deShyper}-\ref{Fig:dSCarPen} at $\eta =0$, $I_+$. Thus the analysis of conformal behavior 
at the ${\mathbb S}^2$ horizon boundary parallels the corresponding late time behavior of the 
FLRW flat ${\mathbb R}^3$ embedding. The scale transformation 
$(\bar x, \bar y) \rightarrow \lambda\, (\bar x, \bar y)$ is analogous to (\ref{dilxa}) 
in the full de Sitter metric (\ref{conftime}) in its Poincar\'e conformal time coordinates. As 
in (\ref{timedep}) this makes possible the study of conformal behavior of fields or correlation 
functions on the horizon boundary by the scaling behavior with $\bar z \propto f^{\frac{1}{2}}$ 
in the corresponding bulk Poincar\'e coordinate. Although the representations of the conformal group 
generated at fixed $r$ need not be simple or irreducible in general, as the horizon boundary
is approached, $\bar z \sim \sqrt{f} \rightarrow 0$, $r \rightarrow r_{\!_H}$, simple power law 
behaviors and definite conformal weights can be produced. This conformal weight of a field 
$\Phi_{w}$ can be obtained by examining its power law behavior \cite{Witten}
\be
\Phi_{w} (r) \propto \bar z^w \propto [f(r)]^{\frac{w}{2}}\qquad {\rm as} \qquad r\rightarrow r_{\!_H}\,,
\label{weightH}
\ee
near the horizon boundary. Further, since the $SO(3,1)$ invariant distance function on \L obachewsky 
space, analogous to (\ref{distconf}) is
\bea
d(x,x') &=& \frac{1}{4\bar z \bar z'} \left[ (\bar z - \bar z')^2 + (\bar x - \bar x')^2 + (\bar y - \bar y')^2\right]\nn
&=& \frac{1}{2 \sqrt{f(r)f(r')}} \left[ 1 - \sqrt{f(r)f(r')} - H^2rr' \,\hat n \cdot \hat n'\right]
\eea
which is conformally related to the de Sitter invariant distance function (\ref{zstat}) with $t=t'$,
we see that positive conformal weight fields with $w > 0$ with scaling behavior (\ref{weightH}) will have
two-point correlation functions on the horizon sphere ${\mathbb S}^2$ with the angular behavior 
\be
\langle \Phi_w (r, \hat n) \Phi_w (r', \hat n') \rangle \sim  \left[d(x,x')\right]^{-w} 
\rightarrow B_w(r,r')\, (1- \hat n \cdot \hat n')^{-w}
\label{angcorsph}
\ee
with
\be
B_w(r,r') \propto [f(r)f(r')]^{\frac{w}{2}}\,,
\ee
analogously to (\ref{Acoeff}). This is a key result which we shall make use of in the next section.

\section{CMB Power Spectrum and Bispectrum on the de Sitter Horizon}
\label{Sec:SpectrumHorizon}

Since the physical origin of fluctuations generated after many $e$-foldings in an inflationary 
epoch on the one hand, and those generated on or near the cosmological horizon on the 
other are quite different, they lead to quite different observational consequences for CMB 
non-Gaussianity. To derive the form of the conformal Ward identities in this second case 
we make use of the representation of the conformal group of the sphere as the
proper Lorentz group in three spatial dimensions described in Appendix \ref{App:Conformal}
In fact, the conformal transformations of the directions on the sphere regarded as light rays 
may be placed in one-to-one correspondence with Lorentz boosts \cite{MazMot}.
The infinitesimal form of this conformal transformation parametrized by the Lorentz boost 
velocity vector $\vec v$ in (\ref{confmn}) or (\ref{confnOm}) is
\be
\delta_{\vec v} \hat n = -\vec v+ (\vec v \cdot \hat n)\, \hat n\,.
\ee
For the two-point correlation function $G_2$ of a conformal field with weight $w$, 
rotational invariance first restricts $G_2$ to be a function only of the invariant
(\ref{zhoriz}). Then the conformal identity is
\be
G_2(z; w) = \langle {\cal O}_{w}(\hat n_1){\cal O}_{w}(\hat n_2)\rangle = \Omega^{w} 
(\hat n_1)\Omega^{w}(\hat n_2)\,\langle {\cal O}_{w}(\hat n_1'){\cal O}_{w}
(\hat n_2')\rangle\,,
\label{G2}
\ee
or using (\ref{conftrans}) in infinitesimal form, we have
\be
\delta_{\vec v} G(z) = \vec v \cdot (\hat n_1w + \hat n_2w) G(z) + 
(\delta_{\vec v} \hat n_1 \cdot \vec\nabla_1 + \delta_{\vec v} \hat n_2 \cdot \vec\nabla_2)\,G(z)\,,
\ee
where $\vec D_1$ denotes the gradient with respect to $\hat n_1$ on the sphere and similarly
$\vec D_2$ denotes the gradient with respect to $\hat n_2$. From (\ref{zhoriz}) on the
de Sitter horizon
\be
z = z_{12} \equiv \frac{1 + \hat n_1 \cdot \hat n_2}{2}\,,
\ee
so that
\be
\vec \nabla_1 G_2(z;w) = \frac{dG_2}{dz}\, \vec\nabla_1 z = \frac{\hat n_2}{2} \frac{dG_2}{dz}\,,
\ee
and similarly for $\vec \nabla_2 G_2(z)$. Hence
\be
\delta_{\vec v} G_2(z) = \vec v \cdot (\hat n_1w + \hat n_2w) G_2(z) - 
\vec v \cdot (\hat n_1 + \hat n_2)\,(1-z) \frac{dG_2}{dz}\,.
\ee
Thus $\delta_{\vec v} G_{w}(z) = 0$ implies that $G_2(z; w)$ satisfies
\be
(1-z)\,\frac{dG_2(z;w)}{dz} = w\, G_2(z;w)\,,
\ee
which has the solution
\be
G_2(z;w) = \frac{a_2(w)}{(1-z)^{w}}\,.
\label{G2sph}
\ee
It is also straightforward to show from the finite transformation (\ref{confnOm}) that on the horizon
\be
[1 - \hat n_1'\cdot \hat n_2'] = \Omega(\hat n_1)\,\left[1- \hat n_1\cdot \hat n_2\right]\,\Omega(\hat n_2)\,,
\ee
with $\Omega(\hat n)$ given by (\ref{conftrans}). When raised to the power $-w$ 
this verifies (\ref{G2}) directly.

Turning next to the three-point function of three fields of equal conformal weight,
\be
G_3 (z_{12}, z_{23}, z_{13}) = \langle {\cal O}_{w}(\hat n_1){\cal O}_{w}(\hat n_2)
{\cal O}_{w}(\hat n_3)\rangle\,,
\ee
invariance under the conformal transformation (\ref{confmn}) implies
\be
G_3(z_{12}, z_{23}, z_{13}; w) = 
\Omega^{w} (\hat n_1)\Omega^{w}(\hat n_2)\Omega^{w}(\hat n_3)\,
\langle {\cal O}_{w}(\hat n_1'){\cal O}_{w}(\hat n_2')
{\cal O}_{w}(\hat n_3')\rangle\,,
\ee
or in infinitesimal form,
\bea
0&=& \delta_{\vec v} G_3 = w \vec v \cdot (\hat n_1 + \hat n_2 + \hat n_3) G_3
+ (\delta_{\vec v} \hat n_1 \cdot \vec\nabla_1 + \delta_{\vec v} \hat n_2 \cdot \vec\nabla_2 
+ \delta_{\vec v} \hat n_3 \cdot \vec\nabla_3)\,G_3\nonumber\\
&=&  w \vec v \cdot (\hat n_1 + \hat n_2 + \hat n_3) G_3
-  \vec v \cdot (\hat n_1 + \hat n_2) (1 - z_{12})\frac{\partial G_3}{\partial z_{12}} \nonumber\\
&& \qquad  -  \vec v \cdot (\hat n_2 + \hat n_3)
(1 - z_{23})\frac{\partial G_3}{\partial z_{23}} -  \vec v \cdot (\hat n_1 + \hat n_3)
(1 - z_{13})\frac{\partial G_3}{\partial z_{13}}\,.
\label{conf3iden}
\eea
This condition can be satisfied for arbitrary $\vec v, \hat n_1, \hat n_2, \hat n_3$ if and only if
\be
(1 - z_{12})\frac{\partial G_3}{\partial z_{12}} = (1 - z_{23})\frac{\partial G_3}{\partial z_{23}}
= (1 - z_{13})\frac{\partial G_3}{\partial z_{13}} = \frac{w}{2}\, G_3\,.
\ee
Therefore, the conformal Ward identity (\ref{conf3iden}) requires
\be
G_3(z_{12}, z_{23}, z_{13}; w) = \frac{a_3(w)}{[(1- z_{12})(1-z_{23})(1-z_{13})]^{\frac{w}{2}}} 
\label{G3sph}
\ee
for some constant $a_3(w)$. This procedure may clearly be continued to higher point 
functions, with the results similar to flat space with $r_{ij}^2 = |x_i - x_j|^2$ replaced by $(1- z_{ij})$.
Again the form of the four point trispectrum is constrained but not uniquely determined
by conformal invariance.

Unlike the case considered previously where the conformal invariance of $N$-point functions
was assumed to pertain on the flat ${\mathbb R}^3$ sections of a FLRW cosmological model,
and has to be translated first to Fourier space and then finally to angular directions on the sky
of the CMB, (\ref{G3sph}) gives the form of the non-Gaussian bispectrum directly in angular
variables, as observed from the origin $r=0$ of the static de Sitter coordinates (\ref{dSstat}).
This corresponds to treating the present universe as approximately de Sitter due to the 
dominance of dark energy today, and the CMB anisotropies as generated by de Sitter 
invariant fluctuations on the cosmological horizon, rather than having re-entered the
Hubble sphere from an earlier inflationary phase. We discuss in the next section a 
possible physical mechanism which could provide these conformal fluctuations on 
the cosmological horizon of de Sitter space. 

If one parallels the discussion in the FRLW case considered previously, one would again start
with energy density fluctuations of conformal weight $w_{\rho} \approx 2$, and then convert these
to fluctuations in the gravitational potentials $\Upsilon_{{\cal A}, {\cal C}}$ in the scalar sector
by solving the linearized Einstein eqs. according to (\ref{linres})-(\ref{Upsolnsumdiff}). An
important difference from the previous case is that whereas the Laplacian $\nabla^2$ is
not a conformal differential operator on ${\mathbb R}^3$, it is conformal when restricted
to ${\mathbb S}^2$. Thus the solution of the linear response eqs. (\ref{linres})-(\ref{Upsolnsumdiff})
will yield conformal scalar potentials that transform under the conformal group of ${\mathbb S}^2$,
$SO(3,1)$ with definite conformal weights, $w_{\Upsilon}$, rather than only under global
dilations only. We will give an explicit example of this in the next section. 

In addition, in a cosmological model of this kind, the observable CMB power spectrum may be obtained 
directly in terms of the angular variables on the sky from conformal field(s) of definite conformal 
weight $w = 0$ (or nearly zero), without any need of the passing through the intermediate step
of spatial Fourier transforms on ${\mathbb R}^3$. Inspection of (\ref{angcorsph}) or (\ref{G2sph}) 
shows that that a weight $w=0$ field on ${\mathbb S}^2$ has an angular correlation function
\be
G_2(z; 0) = C_2 \ln (1 - \hat n\cdot \hat n')  + const.
\label{HZ0}
\ee
which is exactly the angular correlation of the classical Harrison-Zel'dovich CMB spectrum
(\ref{G2log}) in the Sachs-Wolfe regime. Thus, at the level of the two-point power spectrum 
alone one {\it cannot distinguish} this radically different possibility for the origin of the
CMB anisotropies and power spectrum from the standard FLRW cosmological model.
In other words the observable CMB two-point power spectrum is identical in the three 
quite different physial models:
\begin{enumerate}
\vspace{-1mm}
\item The usual slow roll inflation scenario of CMB temperature anisotropies exiting and then 
reentering the cosmological horizon;
\vspace{-3mm}
\item Conformal fluctuations in de Sitter space of a scalar field of nearly zero conformal weight 
on flat FLRW sections (\ref{flatRW}), giving rise by dS/CFT correspondence to conformal weight 
zero fluctuations of the gravitational potential, as in (\ref{confmass}) to (\ref{G2log}); 
\vspace{-3mm}
\item Conformal fluctuations generated {\it on} the cosmological ${\mathbb S}^2$ horizon itself 
as the boundary of \L obachewsky space by scalar fluctuations with conformal weight $w \approx 0$, 
as in (\ref{G2}), (\ref{G2sph}) and (\ref{G2sph}). 
\end{enumerate}
\vspace{-1mm}

A scale invariant Gaussian CMB power spectrum with conformal weight
nearly zero, and logarithmic correlations on ${\mathbb S}^2$ as in (\ref{HZ0}) is obtained
from slow roll inflation, or from the conformal invariance inherited from $SO(4,1)$ de Sitter 
invariance, whether we regard the primordial fluctuations as arising on ${\mathbb R}^3$ 
or ${\mathbb S}^2$ embeddings in a dark energy dominated universe, such as that postulated 
in inflation, or even as applies to the observable universe today for a much lower value of $H$. 
The details of the CMB power spectrum such as the acoustic peaks will be produced in
the interior of the cosmological horizon by the same causal physics in any case, no
matter how the primordial fluctuations were generated, whether on or from beyond the Hubble 
sphere. Thus the two-point power spectrum alone cannot provide definitive evidence of 
detailed inflationary scenarios or the origin of the CMB anisotropies.

When one shifts attention to the non-Gaussian correlations, the situation is quite different.
The bispectrum in the ${\mathbb S}^2$ case (\ref{G3sph})
becomes in limit $w \rightarrow 0$ 
\be
G_3(z_{12}, z_{23}, z_{13}; 0) = C_3 \left[\ln (1 - \hat n_1\cdot \hat n_2) + 
\ln (1 - \hat n_2\cdot \hat n_3)
+ \ln (1 - \hat n_3\cdot \hat n_1)\right]\,,
\label{G3Del0}
\ee
which is completely separable, and quite different again from either (\ref{S2}) or the
single field slow roll inflation prediction (\ref{slowroll}). In fact, in this limit
(\ref{G3Del0}) is {\it ultralocal}, in the sense that the shape function corresponding
to it in the flat Fourier space variables of the previous section has an additional
$\delta-$ function. To see this, note that in flat coordinates
\be
\ln | \vec x_1 - \vec x_2|^2 = \ln (2|\vec x|^2) + \ln (1 - \hat n_1\cdot \hat n_2)
\label{logx}
\ee
if the two points $\vec x_{1,2} = |\vec x|\, \hat n_{1,2}$ from which the CMB photons are emitted
are (neglecting fluctuations) at the same physical distance, as usually assumed in the standard 
cosmological model. Inserting (\ref{logx}) into (\ref{G3Del0}) for each of the factors, and
Fourier transforming with respect to each of the $3$ position variables $\vec x_i$ gives
\be
\tilde G_3(\vec k_1, \vec k_2, \vec k_3; 0) \sim  
\delta^{3}(\vec k_2 + \vec k_3) \frac{\delta^{3}(\vec k_1)}{k_2^3} \ln k_2 +
\delta^{3}(\vec k_1 + \vec k_3) \frac{\delta^{3}(\vec k_2)}{k_3^3} \ln k_3 +
\delta^{3}(\vec k_1 + \vec k_2) \frac{\delta^{3}(\vec k_3)}{k_1^3} \ln k_1 \,,
\ee
where again $k_i \equiv |\vec k_i|$. Extracting an overall $\delta^{3}(\vec k_1 + \vec k_2 + \vec k_3)$ 
and multiplying by $(k_1k_2k_3)^2$ to compare with the dimensionless shape function 
of a conformal weight $2$ field in (\ref{bispectrum}) we find that the result is proportional to
\be
S_3(k_1, k_2, k_3; 0) = k_1^2 \delta^{3}(\vec k_1) \,k_2\ln k_2  
+ k_2^2 \delta^{3}(\vec k_2)\,  k_3 \ln k_3^2 + k_3^2 \delta^{3}(\vec k_3) \, k_1\ln k_1^2\,,
\label{shape3}
\ee
which has support only when one of the $\vec k_i = 0$, {\it i.e.} in the squeezed
configuration. If one takes $w$ slightly greater than $0$ in (\ref{G3sph}), this
extreme concentration on the squeezed configuration is smoothed somewhat,
but should still be distinguishable from the local shape function (\ref{Slocal}).
If non-Gaussianities are detected in the CMB, and in particular the support
of the bispectral shape function is found to be highly peaked on the
squeezed configuration, consistent with (\ref{G3Del0}) or (\ref{shape3}) 
for low $\ell$ multipoles, it would be evidence in favor of conformal
invariant origin of the CMB anisotropies on the cosmological horizon,
as described in this section. If the shape of the primordial bispectrum can be 
determined (irrespective of magnitude), the three different possibilities for the 
spacetime locus and origins of the fluctuations themselves may be distinguished 
by observations. The bispectrum a generic freely falling observer at a point offset 
from the origin would see may be worked out with simple geometric considerations 
and will be presented elsewhere. 

\section{CMB Anisotropies from Cosmological Horizon Modes}
\label{Sec:Horizon}

Up until this point we have simply assumed the existence of scalar fluctuations in
de Sitter space, which couple to the gravitational field, and worked out the
purely kinematic consequences of that assumption. As is well known there
are no scalar fluctuations in the classical Einstein theory. Thus if we are to
do away with the scalar inflaton, the question naturally arises of what supplies
the necessary scalar degree of freedom in its place. In this section we 
discuss a possible quantum origin and physical mechanism for the generation of
scalar fluctuations in de Sitter space, which could give rise to the CMB anisotropy 
and non-Gaussianities on the cosmological ${\mathbb S}^2$ horizon, as discussed 
in the last section. For this we first recall some properties of quantum fields and 
their fluctuations in de Sitter space.

The first observation is that the $O(4,1)$ de Sitter invariant
state with correlation function (\ref{hypergeom}) is in fact a thermal state
with respect to the static time coordinate $t$ of (\ref{dSstat}). Mathematically
this follows from the fact that de Sitter invariant correlator is periodic
in imaginary time $t-t' \rightarrow t-t' + 2 \pi i r_{\!_H}$, as is immediately 
obvious from (\ref{zhoriz}) and the periodicity property of the hyperbolic function,
cosh$[H(t-t' + 2 \pi i r_{\!_H})] = {\rm cosh}[H(t-t')]$. The resulting periodicity of the
correlator (\ref{hypergeom}) is exactly the KMS condition for a propagator in
thermal field theory \cite{KMS}, with the temperature \cite{GibPer}
\be
T_{_H} = \frac{\hbar H}{2 \pi k_{_B}}\,.
\label{HawkdS}
\ee
In the $O(4,1)$ de Sitter invariant equilibrium state the radiation entering the 
horizon volume $r < r_{\!_H}$ is exactly compensated by the radiation leaving it,
so there is no net flux and the state is stationary with respect to the Killing
static time $t$. 

When one considers next fluctuations around the equilibrium state, it is natural 
to consider causal fluctuations in the temperature $T_{_H}$ within one horizon volume. 
At that point one immediately finds a perhaps surprising result: any finite temperature 
fluctuation in the spherical volume enclosed by $r=r_{\!_H}$ away from the equilibrium 
value $T_{_H}$  (no matter how small) produces an {\it infinite} stress-energy tensor at the
horizon. Indeed the expectation value of the renormalized stress-energy tensor 
has the form \cite{AndHis}
\be
\lag T^a_{\ b}\rag_{_R}  \rightarrow \frac{\pi^2}{90} \,\frac{k_{_B}^4}{(\hbar c)^3}\, 
\frac{(T^4 - T_{_H}^4)\,}{\left(1 - H^2r^2\right)^2}\ {\rm diag} \, (-3, 1, 1,1)\,,
\label{dSthermal}
\ee
as $r \rightarrow r_{\!_H}$. The quadratic power divergence in this limit can be understood
from the kinematic blueshift (\ref{blueshift}) and the fact that the stress tensor is
a dimension $4$ operator, and so scales as $\omega_{loc}^4(r)$. The divergence 
in (\ref{dSthermal}) vanishes if and only if $T = T_{_H}$ {\it i.e.} if and only if all 
fluctuations in the temperature within any given horizon volume are identically zero. 
Note that if such a fluctuation does occur it would spontaneously break the
larger de Sitter isometry group $O(4,1)$ to a subgroup $O(3,1)$ or $O(3)$,
selecting from the homogeneous ensemble where any point is equivalent to any 
other, a preferred origin at $r=0$ around which the fluctuation is centered \cite{gstar}.

The quantum matter source on the right side of (\ref{scEin}) that gives rise to these
fluctuations can be described by an effective action functional, the relevant part of 
which can be written in the form,
\be
b'S^{(E)}_{anom}= \frac{b'}{2}\int d^4x\sqrt{-g}
\left\{ - \varphi \Delta_4 \varphi + \left(E - \frac{2}{3}\sq R\right)\varphi\right\}
\label{effact}
\ee
in terms of a new scalar field $\varphi$. Here
\bes\bea
E &\equiv& ^*\hskip-1mmR_{abcd}\,^*\hskip-.1cm R^{abcd} = 
R_{abcd}R^{abcd}-4R_{ab}R^{ab} + R^2 \\
b'&=& -\frac{1}{360 (4 \pi)^2}\, (N_S + 11 N_W + 62 N_V)\,,\\
\Delta_4 &\equiv& \sq^2 + 2 R^{ab}\nabla_a\nabla_b - \frac{2}{3}R \sq +
\frac{1}{3} (\nabla^a R)\nabla_a = -\sq \, (- \sq + 2H^2)\,,
\label{Del4}\eea\ees
in terms of the number of the underlying massless quantum conformal scalar ($N_S$), Weyl 
fermion ($N_W$) and vector ($N_V$) fields contributing to the quantum stress tensor.
The fourth order differential operator $\Delta_4$ is conformally invariant (when multiplied
by $\sqrt{-g}$) and factorizes in de Sitter spacetime as indicated in the last form of (\ref{Del4}).
Let us emphasize that the scalar $\varphi$ due to the trace anomaly is not introduced 
as adding new inflaton degrees of freedom in a model dependent way, but rather is 
the result of quantum fluctuations of Standard Model massless fields (such as the photon, 
or the graviton itself) in curved space with no additional assumptions. The field $\varphi$ is
a purely quantum scalar degree of freedom over and above the classical transverse, traceless
degrees of freedom in General Relativity. It can be understood as describing two-body correlations 
of the underlying quantum theory \cite{Zak}.

In ref. \cite{DSAnom} the solutions of eqs. (\ref{scEin}) with the stress-energy tensor derived from
(\ref{effact}) were studied in both the flat FLRW coordinates (\ref{flatRW}) and the static coordinates 
(\ref{dSstat}). Linearized perturbations of the stress tensor $\delta \lag T^a_{\ b}\rag$ of the form 
(\ref{dSthermal}) were found, corresponding to the fluctuation of the thermal state away from its 
strict Hawking-de Sitter value (\ref{HawkdS}). The large effects on the stress tensor 
at the de Sitter horizon as in (\ref{dSthermal}) are due to the perturbations of the anomaly 
scalar $\varphi$ in the one-loop effective action (\ref{effact}). Note that the stress tensor of 
these fluctuations need not become infinite on the horizon, but only sufficiently large to 
influence the de Sitter geometry through the semiclassical Einstein eqs. (\ref{scEin}), 
which couple the metric perturbation to the fluctuations of the anomaly scalar 
$\varphi \equiv \delta \phi$.

In order to determine how the perturbations of stress tensor due to the trace anomaly
behave under conformal transformations in de Sitter space, and find their conformal
weights, we apply the general method of Sec. \ref{Sec:GravPert} to the these fluctuations
of $\delta \lag T^a_{\ b}\rag$, first in the flat FRLW coordinates (\ref{flatRW}) and then
on the de Sitter horizon sphere. In the coordinates (\ref{flatRW}) the linear response
equations in de Sitter space are of the form (\ref{linres}) with
\be
\delta \rho = 3\, \delta p = \frac{2 H^2 b'}{3} \frac{\vec \nabla^2}{a^2} u
\label{delrho3delp}
\ee
where the quantity $u$ is the effect of the quantum matter source, given in terms of the gauge 
invariant perturbation of the anomaly scalar field $\varphi$ in (\ref{effact}) by the relations \cite{DSAnom}
\bes\bea
u&\equiv& \left(\frac{\partial^2}{\partial\tau^2} + H \frac{\partial}{\partial \tau} 
- \frac{\vec \nabla^2}{a^2}\right)\phi\,,\\
\Delta_4 \phi &=& \left(\frac{\partial^2}{\partial\tau^2} + 5H \frac{\partial}{\partial \tau} + 6H^2 
- \frac{\vec \nabla^2}{a^2}\right) u = 0\,.\label{phiub}
\eea\label{phiu}\ees
The condition (\ref{delrho3delp}) with (\ref{linrestrace}) implies that the linearized perturbation 
in the Ricci scalar vanishes, $\delta R = 0$, which removes the possibility of very high frequency 
Planck scale perturbations from the fourth order operator $\Delta_4$, and guarantees that the 
remaining solutions of (\ref{linres})-(\ref{phiu}) are in the long wavelength regime of validity of the 
semiclassical eqs. far below the Planck scale in energy. Thus (\ref{linres}) with (\ref{phiub}) 
are finally second order equations for $u, \Upsilon_{\cal A}, \Upsilon_{\cal C}$ with no Planck 
scale solutions. The solutions of (\ref{phiub}) in flat FLRW coordinates are
\be
u_{\vec k \pm} = e^{i \vec k\cdot \vec x} \, \frac{1}{a^2}\exp\left(\pm \frac{ik}{Ha} \right)
\label{upm}
\ee
in which the gravitational constant $G$ enters only  through the combination $GH^2 \ll 1$ 
describing the weak coupling of the quantum anomaly to the metric perturbation in (\ref{linres}) 
for a macroscopically large de Sitter space. 

Note that these modes (\ref{upm}) solving the anomaly scalar eqs. (\ref{phiu}) describe an 
additional effective scalar degree of freedom, arising from the purely quantum effects of the
trace anomaly, which act as linearized sources to the usual gravitational potentials 
$(\Upsilon_{\cal A}, \Upsilon_{\cal C})$ in the scalar sector of linearized gravity, without 
introducing an inflaton field. Thus whereas in the purely classical theory with $b'=0$, 
there are no dynamical scalar degrees of freedom and the only solutions of (\ref{linres}) 
with $b' = 0$ are trivial, once $b' \neq 0$ the anomaly scalar degree of freedom drives 
fluctuations in the gravitational potentials much as scalar inflaton fluctuations do in the 
conventional picture. In contrast to the conventional picture these are generated by quantum 
fluctuations described by the anomaly effective action (\ref{effact}), and are intrinsic to de Sitter 
space, requiring no fine tuned inflaton potential or slowly rolling classical expectation value 
$\phi_{cl}$ of a de Sitter breaking inflaton field.

Using the conservation eqs. (\ref{deltacons}) and (\ref{phiub}) we find for the other components 
of the anomaly stress tensor in the scalar sector defined in 
\bes
\bea
&&V = \frac{2H^2b'}{3} \left(\frac{\partial u}{\partial \tau} + 2 H u\right)\,,\\
&&W = - \frac{2 H^2 b'}{3} \frac{\vec \nabla^2}{a^2} u = - \delta \rho\,.
\eea
\ees
Thus using (\ref{Upsolndelrho}) the solution of the linear response eqs. for this anomaly scalar perturbation is
\bes
\bea
&&\hspace{2cm} \Upsilon_{\cal A} + \Upsilon_{\cal C} = - \frac{16\pi G H^2 b'}{3}\, u\,,
\qquad \vec k \neq 0\ \label{sumUps}\\
&&\frac{\vec \nabla^2}{a^2} (\Upsilon_{\cal A} -\Upsilon_{\cal C}) =
16\pi G H^2 b' \left( H \frac{\partial}{\partial \tau} + 2H^2\right)\, u\,,\qquad \vec k \neq 0\,,
\label{diffUpseq}
\eea 
\label{bothUpseq}\ees
for the sum and difference of the gravitational potentials for spatially inhomogeneous perturbations,
in Fourier space $\vec k \neq 0$. Since $u \sim e^{-2H\tau}$ for $H \tau \gg 1$, the first of these
relations show that the sum of the gravitational potentials $\Upsilon_{\cal A} + \Upsilon_{\cal C}$ 
has a scaling dimension $s_+ = 2$ in the conformal scaling region of de Sitter space at late
FRLW time. Thus this combination cannot give rise to the scaling needed (namely, $s=0$) for an
approximately flat Harrison-Zeld'ovich spectrum of CMB fluctuations. For the second linear
combination of potentials, $\Upsilon_{\cal A} - \Upsilon_{\cal C}$ in (\ref{diffUpseq}), inspection
of (\ref{upm}) shows that the differential operator on the right side of (\ref{diffUpseq}) reduces
its behavior to $a^{-3}$ for $H \tau \gg 1$, so that the difference of potentials has the
scaling dimension $s_- = 1$ at late FRLW time, again not the behavior needed for generating
a flat CMB spectrum. Thus we conclude that the scalar degree of freedom present in the quantum
trace anomaly does not generate the observed spectrum in the first realization of conformal
invariance in de Sitter space in the flat FRLW sections (\ref{flatRW}) at late times.

This negative result is a consequence of the special role that the strictly homogeneous mode
with $\vec k =0$ plays in the FRLW case, and the existence of the $5$ conformal Killing 
vectors of de Sitter space, which lead to enhanced gauge symmetry at $\vec k =0$ and to 
non-invertibility of the scalar Laplacian in (\ref{diffUpseq}) in the scaling region, $H\tau \gg 1$. 
The additional coordinate gauge modes at the singular $\vec k=0$ limit in flat FRLW 
coordinates also preclude any use of the homogeneous modes of the gravitational 
potentials to extract their scaling dimensions along the lines of (\ref{timedep}). In the 
second realization of conformal invariance on the de Sitter horizon, as shown explicitly
in Appendix \ref{App:Killingstatic}, the Killing vectors of de Sitter space (\ref{timeT})-(\ref{twosol}) 
do not have non-trivial $\ell =0$ harmonics on ${\mathbb S}^2$, so that the Laplacian
on ${\mathbb S}^2$ is invertible, and the linearized Einstein eq. (\ref{diffUpseq})
for the difference scalar gravitational potentials $\Upsilon_{\cal A} -\Upsilon_{\cal C}$
is solvable (up to a harmless constant). The conformal weight $w_{\Upsilon} =0$
may be also obtained for this linear combination of gravitational potentials by a method
analogous to (\ref{timedep}) in the static coordinates, as we shall now show.

In the de Sitter cosmological horizon limit  $r \rightarrow r_{\!_H}$, we can find the 
conformal weights of $u$ and scaling of $\Upsilon_{\cal A, C}$ by converting the differential 
operators in (\ref{phiub}) and (\ref{diffUpseq}) to static coordinates (\ref{dSstat}), and examining
the functions of $r$ only obtained in the horizon limit. Since $r$ is related to $\bar z$ of the
Poincar\'e coordinates (\ref{Poincoor}) by (\ref{zbardef}), and conformal weights obtained
from (\ref{weightH}) are necessarily tied to the behavior of correlation functions on ${\mathbb S}^2$
by (\ref{angcorsph}) through de Sitter invariance, this is the precise analog of obtaining
conformal weights through the simple time dependence of spatially homogeneous functions
via (\ref{timedep}) in the FRLW case.

Using the explicit forms of the coordinate transformations in Appendix \ref{App:Differential} 
and assuming that $u, \Upsilon_{\cal A}, \Upsilon_{\cal C}$ are now functions
only of the static radius $r$, (\ref{phiub}) becomes
\be
-\frac{1}{r^2} \frac{d}{d r}
\left[r^2 (1-H^2r^2)\frac{du}{d r}\right] + 2H^2r\frac{du }{dr} + 6H^2u = 0\,,
\label{ustatic}
\ee
which has the general solution \cite{Zak,DSAnom}
\be
u = \frac{c_1}{f} + \frac{c_2}{Hrf}\,,
\label{usoln}
\ee
for some constants $c_1$ and $c_2$. Now the scalar Laplacian in (\ref{Upsoln}) is
invertible without any problem (up to constant and $r^{-1}$ solutions, the first of which
is harmless and the latter of which is singular at $r=0$ and hence excluded). Thus 
eqs. (\ref{bothUpseq}) hold and (\ref{diffUpseq}) becomes
\be
-\frac{1}{r^2} \frac{d}{dr}\left(r^2\frac{d}{dr}\right)\left(\Upsilon_{\cal C} - \Upsilon_{\cal A}\right)
=  16\pi G H^4 b' \left(r \frac{d u}{dr}+ 2 u\right) \,,
\ee
which has the general solution 
\be
\Upsilon_{\cal C} - \Upsilon_{\cal A} = 8\pi G H^2 b' \left[\frac{c_1}{Hr} \ln \left(\frac{1-Hr}{1+Hr}\right) 
+ \frac{c_2}{Hr} \ln f\right] \,,
\label{diffUps}
\ee
again up to solutions of the homogeneous eq. $(\nabla^2)^2 \Upsilon_0 = 0$. In static
coordinates these solutions are $\Upsilon_0 = c_0 + c_{-1}r^{-1} + c_1r + c_2r^2$, so that they are 
either singular at $r=0$, or give an irrelevant finite additive constant at $r= r_{\!_H}$.

Now the result (\ref{diffUps}) shows that the difference of gauge invariant gravitational potentials 
$\Upsilon_{\cal C} - \Upsilon_{\cal A}$ behaves {\it logarithmically} on the de Sitter horizon.
Since in the case that all the potentials are functions only of the static radial coordinate $r$,  
and by (\ref{weightH}) conformal weight $w$ fields scale like $f^{\frac{w}{2}}$ as 
$f(r) = 1 - H^2 r^2 \rightarrow 0$, the logarithmic scaling with $f(r)$ in (\ref{diffUps}) 
corresponds to conformal weight 
\be
s= w_{\Upsilon} = 0\,.
\label{s0}
\ee
In the ${\mathbb S}^2$ case the scaling dimension is also the conformal weight under 
the conformal group $SO(3,1)$ on the horizon.

Hence the conformal anomaly scalar in (\ref{effact}), itself a fluctuating scalar
degree of freedom drives the difference of gravitational potentials (\ref{diffUps}) at linear 
order in perturbations around de Sitter space, and provides a possible 
source for zero conformal weight fluctuations that can give rise to the correct scale invariant 
Harrison-Zel'dovich angular correlations at low $\ell$ multipoles according to (\ref{G2log}) 
and (\ref{HZ0}). The result (\ref{sumUps}) for the sum of gravitational 
potentials $\Upsilon_{\cal C} + \Upsilon_{\cal A}$ shows that this linear combination 
gives scaling dimension $s = -2$ according to (\ref{weightH}). A negative conformal 
weight means that this combination cannot give rise to well-defined unitary conformal 
field on the de Sitter horizon. It suggests instead unstable or tachyonic behavior that
requires a study of of the fully time dependent solutions of (\ref{Upsoln}). We note that 
projecting onto the logarithmic weight zero terms requires equal and opposite gravitational 
potentials $\Upsilon_{\cal A} = - \Upsilon_{\cal C}$, which interestingly is just the same 
linear combination generated by hydrodynamic perturbations (with no anisotropic stresses) 
in slow roll inflation models \cite{MukFelBra}. It is the combination $\Upsilon_{\cal C} - \Upsilon_{\cal A}$ 
of zero scaling dimension under global dilations that can give rise to the HZ spectrum. 

The logarithmic power correlation function on the sky (\ref{HZ0}) derives from that of
the anomaly scalar field $\varphi$ of zero conformal weight in (\ref{effact}).
Indeed the propagator for $\varphi$ is the inverse of the $\Delta_4$ operator
defined by (\ref{Del4}). Since in de Sitter space $\Delta_4 = -\sq (-\sq +2H^2)$ factorizes,
this propagator can be written as
\be
D(z(x,x')) = \frac{1}{2H^2} \left [-\frac{1}{\sq} - \frac{1}{-\sq + 2H^2}\right]
= \frac{1}{2H^2} \left [ G_0(z(x,x')) - G_{conf}(z(x,x'))\right]\,,
\label{Del4prop}
\ee
where
\be
G_0(z) = \frac{H^2}{16\pi^2}\, \left[ \frac{1}{1-z}  - 2 \ln (1-z)  + k_0\right] 
\label{minprop}
\ee
is the propagator of a massless, minimally coupled scalar field (on the space of 
non-constant modes) and $G_{conf}$ is the propagator of the massless,
conformally coupled field (\ref{conscalar}). Thus from (\ref{masslessprop}), (\ref{Del4prop}), 
and (\ref{minprop}),
\be
D(z) = - \frac{1}{16\pi^2} \ln \left(1 - z\right) + \frac{k_0}{32 \pi^2}
\label{Del4propz}
\ee
is a pure logarithm, up to an arbitrary additive constant. From (\ref{zhoriz}) on
the ${\mathbb S}^2$ de Sitter horizon,
\be
D(z) \rightarrow - \frac{1}{16\pi^2} \ln \left(1 -  \hat n \cdot \hat n' \right) + d_0\,,
\label{Dlog}
\ee
where $ \hat n \cdot \hat n' $ is the cosine of the angle between the two direction vectors 
on the sky viewed from the origin from which the radiation appears to originate. 
This correlator is that of the {\it two dimensional} conformal Laplacian operator 
on the sphere ${\mathbb S}^2$. This is the conformal weight zero logarithmic correlator 
that is transmitted to the difference of gravitational potentials $\Upsilon_{\cal C} - \Upsilon_{\cal A}$
by the semiclassical Einstein eq. (\ref{scEin}) and therefore to the CMB temperature 
fluctuation power spectrum through the Sachs-Wolfe effect for large angular scales, 
{\it i.e.} low multipole moments. Thus the anomaly scalar fluctuations governed
by (\ref{phiu}) can give rise to the CMB power spectrum in a fully $SO(3,1)$ 
conformally invariant way on the de Sitter horizon, without any need of 
the {\it ad hoc} introduction of a scalar inflaton field. 

Since the form of the two-point correlation function is dependent upon the 
conformal weight of the field giving rise to the fluctuations, and not upon the 
detailed dynamics, observations of the two-point correlations of the CMB 
anisotropies alone cannot distinguish (\ref{HZ0}) arising from fluctuations 
on ${\mathbb R}^3$ entering the horizon with spectral index $n_{_S}=1$ from 
those of (\ref{Del4propz}), arising from fluctuations of the anomaly scalar field 
$\varphi$ and $u$ near the cosmological horizon itself.  These globally radically 
different physical origins lead to the {\it same} gauge invariant gravitational 
potentials, and generate the same Harrison-Zel'dovich large angle CMB power 
spectrum directly on the sky, (\ref{G2log}), (\ref{HZ0}) or (\ref{Dlog}), for an observer 
at or near the origin $r=0$. As shown in Secs. \ref{Sec:Bispectrum} and 
\ref{Sec:SpectrumHorizon} the CMB bispectrum can be used to distinguish 
these different possible origins. The corrections to this angular form from 
displacements from the origin vanish at first order and will be presented 
in a separate publication. 

\section{Gravitational Waves from Conformal Invariance}
\label{Sec:GravWaves}

The considerations of conformal invariance can be extended equally well to
the spectrum of gravitational waves. The fundamental quantity is the
correlation function of transverse traceless tensors,
\be
G_{ijkl}^{(2)}(x,x') = \langle h_{ij}(x) h_{kl}(x') \rangle = 
\tilde C \int d^d p \, e^{i p\cdot(x-x')} \,\tilde P^{(2)}_{ijkl}(p) \,|p|^{n_{_T}-d}
\label{tencor}
\ee
where $\tilde P^{(2)}_{ijkl}(p)$ given by
\bes
\bea
\tilde P^{(2)}_{ijkl}(p) &=&
\frac{1}{2}\big[\Theta_{ik}\Theta_{jl} + \Theta_{il}\Theta_{jk}\big] - \frac{1}{d-1} \Theta_{ij}\Theta_{kl}\,,\\
\Theta_{ij}(p) &\equiv& \delta_{ij} - \frac{p_ip_j}{p^2}
\label{P2k}
\eea
\ees
is the projector onto transverse traceless tensors in flat Euclidean ${\mathbb R}^d$.
Eq. (\ref{tencor}) is the general form of the spin-2 gravitational wave correlation power
allowed by conformal invariance with a general spectral index $n_{_T} -d$ and overall amplitude $\tilde C$, 
which may be obtained by solving the conformal Ward identity relations analogous to 
(\ref{identity}) leading to (\ref{scalingtwo}). Since the projector $\tilde P^{(2)}_{ijkl}(p)$ is invariant 
under scaling $p_i \rightarrow \lambda p_i$, simple power counting in $p$ shows that the spin-2 
correlator in position space (\ref{tencor}) must take the form
\be
G_{ijkl}^{(2)}(x,x') = C\, P^{(2)}_{ijkl}(x,x') \, |x-x'|^{- n_{_T}}
\label{tenpos}
\ee
for some other constant $C$. One can find the full spin-2 conformal correlation function 
$G_{ijkl}^{(2)}$ in position space, obviating the need to construct the projector $P^{(2)}_{ijkl}(x,x')$ 
separately, by following a method for maximally symmetric spaces parallel to that for the 
de Sitter space graviton two-point function \cite{AntMotg}.

To this end note first that translational and rotational invariance in ${\mathbb R}^d$ requires the 
full spin-2 correlator $G^{(2)}_{ijkl}(x,x')$ to be a function only of the unit vector 
\be
n^i=\frac{(x-x')^i}{|x-x'|}\,.
\ee
and the invariant distance
\be
\mu(x,x') \equiv |x-x'|
\ee
Then noting that there are exactly $5$ four-index tensors symmetric under 
interchanges of $i$ and $j$, $k$ and $l$, and the interchange of the pair $(i, j)$ 
and $(k,l)$, that can be constructed out of the unit vector $n^i$ and the flat space 
metric $g_{ij} = \delta_{ij}$, namely
\bes
\bea
&&t^{(1)}_{ijkl} \equiv \delta_{ij}\delta_{kl}\,\\
&&t^{(2)}_{ijkl} \equiv n_in_jn_kn_l\,\\
&&t^{(3)}_{ijkl} \equiv \delta_{ik}\delta_{kl} + \delta_{il}\delta_{jk}\,\\
&&t^{(4)}_{ijkl} \equiv \delta_{ij}n_kn_l + n_in_j \delta_{kl}\,\\
&&t^{(5)}_{ijkl} \equiv -(\delta_{ik}n_jn_l + \delta_{il}n_jn_k + \delta_{jk}n_in_l + \delta_{jl}n_in_k)\,,
\label{t5}
\eea
\ees
one can expand $G_{ijkl}^{(2)} $ in this basis set of $5$ tensors multiplied by arbitrary
scalar functions of $\mu$, 
\be
G_{ijkl}^{(2)} (x,x')= \sum_{i=1}^5 \, G_i(\mu) \,t^{(i)}_{ijkl}\,.
\ee
The minus sign in (\ref{t5}) is in order to agree with the conventions of \cite{AntMotg}. Then using
\be
\partial_i n_j = \frac{1}{\mu} (\delta_{ij} - n_in_j)\,,
\label{diffn}
\ee
and imposing the conditions of tracelessness and transversality,
\be
G_{iikl}^{(2)} = 0\,\qquad {\rm and}\qquad \partial_i G_{ijkl}^{(2)} = 0\,,
\ee
one can derive $4$ independent conditions on the $5$ functions $G_i(\mu)$, {\it viz.}
\bes
\bea
&&dG_1 + 2G_3 + G_4 = 0\,\\
&&G_2 + dG_4 - 4 G_5 = 0\,\\
&&g' + \frac{d}{\mu}\, g - \frac{2d}{\mu}\, h = 0\,\\
&&h' + \frac{d}{\mu}\, h- \frac{d}{2\mu}\, g = -\frac{(d-2)(d+1)}{2\mu}\, G_4
\eea
\label{trtr}\ees
where a prime denotes the derivative with respect to $\mu$ and we have defined the linear combinations
\bes
\bea
&&g \equiv (d-1)G_4 - 2 G_3\,\\
&&h \equiv G_5 - G_3\,.
\eea
\label{gfdef}\ees
Now if we assume the power law dependence,
\be
g(\mu) = C \mu^{-n_{_T}} = \frac{C\quad}{ (x-x')^{2s_{_T}}}
\ee
which defines the tensor scaling exponent $s_{_T}$ in terms of the spectral index,
\be
s_{_T} = \frac{n_{_T}}{2}\,,
\label{wT}
\ee
then (\ref{trtr}) and (\ref{gfdef}) may be solved for
all $5$ scalar functions with the same power law in the form,
\be
G_i(\mu) = A_i \, \mu^{-2 s_{_T}}
\ee
with
\bes
\bea
&&A_1 = -\frac{C}{d} \left[ \frac{4s_{_T}(d-s_{_T})}{(d-2)(d+1)} - 1 \right]\,\\
&&A_2 = - \frac{4C s_{_T}(1+s_{_T})}{d(d+1)}\,\\
&&A_3 = \frac{C}{2} \left[ \frac{4(d-1)s_{_T}(d-s_{_T})}{d(d-2)(d+1)} - 1 \right]\,\\
&&A_4 = - \frac{4C s_{_T}(d-s_{_T})}{d(d-2)(d+1)}\,\\
&&A_5 = \frac{C s_{_T}}{d} \left[ \frac{2(d-1)(d-s_{_T})}{(d-2)(d+1)} - 1 \right]\,.
\eea
\ees
The constant $C$ is then fixed in terms of $\tilde C$ by identifying the $A_3$ coefficient that 
multiplies the tensor $t^{(3)}_{ijkl} = \delta_{ik}\delta_{jl} + \delta_{il}\delta_{jk}$ 
(which does not involve $n^i$) with the same tensor structure in (\ref{tencor}). 
Recalling (\ref{P2k}) this gives
\be
\tilde C = 2\,\frac{B_d(2s_{_T})}{(2\pi)^d}\, A_3 = C\,\frac{B_d(2s_{_T})}{(2\pi)^d}
\left[ \frac{4(d-1)s_{_T}(d-s_{_T})}{d(d-2)(d+1)} - 1 \right]
\ee
where $B_d(2s)$ is defined by 
\be
(x^2)^{-s}= B_d(2s)\int \frac{d^dp}{(2\pi)^d} \,e^{ip\cdot x}\,(p^2)^{s -\frac{d}{2}}  \,,
\label{Bdint}
\ee
so that
\be
B_d(2s_{_T})=\frac{2^{d -2s_{_T}}\pi^{\frac{d}{2}}\,\Gamma\left(\frac{d}{2} - s_{_T}\right)}{\Gamma(s_{_T})} \,.
\label{Bd}
\ee
For $d=3$ we have
\be
\tilde C = C \frac{\Gamma\left(\frac{3}{2} - s_{_T}\right)}{2^{2s_{_T}}\,\pi^{\frac{3}{2}}\,\Gamma(s_{_T})} 
\left[ \frac{2s_{_T}(3-s_{_T})}{3} - 1 \right] \rightarrow -\frac{C}{2\pi}\, s_{_T}\,,\qquad d=3
\ee
as $s_{_T} \rightarrow 0$. As in the scalar case, (\ref{G2log}) the zero weight limit produces a pole
term in position space $C$ if the momentum space normalization $\tilde C$ is held
fixed, and this pole term must be subtracted to obtain the normalization in the
logarithmic case. The tensor correlation (\ref{tencor}) or (\ref{tenpos}) may likewise be converted 
into a correlation of gravitational waves on the directions of the sky by assuming as in (\ref{C2}) that
the emission points are at equal distance  $|\vec x| = |\vec x'|$ from the observer.

As in the case of scalar perturbations, the same result for the two-point correlation 
(\ref{tencor}) or (\ref{tenpos}) is obtained whether one employs the conformal Ward 
identities inherited from $SO(4,1)$ de Sitter invariance to the ${\mathbb R}^3$ spatial 
slices of FRLW cosmology at fixed cosmological time $\tau$, or those of the $SO(3,1)$ 
subgroup of conformal transformations on the de Sitter horizon sphere ${\mathbb S}^2$ 
at fixed de Sitter static time $t$. Either way, once produced, the gravitational waves
will propagate within our cosmological horizon and affect the CMB photon polarization
and particularly the $B$-modes by the same scattering processes as those extensively 
investigated in present models \cite{Durr,Kos}.

Now if one and the same physical mechanism is responsible for the conformal invariance 
of both the scalar and tensor perturbations, as each are components of the same four 
dimensional spacetime linearized metric perturbation $h_{ab}$, one would expect them 
to have the {\it same} conformal scaling dimensions,  
\be
s= s_{_T}\,. 
\label{equalwt} 
\ee
Taking into account the definitions of the scalar and tensor spectral indices,
(\ref{specindex})-(\ref{scals}) and (\ref{wT}) respectively this implies
\be
n_{_T} = n_{_S} - 1\,.
\label{indexST}
\ee
The validity of this relation depends upon the effective theory of the origin 
of the scalar and tensor fluctuations having non-vanishing lowest conformal
weight representations. Thus, we have argued that the lowest conformal
weight scalar density perturbations should be $w_{\rho} = 2$, and these
lead generically to the scalar gravitational potentials having scaling 
dimension $s=0$ according to the linearized Einstein eqs. (\ref{scals}). 
This scaling dimension is actually realized for the scalar gravitational
potentials in the $\mathbb{S}^2$ realization of conformal symmetry in
de Sitter space (\ref{s0}). 

Correspondingly, the metric tensor perturbations in the transverse,
traceless sector are well-known to behave like a massless, minimally
coupled scalar field in de Sitter space, and so according to (\ref{nudef})
and (\ref{timedep}), the tensor modes have $s_{_T} = 0$ as well.
Indeed their propagator in de Sitter space behave logarithmically
with large invariant distance \cite{AntMotg}, just as would be expected
for a $s_{_T}=0$ field. It would be interesting to check this logarithmic
scaling of the tensor two-point function on the cosmological horizon
$\mathbb{S}^2$ realization of conformal symmetry in de Sitter space.
This relation of scalar and tensor spectral indices could furnish an important
check on the proposed origin of the CMB from conformal invariance as we 
have been considering in this paper, and in particular of the realization of 
conformal invariance on the cosmological horizon of de Sitter space
from the quantum fluctuations of the anomaly.

Note that the the equality of spectral weights and the relation (\ref{indexST}) 
between the spectral indices of the scalar and tensor fluctuations expected
from considerations of conformal invariance in de Sitter space differs from 
the predictions of slow roll inflation models where the scalar and tensor spectral 
indices have different dependences on the slow roll parameters \cite{LidL}. 
In practice the deviations of each spectral index ($n_{_S} -1$ in the 
scalar case and $n_T$ in the tensor case) is predicted to be very small, so this may be 
a difficult difference to detect. Conversely, if the spectral indices of the scalar and tensor
perturbations are found to differ measurably from the conformal expectation (\ref{indexST}), 
this would certainly make less plausible a common origin of the scalar and tensor fluctuations 
by conformal invariance. For a treatment of tensor modes in the slow roll model see \cite{Mal}.

As in the scalar case, conformal invariance alone does not predict the amplitude of
the gravitational wave component, provided they are small in magnitude so that
a well-defined geometrical background such as de Sitter space with its classical 
$SO(4,1)$ symmetry group applies. For that one needs a concrete dynamical realization
of conformal symmetry, such as the one proposed in Sec. \ref{Sec:Horizon} \cite{Ruba}.
In the semi-classical limit where fluctuations around the background are small, 
one would also expect the anomalous deviations from naive classical conformal 
scaling dimensions to be be small, and hence $n_{_S} -1= n_{_T} \approx 0$. 

Present observations indicate that that $n_{_S} \simeq 0.97 \pm 0.07$ (from WMAP data alone, 
with the uncertainty estimate reduced to $\pm 0.03$ if all relevant cosmological data is fit \cite{WMAP}). 
Thus a value slightly less that unity is currently preferred, whereas the conformal invariance hypothesis 
would seem to favor positive anomalous dimensions, based on unitarity of quantum field theory in 
de Sitter space. However, two caveats are in order here. First, unitarity of quantum field theory in 
cosmological spacetimes is a largely unexplored issue, and at the moment not even precisely
defined. Thus it is an interesting open question to ask if there are cosmological realizations of 
conformal field theories that possess negative conformal weight representations which could give 
rise to $w_{_{\Upsilon}} = s_{_T} <0$ and hence $n_{_S} -1 = n_{_T} <0$. Second, the observational 
data is processed assuming standard $\Lambda$CDM models, and hence the current quoted 
bounds on the scalar spectral index $n_{_S}$ are dependent upon the model assumptions.
In addition the low $\ell$ CMB data shows some interesting deviations from naive expectations 
which have been the subject of discussion in the literature \cite{CMBanom}. Hence it
might be wise to reserve judgment on the fitting of observations to present model assumptions,
and it might even be necessary to reconsider or substantially revise these assumptions
if dynamical generation of conformal invariance of the CMB by cosmological horizon modes 
and the non-Gaussian bispectrum (\ref{G3sph}) they predict are observed.

\section{CMB Non-Gaussianities as a Probe of Dark Energy Cosmology}
\label{Sec:DarkEnergy}

Our main purpose in this paper has been to show how conformal invariance of the fluctuations 
that give rise to CMB anisotropies may be derived from the embedding of the surface on which 
the fluctuations are generated in full de Sitter space. The mathematical derivation 
of conformal properties of CMB spectral functions from the geometric isometries of de Sitter 
space emphasizes the minimal assumptions and general nature of the results, which
are independent of detailed dynamical realizations of inflation. The essential speculation 
underlying this work is that the observed nearly flat CMB power spectrum may be a hint of a
more general and far-reaching conformal invariance, and that this may have arisen
from the fundamental symmetries of de Sitter space, independently of specific inflationary 
models, or even possibly in a dark energy dominated de Sitter phase similar to the
present epoch. 

The embedding of surfaces in de Sitter space with conformal invariance properties inherited 
from the geometric isometries of de Sitter space may be realized in two distinct and quite 
different ways. The first is the embedding of the usual flat FLRW cosmological space 
in de Sitter spacetime through (\ref{flatRW}) and (\ref{expand}). Conformal invariance is 
obtained because of the isomorphism between the conformal group of ${\mathbb R}^3$ of
flat spatial sections and $SO(4,1)$, the geometric isometry group of de Sitter space. This 
isomorphism is expressed most clearly and explicitly by the one to one mapping of the 
$10$ Killing vectors of de Sitter spacetime to the $10$ conformal Killing vectors of flat 
Euclidean ${\mathbb R}^3$. The representations of the conformal group become 
simple exponentially fast in the number of $e$-foldings of expansion, so that the 
approximation of exact de Sitter space and conformal invariance becomes asymptotically 
exact very rapidly at late times $H\tau \gg 1$, and the lowest conformal weight representation 
is selected in this dS/CFT limit. Conformal invariance in this realization then implies the form 
(\ref{G3corr})-(\ref{bispectrum}) for the non-Gaussian shape function of the CMB bispectrum 
at large angular separations.

Physically, this corresponds to requiring that the fluctuations which give
rise to the CMB anisotropies arose in a de Sitter inflation phase, similar to the 
standard picture, except that they must be generated intrinsically in de Sitter space from 
some degrees of freedom with conformal properties close to those of a massless, 
minimally coupled scalar field. This could be the inflaton of many currently popular 
scenarios for generating the initial fluctuations responsible for CMB anisotropies,
but it need not be. The important feature is only that the energy density fluctuations
of this scalar should have conformal weight $w_{\rho} \approx 2$, in order to give rise
to a Harrison-Zel'dovich energy fluctuation spectrum, and a flat CMB power spectrum
consistent with observations. The weight $w_{\rho} =2$ seems to be the minimal one
allowed in unitary scalar field theories in de Sitter space. In this sense the
realization of conformal invariance is very general. Any fluctuations with the 
correct conformal weight in de Sitter space, the minimal one, will produce the same 
result for the power spectrum, no matter how they are generated.

We have pointed out several physical differences between our approach and the more 
conventional slow roll models. We do not assume any scalar field expectation value
whose slow roll breaks de Sitter and conformal invariance. Yet, once propagating 
within our cosmological horizon these primordial perturbations  will be subject to the 
same matter interactions and pressure effects as in more conventional scenarios. 
Thus the qualitative behavior of the acoustic peaks at larger $\ell$ in the CMB 
spectrum should be substantially the same as well. For this reason physical
mechanisms for the origin of CMB anisotropies cannot be easily determined 
from the scalar power spectrum alone. 

In any approach based on the fundamental or intrinsic symmetries of de Sitter space 
giving rise to conformally invariant fluctuations, the form of the bispectral shape function 
(\ref{G3corr}) is fully determined, although its amplitude cannot be determined by 
conformal invariance alone and requires additional dynamical assumptions. It is 
important that this form of the non-Gaussian shape function, given by 
(\ref{G3corr})-(\ref{bispectrum}) is quite different and observationally distinguishable 
in principle from that of slow roll inflaton models, which rely upon departures from 
de Sitter space to produce non-Gaussianities proportional to small slow roll parameters. 
In the more general setting we have described there are no slow roll parameters, and 
the fluctuations {\it intrinsic} to de Sitter space do not have to be small (or large) {\it a priori}. 
Observation of the bispectral shape function (\ref{bispectCMB})-(\ref{SCMB}) in the CMB 
by WMAP or Planck would provide a strong indication of dynamics which is quite 
different than that assumed in slow roll inflation models and the de Sitter phase during 
which such fluctuations are assumed to have been generated. The linearized Einstein 
eqs. for the gauge invariant gravitational potentials in the scalar sector were also solved 
for the general stress energy perturbation in Sec. \ref{Sec:GravPert}, in de Sitter space and 
may be used to compute the gravitational perturbations in the scalar sector in the general case.

In Secs. \ref{Sec:Static}-\ref{Sec:Horizon} we have described a second embedding 
by which conformal invariance is naturally derived from the geometric isometries
of de Sitter spacetime, which relies explicitly on the existence of its cosmological 
${\mathbb S}^2$ horizon, evident in the static frame (\ref{dSstat}). Mathematically 
in this case the conformal group of ${\mathbb S}^2$ is $SO(3,1)$, the proper Lorentz 
group, whose conformal generators also may be mapped from the those of the $SO(4,1)$ 
isometry group of de Sitter space. This projective mapping of Killing vectors also leads 
to conformal invariance of correlation functions,  and a quite different form again
of the CMB bispectrum given by (\ref{G3sph}) or (\ref{G3Del0}), directly in terms 
of the angular variables on the sky.

In this second realization of conformal invariance on ${\mathbb S}^2$ we have 
presented a physical mechanism of quantum origin for fluctuations coupling to 
the scalar gravitational potentials, arising from the conformal trace anomaly of massless 
quantum matter or radiation in curved spacetime \cite{Zak,MotVau,AntMot,NJP}. These 
scalar fluctuations, not present in the classical Einstein theory but predicted by Standard 
Model physics are intrinsic to de Sitter space and can have significant effects on or near 
its cosmological horizon. As discussed in Sec. \ref{Sec:Horizon}, linear perturbation 
theory about de Sitter space couples these scalar anomaly fluctuations to the gauge 
invariant gravitational potentials of the metric, with one linear combination having 
exactly the correct zero conformal weight needed to reproduce the flat CMB 
Harrison-Zel'dovich power spectrum at large angular separations. Thus this is a 
possible mechanism for generating the CMB power spectrum and non-Gaussian 
correlations from quantum effects in de Sitter space that do not require slow roll or breaking 
of $SO(4,1)$ invariance by a classical inflaton field and potential as usually assumed.
Certainly this realization of conformal invariance on the cosmological horizon of
de Sitter space would lead to radically different cosmologies.
Our purpose here has not been to present such alternative cosmological models,
but rather to follow the consequences of the hypothesis of conformal invariance,
and determine how to test this hypothesis by forthcoming observations of the CMB.

In Sec. \ref{Sec:GravWaves} we have extended our conformal analysis of correlation
functions to the spin-2 case of tensor perturbations. Since both tensor and scalar
perturbations are part of the same four dimensional linearized gravitational perturbations,
in a generally covariant effective theory such as Einstein's theory, one would expect  
the scalar and tensor perturbations to have same scaling dimensions under
global dilations, even in the absence of full conformal invariance of the
gravitational fluctuations. This expectation relies upon the lowest allowable weight
of energy density fluctuations in the scalar sector, namely $w_{\rho} =2$ being realized,
so that $s=0$ for the corresponding potential, to go with the gravitational waves in the 
tensor sector, which also behave logarithmically in de Sitter space in the scaling region.
This expectation of the relation (\ref{indexST}) between the spectral indices of scalar and 
tensor modes is explicitly satisfied by the difference of scalar gravitational potentials 
$\Upsilon_{\cal A} - \Upsilon_{\cal C}$ in the second realization of conformal invariance on
the de Sitter horizon in (\ref{diffUps}), with the fundamental fluctuations of the quantum
trace anomaly stress tensor as a source. In this realization the scalar fluctuations are
generated by the trace anomaly terms intrinsic to matter in curved space, with no
inflaton needed at all.

We have not considered fundamental non-scalar fields in detail, but as in the discussion
of tensor modes most of our considerations could easily be extended to higher spin fields. 
In general, one could have a non-trivial 3D CFT with  a conformal weight for the energy 
density near to $2$, not necessarily the stress tensor of a free scalar field, giving rise 
to gravitational potential perturbations of conformal weight near to zero, and the power 
spectrum would be the same. One could try to describe such a theory by a non-trivial bulk 
QFT around de Sitter space. Finding additional non-trivial examples of dS/CFT correspondence 
is an open question which could have additional consequences for cosmology. Some related 
ideas have been discussed in Refs. \cite{Mal,Ruba,Hamada,McFSkend}.

That the same power spectrum can be produced by this very different mechanism at the 
cosmological de Sitter horizon emphasizes the need for more detailed measurements of 
CMB properties and correlations to determine the physical origins of the anisotropies. Since 
the form of the non-Gaussian bispectrum in Secs. \ref{Sec:Bispectrum} and \ref{Sec:Static} 
are quite  different from each other, and each is different from the predictions of slow roll 
inflation, any detection of CMB non-Gaussianity would allow the possibility of distinguishing 
physically very different possible origins of the fluctuations in the CMB.

As discussed elsewhere \cite{Moriond,Zak,AntMot,AMM,NJP,QFEXT09}, the 
fluctuations of the scalar degrees of freedom in the effective action of the trace 
anomaly (\ref{effact}) leads to the infrared running of the cosmological vacuum 
energy to smaller values, and hence to cosmological models which are globally 
quite different from the standard FLRW paradigm. The results of this paper indicate that
observational evidence for such a new cosmological model involving dynamical 
dark energy may be sought and possibly discovered first in the angular form of 
the non-Gaussian CMB bispectrum. If the particular form of the angular correlations 
of the bispectrum (\ref{G3sph}) or (\ref{G3Del0}), derived from conformal invariance 
on ${\mathbb S}^2$ are observed in the CMB non-Gaussian signal by WMAP
or Planck, it would imply both a completely different locus and physical mechanism 
for the generation of the CMB than in current cosmological models, with possibly
far ranging implications for the physics of cosmological dark energy and the
large scale structure of the universe.

\section*{Acknowledgments}
\vspace{-4mm}

E. M. wishes to thank the CERN Theory Division for its hospitality 
and the Scientific Associateship Oct., $2009$-Apr., $2010$ during which this
work was initiated. E. M. and I. A. acknowledge useful conversations
with A. Riotto. P. O. M. acknowledges the assistance of P. J. Moraviec
in the initial preparation of the plots of the bispectral shape function in Figs. 
\ref{Fig:Bispectrum96}-\ref{Fig:HZContourplot}, and the hospitality of
C. Lawrence and K. M. G\'orski during an extended visit to 
the Jet Propulsion Laboratory, Pasadena CA, and to the Ec\'ole 
Polytechnique, Palaiseau, France during his sabbatical year. Work 
supported in part by the European Commission under the ERC Advanced 
Grant 226371 and the contract PITN-GA-2009-237920 and in part by the 
CNRS grant GRC APIC PICS 3747.

\appendix

\section{Geometry, Coordinates and dS/CFT Correspondence of de Sitter Space}
\label{App:Geometry}

We collect in this Appendix some well-known properties of de Sitter space and
show how the conformal group of both the ${\mathbb R}^3$ spatial sections
and the ${\mathbb S}^2$ sphere of directions on the sky are contained within
the de Sitter group $O(4,1)$.

The de Sitter manifold is defined as the single sheeted hyperboloid \cite{HawEll}, {\it c.f.} 
Fig. \ref{Fig:deShyper}
\be
\eta_{_{AB}}\,X^{\scriptscriptstyle A}\, X^{\scriptscriptstyle B} \equiv -(X^0)^2 + X^iX^i + (X^4)^2
\equiv -T^2 + X^2 + Y^2 + Z^2  + W^2 = \frac{1}{H^2}  \equiv r_{\!_H}^2
\label{TWXYZcond} 
\ee
in five dimensional flat Minkowski space,
\be
ds^2 = \eta_{_{AB}}\,dX^{\scriptscriptstyle A}\,dX^{\scriptscriptstyle B} =
 -dT^2 + dW^2 + dX^2 + dY^2 + dZ^2\,,
\label{flat5}
\ee
This manifold has the isometry group $O(4,1)$ which is the maximal
possible for any solution of the vacuum Einstein field equations,
\be
R^{a}_{\ b} - \frac{R}{2} \,\delta^{a}_{\ b} + \Lambda \,\delta^{a}_{\ b} = 0
\label{Ein}
\ee
in four dimensions. The curvature tensor of de Sitter space satisfies
\bes
\bea
R^{ab}_{\ \ cd} &=& H^2 \left(\delta^a_{\ c}\,\delta^b_{\ d} - \delta^a_{\ d}\,\delta^b_{\ c}\right)\,,\\
R^a_{\ b} &=& 3 H^2\, \delta^a_{\ b}\,,\\
R &=& 12 H^2\,,
\eea
\ees
where the Hubble constant is related to $\Lambda$ by
\be
H = \sqrt{\frac{\Lambda}{3}}\,.
\ee
Here and henceforward we set the speed of light $c=1$.

\begin{figure}
\begin{center}
\includegraphics[height=6cm,width=6cm]{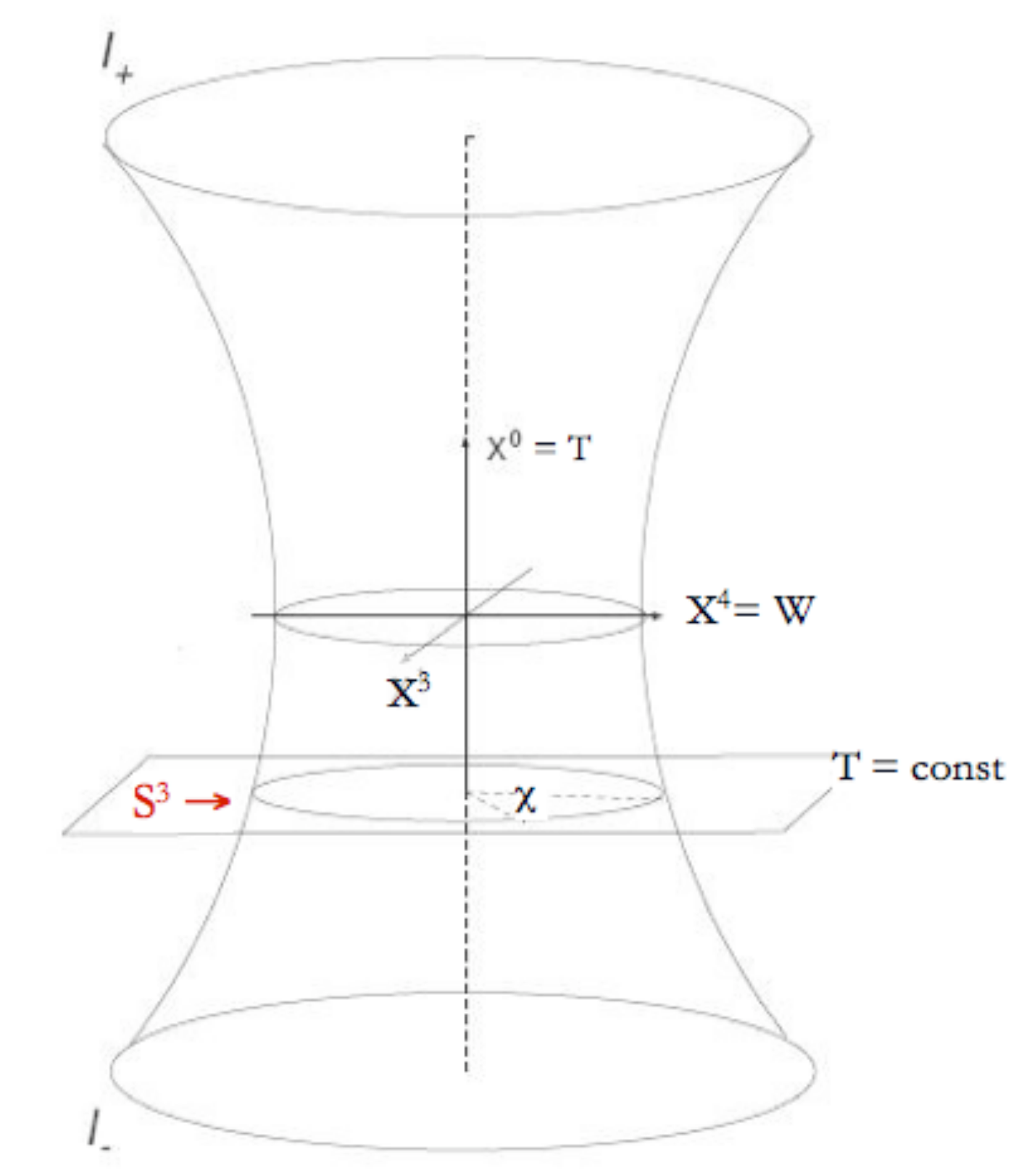}
\caption{The de Sitter manifold represented as a single sheeted hyperboloid of revolution
about the $T$ axis, in which the $X^1$, $X^2$ coordinates are suppressed. The hypersurfaces 
at constant $T$ are three-spheres, $\mathbb{S}^3$. The three-spheres at $T= \pm \infty$ are 
denoted by $I_{\pm}$.}
\label{Fig:deShyper}
\end{center}
\vspace{-7mm}
\end{figure}

The Lie algebra $so(4,1)$ is generated by the generators of the Lorentz group
in the $4+1$ dimensional flat embedding spacetime (\ref{flat5}). In the coordinate
(adjoint) representation the $10$ anti-Hermitian generators of this symmetry are
\be
L_{\scriptscriptstyle AB} = X_{\scriptscriptstyle A} \frac{\partial}
{\partial X^{\scriptscriptstyle B}} - X_{\scriptscriptstyle B}
\frac{\partial}{\partial X^{\scriptscriptstyle A}} = - L_{\scriptscriptstyle BA}\,,
\ee
with the indices $A, B = 0, 1, 2, 3, 4$ raised and lowered with the five dimensional
Minkowski metric $\eta_{_{AB}} =$ diag $(-1, 1, 1, 1, 1)$ of (\ref{flat5}).
These $10$ generators satisfy
\be
[L_{\scriptscriptstyle AB}, L_{\scriptscriptstyle CD}] = 
-\eta_{\scriptscriptstyle AC} \, L_{\scriptscriptstyle BD} + 
\eta_{\scriptscriptstyle BC} \, L_{\scriptscriptstyle AD} - 
\eta_{\scriptscriptstyle BD} \, L_{\scriptscriptstyle AC}  +
 \eta_{\scriptscriptstyle AD} \, L_{\scriptscriptstyle BC} \,.
\ee
de Sitter space has $10$ Killing vectors corresponding  to these $10$ generators, {\it i.e.}
there are $10$ linearly independent vector fields $K^{(\alpha)}_{\mu}$ which
are solutions of the Killing eq. (\ref{Kil}) that leave the de Sitter metric invariant. 
We find the the explicit form of these $10$ Killing solutions in various coordinate
systems below.

The hyperbolic coordinates of de Sitter space are defined by
\bes
\bea
T &=& \frac{1}{H}\, \sinh u\,,\\
X^i &=& \frac{1}{H}\, \cosh u\,\sin\chi\, \hat n^i\,,\qquad i =1, 2, 3\,\\
W &=&  \frac{1}{H}\, \cosh u\,\cos\chi\,,
\eea
\label{hypercoor}\ees
where
\be
\hat n = (\sin\theta\,\cos\phi\,,\sin\theta\,\sin\phi\,,\cos\theta)
\ee
is the unit vector on $\mathbb{S}^2$, and cast the de Sitter line element in the form
\be
ds^2 = \frac{1}{H^2} \left[ -du^2 + \cosh^2 u\, (d\chi^2 + \sin^2\chi\, d\Omega^2)\right]\,.
\label{hypermet}
\ee
The quantity in round parentheses is
\be
d\Omega_3^2 \equiv [d (\sin\chi\, \hat n^i)]^2 + [d (\cos\chi)]^2 
= d\chi^2 + \sin^2\chi\, d\Omega^2\,,
\ee
the standard round metric on $\mathbb{S}^3$. Hence in the geodesically complete
coordinates of (\ref{hypercoor}) the de Sitter line element (\ref{hypermet}) is a
hyperboloid of revolution whose constant $u$ sections are three spheres, 
represented in Fig. \ref{Fig:deShyper}, which are invariant under the $O(4)$ 
subgroup of $O(4,1)$. 

In cosmology it is more common to use instead the Friedmann-Lema\^itre-Robertson-Walker 
(FLRW) line element with flat $\mathbb{R}^3$ spatial sections, {\it viz.}
\bea
ds^2 &=& - d\tau^2 + a^2(\tau)\, d\vec x\cdot d\vec x = 
- d\tau^2 + a^2(\tau)\, (dx^2 + dy^2 + dz^2)\nn\\
&=& - d\tau^2 + a^2(\tau)(d\varrho^2 + \varrho^2 d\Omega^2)\,.
\label{FLRW}
\eea
De Sitter space can be brought in the FLRW form by setting
\bes
\bea
T &=& \frac{1}{2H}\left ( a - \frac{1}{a} \right) + \frac{Ha}{2}\, \varrho^2\,,\\
X^i &=& a \,\varrho \,\hat n^i \,,\\
W &=&  \frac{1}{2H}\left ( a + \frac{1}{a} \right) - \frac{Ha}{2}\, \varrho^2\,,
\eea\label{RWcoor}\ees
with
\bes
\bea
a (\tau) &=& e^{H\tau} \label{deSexp}\\
\varrho &=& |\vec x| = \sqrt{x^2 + y^2 + z^2}\,.
\eea\label{rhotau}\ees
From (\ref{RWcoor}) and (\ref{rhotau}), $T + W \ge 0$ in these coordinates. 
Hence the flat FLRW coordinates cover only one half of the full de Sitter 
hyperboloid, with the hypersurfaces of constant FLRW time $\tau$ slicing the 
hyperboloid in Fig. \ref{Fig:deShyper} at a $45^{\circ}$ angle.

The change of time variable to the conformal time coordinate,
\be
\eta = - H^{-1} e^{-H\tau} = -\frac{1}{Ha}\,,\qquad a(\tau) = \Omega(\eta) = - \frac{1}{H\eta}\,,
\ee
is also often used to express the de Sitter line element in the conformally flat Poincar\'e form
\be
ds^2 = a^2 \, (- d\eta^2 + d\vec x^2) = \frac{1}{H^2\eta^2}\, \left(- d\eta^2 + d\vec x^2\right)\,,
\label{conflat}
\ee
which is (\ref{conftime}) of the text. From (\ref{hypercoor}) and (\ref{RWcoor}), 
\bes
\bea
\cosh u\,\sin\chi = H\varrho\, a &=&  -\frac{\varrho}{\eta}\,,\\
\sinh u + \cosh u\,\cos\chi = a &=&  -\frac{1}{H\eta}\,,
\eea
\ees
\noindent
which gives the direct relation between hyperbolic coordinates and flat FLRW coordinates.
The conformal time representation of the de Sitter line element (\ref{conflat}) is of the 
Fefferman-Graham (FG) form \cite{FG}, 
\be
(ds^2)_{FG} = \ell^2 \left[ \pm \frac{d\eta^2}{\eta^2} + \frac{g_{ij}( \vec x, \eta)}{\eta^2} dx^i dx^j \right]
\label{FG}
\ee
where $g_{ij} (\vec x, \eta)$ is required to possess a regular Taylor series expansion in $\eta$ as 
$\eta \rightarrow 0$, certainly satisfied in this case since $g_{ij} = \delta_{ij}$ for flat
${\mathbb R}^3$ spatial sections, and the minus sign in (\ref{FG}) applies since real de Sitter 
spacetime has a Lorentzian signature. The FG form (\ref{FG}) is useful for determining the conformal 
scaling dimensions of fields in the dS/CFT correspondence which applies to conformal fields
defined on the Euclidean ${\mathbb R}^3$ at the asymptotic limit $\eta \rightarrow 0$ \cite{MazMot,Skend}. 

The de Sitter static coordinates $(t, r, \theta, \phi)$ are defined by
\bes
\bea
T &=& \frac{1}{H}\, \sqrt{1 - H^2r^2}\,\sinh(Ht)\,,\\
X^i &=& r\, \hat n^i\,,\\
W &=& \frac{1}{H}\, \sqrt{1 - H^2r^2}\,\cosh(Ht)\,.
\eea\label{staticoor}\ees
They bring the line element (\ref{flat5}) into the static, spherically symmetric 
form,
\be
ds^2 = -f(r)\, dt^2 + \frac{dr^2}{f(r)} + r^2 d\Omega^2  = f(r) ds^2_{opt}\,,\qquad f(r) \equiv 1 - H^2r^2
\label{static}
\ee
where $ds^2_{opt}$ is called the four dimensional ``optical" metric. From (\ref{staticoor}), real 
static $(t, r)$ coordinates cover only the quarter of the de Sitter manifold where $W \ge 0$ and 
both $W \pm T \ge 0$. This quarter is represented as the rightmost wedge of the Carter-Penrose 
conformal diagram of de Sitter space in Fig. \ref{Fig:dSCarPen}. Comparing $W+T$ in the
spatially flat FRLW coordinates (\ref{RWcoor}) with $W+T$ expressed in static coordinates 
(\ref{staticoor}) and using (\ref{deSexp}), we obtain the direct relation (\ref{tauHrt}) of the text
between the two coordinate charts in the wedge where they both apply.

\begin{figure}
\begin{center}
\includegraphics[height=6cm,width=6cm]{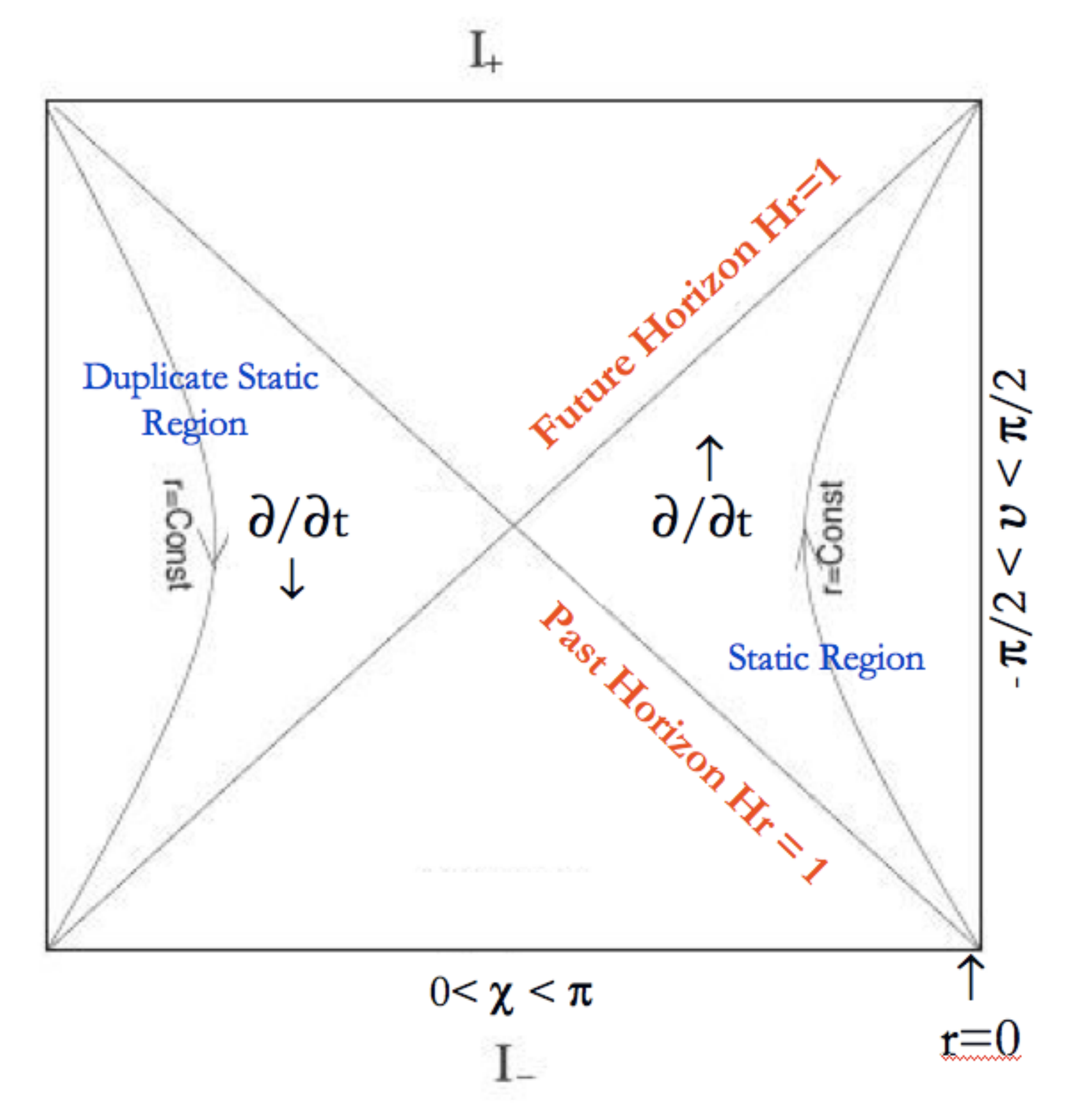}
\caption{The Carter-Penrose conformal diagram for de Sitter space. Future and past infinity 
are at $I_{\pm}$. Only the quarter of the diagram labeled as the static region are covered by
the static coordinates of (\ref{dSstat}). The orbits of the static time Killing field
$\partial/\partial t$ are shown. The angular coordinates $\theta,\phi$ are again suppressed.}
\label{Fig:dSCarPen}
\end{center}
\vspace{-7mm}
\end{figure}

The Regge-Wheeler radial coordinate $r^*$ can be defined in the static frame by
\bes
\bea
r^* &=& \frac{1}{2H}\, \ln\left(\frac{1+ Hr}{1 - Hr}\right) = \frac{1}{H} \, \tanh^{-1} (Hr)\,,\\
r &=&  \frac{1}{H} \, \tanh (Hr^*)\,\qquad {\rm so\ that}\\
dr^* &=& \frac{dr}{1-H^2r^2}\,,\qquad \sqrt{1 - H^2r^2}=  {\rm sech} (Hr^*)\,,
\eea
\label{RegWhdS}
\ees
and
\bea
ds^2 &=&{\rm sech}^2 (Hr^*)\, (-dt^2 + dr^{*\,2}) 
+ \frac{1}{H^2} \tanh^2(Hr^*)\, d\Omega^2\nn\\
&=&{\rm sech}^2 (Hr^*)\,\left[ -dt^2 + dr^{*\,2} + 
\frac{1}{H^2}\sinh^2(Hr^*) \,d\Omega^2\right].\qquad
\label{optmet}
\eea
Note that the horizon at $r = H^{-1} = r_{\!_H}$ is mapped to $r^* = \infty$ in these coordinates, and 
that the spatial part of the ``optical" metric with $dt =0$ is
\be
d \ell^2_{\mbox{\tiny \L}} = dr^{*\,2} + \frac{1}{H^2}\sinh^2(Hr^*) \,d\Omega^2 
=4\,r_{\!_H}^2\,\frac{d\vec {\sf y} \cdot d\vec {\sf y}}{(1 - |\vec{\sf y}|^2)^2} 
\label{Lob}
\ee
where the second form is obtained by defining
\bes
\bea
\vec {\sf y}&\equiv&  \,|\vec {\sf y}|\,  \hat n\,,\\
|\vec {\sf y}| &\equiv & \tanh \left(\frac{Hr^*}{2}\right) = \frac{Hr}{1 + \sqrt{1-H^2r^2}} 
\quad {\rm so\  that}\\
r&= &\frac{2}{H}\frac{|\vec{\sf y}|}{1+ |\vec {\sf y}|^2}\,,\qquad r^* = \frac{1}{H} \,
\ln \left(\frac{1 + |\vec {\sf y}|}{1 - |\vec {\sf y}|}\right)\,.
\eea
\ees
Eq. (\ref{Lob}) is a standard form of the line element of three dimensional \L obachewsky 
(hyperbolic or Euclidean anti-deSitter) space ${\mathbb H}^3$. Thus, one expects conformal 
field theory (CFT) behavior at the horizon boundary, $|\vec {\sf y}|=1$, $r= r_{\!_H}$, namely
on the sphere ${\mathbb S}^2$ of directions $\hat n$ on the horizon \cite{MazMor}. 

This conformal behavior can be seen most clearly by introducing Poincar\'e coordinates 
(\ref{Poincoor}). Alternatively, the change of coordinates,
\be
\zeta \equiv  2\, \frac{1 - |\vec{\sf y}|}{1 + |\vec{\sf y}|} = 2\, e^{-Hr^*} =  2\, \sqrt{\frac{1-Hr}{1+Hr}} 
= \frac{2}{1 + Hr}\,f^{\frac{1}{2}}
\label{zetadef}
\ee
brings the three dimensional  \L obachewsky metric (\ref{Lob}) into the
Fefferman-Graham form (\ref{FG})
\be
d \ell^2_{\mbox{\tiny \L}} = r_{\!_H}^2 \left[\frac{d\zeta^2}{\zeta^2} + \frac{1}{\zeta^2} 
\left(1 - \frac{\zeta^2}{4}\right)^2 d\hat n \cdot d \hat n\right]
\label{optFG}
\ee
where in (\ref{FG}) the spatial variable $\zeta$ is substituted for $\eta$, the upper (positive)
sign applies and in this case $g_{ab}(\hat n, \zeta) = \gamma_{ab} (1 - \zeta^2/4)^2$, 
with $\gamma_{ab}$, the two dimensional standard round metric on ${\mathbb S}^2$. 
As in the case of the dS/CFT $SO(4,1)$ correspondence in the conformal coordinates 
(\ref{conftime}) as $\eta \rightarrow 0$ at the $I_+$ at the top of the de Sitter hyperboloid 
in Figs. \ref{Fig:deShyper} and \ref{Fig:dSCarPen}, the Fefferman-Graham form of the three 
dimensional optical metric (\ref{optFG}) permits  easy identification of conformal fields of 
definite conformal weight with respect to $SO(3,1)$, the conformal group of ${\mathbb S^2}$ 
now on the de Sitter horizon by counting powers of $\zeta \rightarrow \sqrt f \rightarrow 0$ 
as $r \rightarrow r_{\!_H}$. 

\section{Killing Vectors of de Sitter Space in Flat FLRW Coordinates}
\label{App:KillingFlat}

The solutions of (\ref{KilldS}) can be catalogued as follows.
For $K_{\tau} =0$ we have the three translations,
\be
K_{\tau}^{(Tj)} = 0\,,\qquad K_i^{(Tj)} = a^2 \delta_i^{\ j}\,,\qquad j = 1, 2, 3\,,
\ee
and the three rotations,
\be
K_{\tau}^{(R\ell)} = 0\,,\qquad K_i^{(R\ell)} = a^2 \epsilon_{i\ell m}x^m\,,\qquad \ell = 1, 2, 3\,.
\ee
The spatial ${\mathbb R}^3$ sections also have four conformal Killing vectors  which satisfy
(\ref{KilldS}) with $K_{\tau} \neq 0$. They are the three special conformal transformations
of ${\mathbb R}^3$,
\be
K_{\tau}^{(Cn)} = -2H x^n\,,\qquad K_i^{(Cn)} = H^2a^2( \delta_i^{\ n} \delta_{jk}x^jx^k 
- 2 \delta_{ij}x^jx^n)
 - \delta_i^n\,,\qquad n = 1, 2, 3\,,
 \label{speconformal}
 \ee
 and the dilation,
 \be
 K_{\tau}^{(D)} = 1\,,\qquad K_i^{(D)} = H a^2\,\delta_{ij} x^j\,.
 \label{dilation}
 \ee
This last dilational Killing vector is the infinitesimal form of the
finite dilational symmetry,
\bes
\bea
&&\vec x \rightarrow \lambda \vec x\,,\\
&& \eta \rightarrow \lambda \eta\,\\
&& a(\tau) \rightarrow \lambda^{-1} a(\tau)\,,\\
&&\tau \rightarrow \tau - H^{-1}\ln \lambda
\eea
\ees of de Sitter space. Since the maximum number of Killing isometries
in four dimensions is $10$, there are no other solutions of (\ref{KilldS}), and
de Sitter space, being a fully symmetric space, possesses the maximum number 
of symmetries. Being conformally flat, de Sitter space also possess $5$ conformal
Killing vector fields $V_{\mu}$, satisfying $\nabla_{\mu} V_{\nu} + \nabla_{\nu} V_{\mu}
= \frac{1}{2} g_{\mu\nu} \nabla^{\lambda} V_{\lambda}$ whose explicit forms may be
found as well, but which we do not require and omit here.

\section{Killing Vectors of de Sitter Space in Static Coordinates}
\label{App:Killingstatic}

In static coordinates (\ref{dSstat}) the Killing  eq. (\ref{Kil}) gives the following $10$ eqs. 
for the components $K_a$:
\bes
\bea
&&\partial_t K_t = \frac{ff'}{2}\, K_r\,\\
&&\partial_r K_r = - \frac{f'\,}{2f}\, K_r\,\\
&&\partial_{\theta} K_{\theta} = -r f\, K_r\,  \label{thetatheta}\\
&&\partial_{\phi} K_{\phi} = -rf\,\sin^2\theta\,K_r - \sin\theta\cos\theta\,K_{\theta}\, \label{phiphi}\\
&&\partial_t K_r + \partial_rK_t = \frac{f'}{f}\, K_t\,\\
&&\partial_t K_{\theta} + \partial_{\theta} K_t = 0\,\\
&&\partial_t K_{\phi} + \partial_{\phi} K_t = 0\,\\
&&\partial_r K_{\theta} + \partial_{\theta} K_r = \frac{2}{r}\,K_{\theta}\,\\
&&\partial_r K_{\phi} + \partial_{\phi} K_r = \frac{2}{r}\,K_{\phi}\,\\
&&\partial_{\theta} K_{\phi} + \partial_{\phi} K_{\theta} = 2\cot\theta\,\,K_{\phi} \label{thetaphi}\,.
\eea
\label{KillstaticdS}
\ees
This system of eqs. is solved by the static time translation Killing vector,
\be
K^t = 1\, \quad {\rm or} \quad  K_t = g_{tt} = - f(r)= 1-H^2r^2\, \qquad
{\rm with}\qquad K_r = K_{\theta} = K_{\phi} =0\,,
\label{timeT}
\ee
which corresponds to the symmetry $t \rightarrow t + const.$ under static time 
translation, and the $3$ rotations, 
\bes
\bea
&&(i)\quad K_t = K_r = K_{\theta} = 0\,, K_{\phi} 
= r^2\,\sin^2\theta\,,\\
&&(ii)\ \quad K_t = K_r = 0\, ,\quad K_{\theta} = r^2\sin\phi\,,\qquad K_{\phi} 
= r^2\cos\phi\sin\theta\cos\theta\,,\\
&&(iii)\quad K_t = K_r = 0\, ,\quad K_{\theta} =-r^2\cos\phi\,, \quad\, K_{\phi} 
= r^2\sin\phi\sin\theta\cos\theta\,,
\eea
\label{rotat}
\ees
which leave the (arbitrary) origin at $r=0$ fixed. 

Then we have two pair ($4$ total) solutions of the form,
\bea
K_t = -r f^{\frac{1}{2}}\,\dot A(t)\sin\theta\,B(\phi)\,,\quad K_r = 
f^{-\frac{1}{2}} \,A(t) \,\sin\theta\, B(\phi)\,,
\nonumber\\
\ K_{\theta} = r f^{\frac{1}{2}}\, A(t)\, \cos\theta\, B(\phi)\,,\quad K_{\phi} = 
r f^{\frac{1}{2}}\, A(t) \sin\theta \,B'(\phi)\,,
\label{foursol}
\eea
where both $A(t)$ and $B(\phi)$ each have two linearly independent solutions,
\be
A(t) = \left(\begin{array}{c}{\rm sinh}\,Ht\\{\rm cosh}\,Ht\end{array}\right)\qquad {\rm and}\qquad
B(\phi) = \left(\begin{array}{c}{\rm sin}\,\phi\\{\rm cos}\,\phi\end{array}\right)\,,
\ee
and $\dot A \equiv \frac{dA}{dt}$, $B' \equiv \frac{dB}{d\phi}$. Finally we have two solutions of the form,
\be
K_t = -r f^{\frac{1}{2}}\,\dot A(t)\cos\theta\,,\qquad K_r = f^{-\frac{1}{2}} \,A(t) \,\cos\theta\,,
\qquad K_{\theta} = r f^{\frac{1}{2}}\, A(t)\, \sin\theta\,,\qquad K_{\phi} = 0\,,
\label{twosol}
\ee
with $A(t)$ again either sinh\,$Ht$ or cosh\,$Ht$.

The purely angular part of these last $6$ solutions in fact are proportional to the three conformal
Killing vectors of ${\mathbb S}^2$. This follows from the fact that the substitution of (\ref{thetatheta}) into
(\ref{phiphi}) and eq. (\ref{thetaphi}) give
\bes
\bea
&&\partial_{\phi} K_{\phi} = \sin^2\theta\ \partial_{\theta} K_{\theta} - \sin\theta\cos\theta\,K_{\theta}\\
&&\partial_{\theta} K_{\phi} + \partial_{\phi} K_{\theta} = 2\cot\theta\,\,K_{\phi}
\eea\label{Kphitheta}
\ees
Comparing (\ref{Kphitheta}) with 
\be 
D_a\upsilon_b + D_b\upsilon_a= {\gamma}_{ab}\,D_c\upsilon^c\,,
\label{consphere}
\ee
with $a,b = \theta, \phi$, ${\gamma}_{ab}$ the standard round metric on ${\mathbb S}^2$ 
and $D_a$ the covariant derivative with respect to this metric, when written explicitly in 
polar coordinates
\bes
\bea
&&\partial_{\phi} \upsilon_{\phi} +  \sin\theta\cos\theta\,\upsilon_{\theta}
= \sin^2\theta\ \partial_{\theta} \upsilon_{\theta} 
\label{sepKil}\\
&&\partial_{\theta} \upsilon_{\phi} + \partial_{\phi} \upsilon_{\theta} = 2\cot\theta\,\,\upsilon_{\phi}
\eea
\label{Kilsphere}\ees\noindent 
and with
\be
D_c \upsilon^c = \partial_{\theta}\upsilon_{\theta}
 + \frac{1}{\sin^2\theta} \,\partial_{\phi} \upsilon_{\phi} + \cot\theta\, \upsilon_{\theta}\,.
\ee
shows that (\ref{Kphitheta}) and (\ref{Kilsphere}) coincide. The solutions of the Killing eq. in 
de Sitter space are simply related to those on the sphere by taking their $\theta, \phi$ components 
and dependences on these angles.

The solutions of (\ref{Kilsphere}) from (\ref{rotat}) are the $3$ rotational Killing vector fields
\bes
\bea
\upsilon_{\theta} = 0\,,\quad &&\upsilon_{\phi} = \sin^2\theta\,, \qquad {\rm and}\\
\upsilon_{\theta} = B(\phi)\,,\quad && \upsilon_{\phi} = \sin\theta\,\cos\theta\, B'(\phi)\,,
\eea
\label{rotatsph}
\ees and from (\ref{foursol}) and (\ref{twosol}) the $3$ conformal Killing vector fields
\bes
\bea
\upsilon_{\theta} = \sin\theta\,,\quad && \upsilon_{\phi} =0\,, \qquad {\rm and}\\
\upsilon_{\theta} = \cos\theta\, B(\phi)\,,\quad && \upsilon_{\phi} =\sin\theta \,B'(\phi)\,,
\eea
\label{confKsph}\ees
where in each case again $B(\phi)$ is either $\sin \phi$ or $\cos\phi$. Thus the $6$ de Sitter Killing 
vectors (\ref{foursol}) and (\ref{twosol}) have angular components which are just proportional to 
{\it two copies} of the $3$ conformal Killing transformations of ${\mathbb S}^2$ (if the differing $t$ 
dependence is not considered). The $3$ rotational Killing vectors of ${\mathbb S}^2$ given in (\ref{rotatsph}) 
satisfy $D_c \upsilon^c = 0$, while the $3$ conformal Killing vectors of ${\mathbb S}^2$ given by 
(\ref{confKsph}) have $D_c \upsilon^c \neq 0$. These latter $3$ conformal Killing vectors are linear 
combinations of $\partial_{a} Y_{1,m}(\theta, \phi)$ where $Y_{1,0} \sim \cos\theta$ and 
$Y_{1, \pm 1} \sim \sin\theta e^{\pm i\phi}$ are the $3$ (scalar) spherical harmonics for $\ell = 1$. 
The $3$ rotational transverse Killing vectors (\ref{rotatsph}) may be expressed in terms 
of the $3$ transverse vector harmonics $\vec L \,Y_{1, m}$. 

This demonstrates that in addition to the one static time translation (\ref{timeT}), the de Sitter group 
induces the conformal group $SO(3,1)$ on the sphere ${\mathbb S}^2$ of constant $r$, 
with the three special conformal Killing vectors appearing twice, because of the two
possibilities for the time dependent function $A(t)$, giving $1 + 3 + 2\times 3 = 10$ 
de Sitter Killing vectors in all. We also observe that as $r \rightarrow r_{\!_H}$,
the solutions (\ref{foursol}) and (\ref{twosol}) have $K_{\theta}, K_{\phi} \propto f^{\frac{1}{2}} 
\rightarrow 0$. In the near horizon limit  $r \rightarrow r_{\!_H}$ the scaling behavior of the 
conformal factor
\be
D_a K^a \propto f^{\frac{1}{2}}  \propto \bar z \propto \zeta\,,
\ee
where $\bar z$ is defined by (\ref{Poincoor}) and $\zeta$ defined by (\ref{zetadef}). The special 
conformal vectors (\ref{confKsph}) of ${\mathbb S}^2$ when viewed as embedded in the three 
dimensional optical metric of (\ref{optFG}) of de Sitter space correspond to the de Sitter symmetry 
transformations (\ref{foursol}), (\ref{twosol}). This allows one to identify the conformal weight of 
fields on the horizon, as in (\ref{weightH}) by the power of $f^{\frac{1}{2}}$ the $r$ dependent
solution contains as $r \rightarrow r_{\!_H}$.

\section{Invariant Distance and Correlation Functions}
\label{App:Distance}

Since de Sitter spacetime is geometrically a hyperboloid of revolution, the invariant
distance can be defined from analytic continuation of the invariant geodesic distance
$\sigma (x, x')$ between two points on ${\mathbb S}^4$. For spacelike separations 
this is defined by
\be
\sigma (x, x') = r_{\!_H}\, \cos^{-1} \, \left[H^2 (-TT' + XX' + YY' + ZZ' + WW')\right]\,.
\ee
It is also convenient to define the function of $\sigma(x,x')$,
\bes
\bea
z(x,x') &\equiv& \frac{1 + \cos \left(H\sigma(x,x')\right)}{2} = \cos^2 \left(\frac{H\sigma (x,x')}{2}\right)\,,\\
1- z(x,x') &=&  \sin^2 \left(\frac{H\sigma (x,x')}{2}\right)\,.
\eea
\ees
For spacelike separated points, $\sigma (x, x')$ is real and $z < 1$, while for timelike 
separated points, $\sigma (x, x')$ is replaced by $i |\sigma(x,x')|$ and $z > 1$.

In the conformally flat coordinates (\ref{conftime}), the invariant function $1- z(x,x')$ is given
by (\ref{distconf}) of the text, and the two-point propagator function for a conformally 
invariant scalar field in the Bunch-Davies state in de Sitter space is given by 
(\ref{conscalar}) by a conformal transformation of a weight one field from flat space. 
Because this conformal transformation is time dependent, the Bunch-Davies state 
is not a ``vacuum" state and corresponds instead to a state thermally populated with matter
at the Hawking-de Sitter temperature $T_{_{H}} = \frac{\hbar H}{2 \pi k_{_B}}$
in the static time vacuum defined in coordinates (\ref{staticoor}). The function $G_{conf}(z)$ of
(\ref{conscalar}) satisfies the wave equation for a massless, conformally coupled
scalar field,
\be
\left(-\sq + 2H^2\right) G_{conf}(z(x,x')) = -H^2 \left[ z(1-z) \frac{d^2}{dz^2} + 2 (1 -2z) \frac{d}{dz} -2\right]
G_{conf}(z) = 0\,,\quad x \neq x'\,,z\neq 1\,.
\label{masslessprop}
\ee 
At $x=x'$ it is correctly normalized to give $\delta^4(x,x')$. For arbitrary mass $M$, the de Sitter 
invariant correlation function is given by (\ref{hypergeom}) of the text. In each case the Feynman
function in real time is defined by replacing $1- z$ in (\ref{conscalar}) or (\ref{hypergeom}) by 
$1 - z(x,x') - i \epsilon$ for timelike separated points, for which $z >1$. The solution 
(\ref{hypergeom}) may also be written in terms of a Legendre function \cite{BD}. 

\section{De Sitter Space and the $SO(3,1)$ Conformal Group of the Sphere}
\label{App:Conformal}

The massive wave eq. (\ref{massiveprop}) can be separated in static coordinates (\ref{dSstat})
with solutions of the form
\be
\Phi \sim  e^{-i\omega t}\, \frac{\psi_{\omega \ell}}{r} \, Y_{\ell m}(\theta, \phi)\,,
\ee
where the radial function $\psi_{\omega \ell}$ satisfies
\be
\left[- \frac{d^2\ }{dr^{*\,2}} + V_{\ell}\right] \,\psi_{\omega \ell} = \omega^2\, \psi_{\omega \ell}\,,
\label{psieq}
\ee
with the potential.
\bea
V_{\ell} &=&(1- H^2 r^2) \left[ \frac{\ell(\ell + 1)}{r^2} + M^2 - 2H^2\right] \nonumber\\
&= & H^2 \ell(\ell +1)\, {\rm csch}^2(Hr^*) + (M^2 - 2H^2)\, {\rm sech}^2(Hr^*)\,.
\label{potent}
\eea
As $r \rightarrow r_{\!_H}$, $r^* \rightarrow \infty$, all finite mass terms drop out, 
$V_{\ell} \rightarrow 0$ and the wave eq. (\ref{massPhi}) or (\ref{psieq}) becomes identical
to that of massless conformal field, with $\psi_{\omega \ell} \sim e^{\pm i \omega r^*}$ 
as the horizon is approached. This is a consequence of the gravitational blueshift
(\ref{blueshift}).

This conformal behavior on the cosmological horizon is described
by the mapping between the Killing vectors of de Sitter space and the conformal
Killing vectors of ${\mathbb S}^2$, with two copies of the the special conformal 
transformations of the sphere. In general, the $SO(d+1,1)$ Lorentz group can be 
mapped to the conformal group of null directions on the $d$ dimensional sphere
${\mathbb S}^d$. This mapping may be made as follows \cite{MazMot}.
The four vector $X^\mu$ in flat four dimensional Minkowski spacetime transforms linearly 
under the $SO(3,1)$ homogeneous Lorentz group
\be
X^\mu \rightarrow \Lambda^\mu_{\ \nu}(\vec v)\, X^\nu\,,\qquad \mu,\nu = 0, 1, 2, 3\,.
\label{Lorentz}
\ee
where $\Lambda^\mu_{\ \nu}(\vec v)$ is the usual Lorentz transformation matrix for boost
velocity $\vec v$. This is a subgroup of the $SO(4,1)$ de Sitter group in which the fifth coordinate 
$W$ does not participate. The (future) light cone in Minkowski space is defined by
\be
X^\mu\,X^\nu\,\eta_{\mu\nu} = -T^2 + X^iX^i = 0\,,\qquad T\equiv X^0 > 0\,,.
\ee
Then
\be
\hat n^i = \frac{X^i}{X^0}
\ee
is a unit vector which defines the direction of the outgoing light ray. The four
dimensional flat Minkowski line element becomes
\be
ds^2_0 = (X^0)^2\, d\hat n^i d\hat n^i
\ee
when restricted to the light cone.

Under the Lorentz transformation (\ref{Lorentz}) $\hat n^i$ transforms into
\be
\hat n^{\prime i} = \frac{\Lambda^i_{\ 0} + \Lambda^i_{\ j}\hat n^j}{\Lambda^0_{\ 0} 
+ \Lambda^0_{\ k} \hat n^k} = \frac{\hat n^i + \hat v^i (\hat v\cdot \hat n)(\gamma - 1)
 -\gamma v^i}{\gamma\, (1 - \vec v \cdot \hat n)}\,,
\label{confmn}
\ee
with as usual $\gamma \equiv (1-v^2)^{-\frac{1}{2}}$. On the ${\mathbb S}^2$ space of directions, this 
transformation is a conformal transformation since by the Lorentz invariance of the light cone,
\be
d \hat n^i d\hat n^i \rightarrow \left(\frac{X^0}{X^{\prime 0}}\right)^2\, d\hat n^i d\hat n^i
= \Omega^2(\hat n)\,d\hat n^i d\hat n^i\,,
\ee
with
\be
\Omega(\hat n) = \left[ \gamma\,(1- \vec v \cdot \hat n)\right]^{-1} = 1 + \vec v \cdot \hat n + {\cal O}(v^2)\,,
\label{conftrans}
\ee
so that (\ref{confmn}) may also be written in the form
\be
\hat n \rightarrow \hat n'= \Omega(\hat n) \left[\hat n + \hat v (\hat v\cdot \hat n)(\gamma - 1) -\gamma \vec v\right]
\label{confnOm}
\ee
in terms of $\Omega(\hat n)$. This shows that the $SO(3,1)$ subgroup of the de Sitter isometry
group $SO(4,1)$ is the group of conformal transformations on the sphere ${\mathbb S}^2$.
From (\ref{confnOm}) it is straightforward to verify that for any two unit vectors on ${\mathbb S}^2$,
$\hat n_1$ and $\hat n_2$,
\be
1 - \hat n_1 \cdot \hat n_2 \rightarrow \Omega(\hat n_1)\, [1 - \hat n_1 \cdot \hat n_2]\, \Omega(\hat n_2)\,,
\label{confn1n2}
\ee
so that this quantity transforms under $SO(3,1)$ as a conformally covariant measure of
distance between $\hat n_1$ and $\hat n_2$. By  (\ref{zstat}) the de Sitter invariant distance
$1 - z(x,x')$ reduces to (\ref{confn1n2}) on the de Sitter horizon sphere. Since any
de Sitter invariant function of $z(x,x')$ can be expanded in powers of $1-z(x,x')$, it
follows that it can always be decomposed into a linear superposition of representations of the
conformal group $SO(3,1)$, transforming with definite conformal weights on the horizon. 
However, since unlike at spacelike infinity $I_+$ where $1- z(x,x')$ becomes arbitrarily
large as $\tau \rightarrow \infty$, on the horizon (\ref{zstat}) or (\ref{confn1n2}) remains
finite in the physical de Sitter metric. Hence simple irreducible representations 
of $SO(3,1)$ are induced on the horizon only by fields with simple local Weyl
invariant transformation properties.

An example of a correlation function with simple local Weyl transformation properties
is the Green's function of the fourth order differential operator $\Delta_4$ defined by (\ref{Del4}),
given by (\ref{Del4prop})-(\ref{Dlog}). This is the same (up to normalization) as the Green's function 
of the scalar Laplacian on ${\mathbb S}^2$, namely 
\be 
L^2\left[\frac{1}{4\pi} \ln (1 - \hat n \cdot \hat n')\right] = \delta^2(\hat n, \hat n') - \frac{1}{4\pi}\,,
\label{sphprop}
\ee
with $L^2$ defined by (\ref{lapl}). That is, the second order differential operator of a conformal
weight zero scalar on ${\mathbb S}^2$ has the same correlation function (up to normalization)
as that of a the fourth order differential operator of a conformal weight zero field in de Sitter space.
The constant $-\frac{1}{4\pi}$ on the right side of (\ref{sphprop}) reflects the fact that $L^2$ 
can be inverted on ${\mathbb S}^2$ only on the space of non-zero spherical harmonics, {\it i.e.}
\be
\sum_{\ell = 1}^{\infty}\sum_{m=-\ell}^{\ell} \frac{Y_{\ell m}(\hat n)\, Y^*_{\ell m}(\hat n')} {\ell (\ell + 1)}
= -\frac{1}{4\pi} \,[\ln (1-z) +1] = -\frac{1}{4\pi} \ln \left(1 - \hat n \cdot \hat n'\right) 
+ \frac{ \ln 2 -1}{4\pi}\,,
\ee
with the $\ell = 0$ constant mode excluded from the sum. Thus the fields described in the
full de Sitter space by the correlation function (\ref{Dlog}) as zero conformal weight fields
are also zero conformal weight fields with respect to the conformal group on ${\mathbb S}^2$,
when restricted to the cosmological horizon of de Sitter space.

\section{Exact Formulae for the Bispectral Shape Function (\ref{bispectrum})}
\label{App:Bispectral}

The Fourier transform  of the three-point function (\ref{G3form}) is
\be
\tilde G_3({\vec k_1},{\vec k_2},{\vec k_3};w) = C_3(w)\int d^3\vec x_1\,d^3\vec x_2\,d^3\vec x_3 \,
e^{i({\vec k_1}\cdot{\vec x_1} + {\vec k_2}\cdot{\vec x_2} + {\vec k_3}\cdot{\vec x_3})}
|{\vec x_1}-{\vec x_2}|^{-w} |{\vec x_2}-{\vec x_3}|^{-w} |{\vec x_3}-{\vec x_1}|^{-w}\,. \label{G3Fourier}
\ee
\noindent 
Using the Fourier representation of $|{\vec x}|^{-w}$ in $d=3$ dimensions (for $w \neq 3$)
\be
|{\vec x}|^{-w} = B_3(w)(2\pi)^{-3}\int d^3\vec p\, |{\vec p}|^{w - 3} e^{-i{\vec p}\cdot{\vec x}}\,,\qquad
{\rm with}\qquad B_3(w) = 2^{3-w}\pi^{\frac{3}{2}} \frac{\Gamma(\frac{3}{2} -\frac{w}{2})}{\Gamma(\frac{w}{2})}\,,
\label{xtow}
\ee
defined by (\ref{Bdint})-(\ref{Bd}) of the text, we may easily perform all three of the $\vec x_i$ integrals and two 
of the momentum integrals to obtain
\be
\tilde G_3({\vec k_1},{\vec k_2},{\vec k_3};w) = C_3(w)\,\left[B_3(w)\right]^3\,\delta^{(3)}({\vec k_1}+{\vec k_2}+{\vec k_3})
 \int d^3p\, |{\vec p}|^{w-3} |{\vec p}-{\vec k_1}|^{w-3} |{\vec p}+{\vec k_2}|^{w-3}\,. 
\label{G3cora}
\ee
which is (\ref{G3corr}) of the text.  A closed form expression for the momentum integral in (\ref{G3cora}) 
may be given in terms of a generalized hypergeometric function of two variables $F_4$, known as Appell's 
fourth function, which is defined by the double series \cite{Bailey}
\be
F_4(\alpha, \beta; \gamma, \gamma'; X, Y)  = \sum_{m=0}^{\infty} \sum_{n=0}^{\infty} 
\frac{(\alpha)_{m + n} (\beta)_{m+n}}{(\gamma)_m\, (\gamma')_n}\, \frac{X^m}{m!}\, \frac{Y^n}{n!}\,,
\label{F4def}
\ee
for $|X|^{\frac{1}{2}} + |Y|^{\frac{1}{2}} < 1$, and for other values by analytic continuation. In (\ref{F4def})
\be
(a)_m \equiv \frac{\Gamma(a+m)}{\Gamma(a)}
\ee
denotes the Pochhammer symbol. Making use of the result of ref. \cite{Davydychev} and
taking into account the factors in (\ref{G3corr}), the non-Gaussian shape function
defined there may be written in closed form as 
\bea
&&\hspace{1cm} S( X, Y; w) = S_1(w)\left\{ 
\Gamma\left(3 \! - \! \frac{3w}{2}\right)\, 
F_4\!  \left(\frac{3}{2}\!  - \! \frac{w}{2}, 3 \! - \! \frac{3w}{2}; \frac{5}{2}\!  - \! w, \frac{5}{2}\!  - \! w; X, Y\right) \right. +\nn
&& \hspace{-2mm}\left.\Gamma\! \left(w\! -\!  \frac{3}{2}\right) X^{w \! - \! \frac{3}{2}}\, 
F_4\! \left(  \frac{3}{2}\!  - \! \frac{w}{2}, \frac{w}{2}; w\! - \! \frac{1}{2}, \frac{5}{2}\!  - \! w; X, Y\! \right)
+\Gamma\! \left(\frac{3}{2}\! -\! w\! \right) Y^{w\!  -\!  \frac{3}{2}} \,
F_4 \! \left(\frac{3}{2} \! - \! \frac{w}{2}, \frac{w}{2}; \frac{5}{2} \! - \! w, w\!  -\!  \frac{1}{2}; X, Y\! \right)\! \right\}\nn
&&\hspace{3cm} +\ S_2(w)\,  (XY)^{w - \frac{3}{2}}\,
F_4\!  \left(\frac{3w}{2}\!  - \! \frac{3}{2}, \frac{w}{2}; w\!  - \! \frac{1}{2}, w \! - \! \frac{1}{2}; X, Y\right)\,,
\label{SF4}
 \eea
with
\be
S_1(w) \equiv \frac{\pi^3 \left[\Gamma\left(w - \frac{3}{2}\right)\right]^2 \left[\Gamma\left(\frac{3}{2}- w\right)\right]^3}
{\Gamma\left(\frac{w}{2}\right)\, \left[\Gamma\left(\frac{3}{2} - \frac{w}{2}\right)\right]^2 \,\Gamma\left(\frac{3w}{2} - \frac{3}{2}\right)}\,,
\qquad\quad S_2(w) \equiv  \frac{\pi^3 \left[\Gamma\left(\frac{3}{2}- w\right)\right]^5}{\left[\Gamma\left(\frac{3}{2}- \frac{w}{2}\right)\right]^3} \,.
\ee
From this expression it is clear that only the first term has a pole singularity as $w\rightarrow 2$, the remaining three terms 
in (\ref{SF4}) being finite in that limit.

An expression for the momentum integral in (\ref{G3cora}) and hence for the shape function in terms 
of Feynman parameter integrals which is potentially more useful for numerical evaluation
may be found as follows. Noting that products of arbitrary powers of different 
factors can be represented as integrals over the Feynman parameters $u,v$ by
\bea
&&A^{-\alpha}B^{-\beta}C^{-\gamma} =
\frac{\Gamma(\alpha+\beta+\gamma)}{\Gamma(\alpha)\Gamma(\beta)\Gamma(\gamma)}
\int_0^1du \int_0^1 dv\, \times \nonumber\\ 
&& \quad\quad\quad\quad\quad\quad\quad u^{\alpha-1}(1-u)^{\beta -1}v^{\alpha+\beta-1}(1-v)^{\gamma-1}
[uvA+(1-u)vB+(1-v)C]^{-\alpha-\beta-\gamma}\,, 
 \label{Feynman3}
\eea 
and setting $A=|{\vec p}|^2$,  $B=|{\vec p}-{\vec k_1}|^2$,  $C=|{\vec p}+{\vec k_2}|^2$, and 
$\alpha=\beta=\gamma=s\equiv\frac{3-w}{2}$, the integral in (\ref{G3cora}) is
\be
\frac{\Gamma(3s)}{[\Gamma(s)]^3}\int_0^1\int_0^1du\,dv\, [u(1-u)(1-v)]^{s-1}v^{2s-1} 
I(u,v; {\vec k_1},{\vec k_2})\,,
\label{intinG3}
\ee
with
\be
I(u,v; {\vec k_1},{\vec k_2})\equiv \int d^3\vec p \,
\left[uv|\vec p|^2+(1-u) v|{\vec p}-{\vec k_1}|^2+(1-v)|{\vec p}+{\vec k_2}|^2\right]^{-3s}\,. 
\label{Fuv}
\ee
By shifting the momentum $\vec p$ this latter integral may be reduced to the form 
\be
I(u,v; {\vec k_1},{\vec k_2})=\int \frac{d^3\vec p}{(p^2 + M^2)^{3s}} = {\pi}^{\frac{3}{2}}
\frac{\Gamma(3s-\frac{3}{2})}{\Gamma(3s)}(M^2)^{-3s+\frac{3}{2}}\,,
\label{intp}
\ee
\noindent where 
\be
M^2 \equiv (1-u)v[1-(1-u)v]k_1^2 + v(1-v)k_2^2 + 2(1-u)v(1-v){\vec k_1}\cdot{\vec k_2}\,.
\label{defa2} 
\ee
\noindent 
Thus, using $\vec k_1 + \vec k_2 = -\vec k_3$, (\ref{intinG3}) becomes
\bea
&&\pi^{\frac{3}{2}}\frac{\Gamma(3s-\frac{3}{2})}{[\Gamma(s)]^3}\int_0^1du \int_0^1dv
[u(1-u)(1-v)]^{s-1}v^{\frac{1}{2}-s}\, \times \nonumber\\ 
&& \quad\quad\quad\quad\quad\left[uv(1-u)k_1^2 + u(1-v)k_2^2 + (1-u)(1-v)k_3^2\right]^{-3s + \frac{3}{2}}\,,
\label{intresult}
\eea
Changing variables $u\rightarrow 1-u$, $v\rightarrow 1-v$, inserting the result into (\ref{G3cora}) and recalling
the definition of $s=\frac{3-w}{2}$ we finally obtain the second form of (\ref{G3corr}) with the shape
function (\ref{bispectrum}). This form corrects two misprints in Ref. \cite{sky}.

\section{Differential Operators in de Sitter Space}
\label{App:Differential}

The explicit coordinate transformation between flat FLRW coordinates 
(\ref{flatRW}) and static coordinates (\ref{dSstat}) of de Sitter space is \cite{HawEll}
\bes
\bea
\tau &=& t +  \frac{1}{2H}\, \ln \left(1 - H^2r^2\right)\,\\
\varrho &\equiv &|\vec x | = \frac{r\, e^{-Ht}}{\sqrt{1-H^2r^2}}\,,
\eea 
\label{FLRWtostatic}\ees
with the inverse transformation
\bes
\bea
&& t = \tau - \frac{1}{2H} \ln \left(1 - H^2 \varrho^2 e ^{2H\tau}\right)\,,\\
&& r = a \varrho = \varrho\, e^{H \tau}\,.
\eea
\label{statictoFLRW}\ees
The Jacobian matrix of this $2 \times 2$ transformation of spacetime coordinates is
\be
\frac{\partial (t,r)}{\partial(\tau,\varrho)} \equiv \left( \begin{array}{cc}
\frac{\partial t}{\partial \tau} & \frac{\partial t}{\partial \varrho}\\
\frac{\partial r}{\partial \tau} & \frac{\partial r}{\partial \varrho}
\end{array}\right) = \left(\begin{array}{cc}
\frac{1}{1-H^2r^2} & \frac{Hr^2}{\varrho(1-H^2r^2)}\\
Hr & r/\varrho \end{array}\right)
\label{Jacob}
\ee
From these relations the transformations from flat FLRW to static de Sitter coordinates
of the various differential operators appearing in the text are easily found. We have
\bes\bea
&& \hspace{3cm}\frac{\partial}{\partial \tau} = \frac{1}{1-H^2r^2} \frac{\partial}{\partial t}
+ Hr \frac{\partial}{\partial r} \,,\\
&& \hspace{3cm}\frac{1}{a}\frac{\partial}{\partial \varrho} = \frac{H r }{1-H^2r^2} \,\frac{\partial}{\partial t}
+ \frac{\partial}{\partial r}\,,\\
&& \frac{\partial^2}{\partial \tau^2} = \frac{1}{(1-H^2r^2)^2} \frac{\partial^2}{\partial t^2}
+ \frac{2Hr}{1- H^2r^2} \frac{\partial^2}{\partial t \partial r} 
+ \frac{2H^3r^2}{(1-H^2r^2)^2} \frac{\partial}{\partial t}
+ H^2r^2 \frac{\partial^2}{\partial r^2} + H^2r \frac{\partial}{\partial r}\,,\\
&& \frac{\vec\nabla^2}{a^2} = \frac{1}{a^2}\left[\frac{1}{\varrho^2} \frac{\partial}{\partial \varrho}
\left( \varrho^2 \frac{\partial}{\partial \varrho}\right) - \frac{L^2}{\varrho^2}\right]
= \frac{H}{(1-H^2r^2)^2} \left[ Hr^2 \frac{\partial^2}{\partial t^2} 
+ (3-H^2r^2) \frac{\partial}{\partial t}\right]\nn
&& \hspace{2cm} +\ \frac{2Hr}{1-H^2r^2} \frac{\partial^2 }{\partial t\partial r}
+ \frac{1}{r^2} \frac{\partial}{\partial r} \left(r^2 \frac{\partial}{\partial r}\right)
-\frac{L^2}{r^2}\,,
\eea \label{difftrans} \ees
where
\be
-L^2 \equiv \frac{1}{\sin\theta} \frac{\partial}{\partial\theta}
\left(\sin\theta\frac{\partial}{\partial\theta}\right)
+ \frac{1}{\sin^2\theta}\frac{\partial^2}{\partial\phi^2}
\label{lapl}
\ee
is the scalar Laplacian on ${\mathbb S}^2$, with eigenvalues $-\ell(\ell+1)$.
It follows then that
\bes\bea
&&\hspace{7mm}\frac{\partial^2}{\partial \tau^2} + H\frac{\partial}{\partial \tau}
- \frac{\vec \nabla^2}{a^2} =\frac{1}{1-H^2r^2}\left( \frac{\partial}{\partial t} 
- 2H\right)\frac{\partial}{\partial t} - (1-H^2r^2)\frac{1}{r^2} \frac{\partial}{\partial r}
\left(r^2 \frac{\partial}{\partial r}\right) + \frac{L^2}{r^2}\,,\hspace{2.8cm}\\
&& \hspace{1cm}- \sq = \frac{\partial^2}{\partial \tau^2} + 3H\frac{\partial}{\partial \tau}
- \frac{\vec \nabla^2}{a^2} =
\frac{1}{1-H^2r^2}\frac{\partial^2}{\partial t^2} - \frac{1}{r^2} \frac{\partial}{\partial r}
\left[r^2(1-H^2r^2)\frac{\partial}{\partial r}\right] + \frac{L^2}{r^2}\,,\\
&&\frac{\partial^2}{\partial \tau^2} + 5H\frac{\partial}{\partial \tau}
- \frac{\vec \nabla^2}{a^2} =
\frac{1}{1-H^2r^2}\left( \frac{\partial}{\partial t} 
+  2H\right)\frac{\partial}{\partial t} - \frac{1}{r^2} \frac{\partial}{\partial r}
\left[r^2 (1-H^2r^2)\frac{\partial}{\partial r}\right] + 2H^2r\frac{\partial }{\partial r} 
+ \frac{L^2}{r^2},
\eea\label{diffopsa}\ees
and also
\bes\bea
\left(H\frac{\partial}{\partial \tau} + 2H^2 + \frac{1}{3}\frac{\vec\nabla^2}{a^2}\right) u
&=& \frac{1}{3r^2}\frac{d}{dr}\left(r^2\frac{du}{dr}\right)  + H^2r \frac{du}{dr} + 2H^2 u\,,\\
\left(\frac{\partial^2}{\partial \tau^2} + 4H \frac{\partial}{\partial \tau} + 4H^2
- \frac{2}{3}\frac{\vec \nabla^2}{a^2}\right) u
&=& -\frac{\, 2\ }{3r^2}\frac{d}{dr}\left(r^2\frac{du}{dr}\right) + 
H^2r^2 \frac{d^2 u}{dr^2} + 5 H^2 r \frac{d u}{dr}+ 4H^2u\,,\hspace{1cm}
\eea\label{solopsa}\ees
when restricted to functions $u = u(r)$ that depend only on the static radial coordinate $r$.

\end{document}